\documentclass[aps,prx,twocolumn,groupedaddress,longbibliography]{revtex4-1}
\usepackage{amsmath}
\usepackage{amssymb}
\usepackage{mathrsfs}
\usepackage[pdftex,colorlinks=true,urlcolor=blue,linkcolor=blue,citecolor=blue,breaklinks=true]{hyperref}
\usepackage{graphicx}
\usepackage{float}
\usepackage[abs]{overpic}
\usepackage[authormarkup=none,final
]{changes}
\definechangesauthor[name={ShovanA},color=blue]{SA}
\definechangesauthor[name={ShovanD},color=red]{SD}

\begin{document}

\title{Coherent generation of photonic fractional quantum Hall states in a cavity \\and the search for anyonic quasiparticles}

\author{Shovan Dutta}
\email[E-mail: ]{sd632@cornell.edu}
\author{Erich J. Mueller}
\email[E-mail: ]{em256@cornell.edu}
\affiliation{Laboratory of Atomic and Solid State Physics, Cornell University, Ithaca, New York 14853, USA}

\date{\today}

\begin{abstract}
We present and analyze a protocol in which polaritons in a noncoplanar optical cavity form fractional quantum Hall states. We model the formation of these states and present techniques for subsequently creating anyons and measuring their fractional exchange statistics. In this protocol, we use a rapid adiabatic passage scheme to sequentially add polaritons to the system, such that the system is coherently driven from $n$- to $(n+1)$-particle Laughlin states. Quasiholes are created by slowly moving local pinning potentials in from outside the cloud. They are braided by dragging the pinning centers around one another, and the resulting phases are measured interferometrically. The most technically challenging issue with implementing our procedure is that maintaining adiabaticity and coherence requires that the two-particle interaction energy $V_0$ be sufficiently large compared to the single-polariton decay rate~$\gamma$, $V_0 /\gamma \gg 10 N^2 \ln N$, where $N$ is the number of particles in the target state.\deleted[id=SD]{ Current experiments have $V_0 /\gamma\sim 50$.}\added[id=SA]{ While this condition is very demanding for present-day experiments where $V_0 /\gamma\sim 50$, our protocol presents a significant advance over the existing protocols in the literature.}
\end{abstract}

\maketitle

\section{\label{intro}Introduction}
Fractional quantum Hall (FQH) states are the iconic examples of strongly correlated topological phases. They arise from a delicate interplay between interactions and magnetic field in a two-dimensional (2D) electron gas~\cite{stormer1999fractional, stormer1999nobel, wen1995topological}. Both theory~\cite{arovas1984fractional, halperin1984statistics, stern2008anyons} and experiments~\cite{camino2005realization, willett2010alternation, an2011braiding} suggest that they possess ``anyonic'' quasiparticle excitations with fractional statistics, which could provide the building blocks for fault-tolerant quantum computation~\cite{kitaev2003fault, nayak2008non}. In recent years, synthetic quantum materials~\cite{georgescu2014quantum, bloch2008many, lewenstein2007ultracold, bloch2012quantum, carusotto2013quantum, noh2016quantum, hartmann2016quantum, houck2012chip} have rapidly emerged as a promising platform to engineer FQH states, especially bosonic Laughlin states~\cite{laughlin1983anomalous, wilkin2000condensation, paredes20011, kapit2012non, cooper2001quantum}. Two leading platforms are ultracold neutral atoms~\cite{wilkin2000condensation, paredes20011, kapit2012non, cooper2001quantum, paredes2003fractional, popp2004adiabatic, chang2005composite, baranov2005fractional, sorensen2005fractional, palmer2006high, hafezi2007fractional, bhat2007hall, baur2008stirring, moller2009composite, gemelke2010rotating, kapit2010exact, roncaglia2011rotating, julia2012fractional, nielsen2013local, cooper2013reaching, zhang2016creating, he2017realizing} and cavity photons~\cite{cho2008fractional, hayward2012fractional, maghrebi2015fractional, anderson2016engineering, umucalilar2013many, umucalilar2014probing, umucalilar2012fractional, hafezi2013non, umucalilar2017generation, kapit2014induced, grusdt2014topological}. Unfortunately, as we describe below, technical issues have so far prevented the realization of these aspirations. Here we describe a simple protocol which overcomes many of the hurdles. It will allow experimentalists to coherently produce particle-number-resolved $\nu=1/2$ Laughlin states in a high-finesse optical cavity using techniques that have already been demonstrated~\cite{ozawa2018topological, lu2014topological, chang2014quantum, schine2015synthetic, jia2017strongly, ningyuan2017photons}. We additionally show how one can create quasiholes in a Laughlin state that are bound to external laser potentials. We model a scheme for interferometrically measuring the braiding phase when two such quasiholes are moved around one another~\cite{grusdt2016interferometric}. This procedure not only yields the quasiparticle exchange statistics, but is also a prototype of the externally controlled braiding needed for topological quantum computation.

Photonic systems offer unique features particularly suited for quantum information processing---fast dynamics, long coherence times, versatile optical in-out coupling, and ease of transmission over communication channels~\cite{northup2014quantum, kimble2008the, kok2007linear, monroe2002quantum}. These features are also useful for preparing interesting many-body states. However, in conventional nonlinear media, photons interact too weakly with one another to establish strong enough correlation to produce FQH states. Nonetheless, some nontrivial quantum states, such as a thermal Bose-Einstein condensate, have been produced~\cite{klaers2010bose, marelic2015experimental, damm2016calorimetry, marelic2016spatiotemporal, damm2017first}. The strong-coupling limit can be reached by resonantly coupling the light to matter and using the matter-matter interactions to mediate the photon-photon interactions~\cite{hartmann2016quantum, chang2014quantum, lukin2003colloquium}. Such mediated interactions have been demonstrated in both optical and microwave domains. Optical experiments have confined the light via macroscopic cavities~\cite{jia2017strongly, birnbaum2005photon} or photonic structures~\cite{bajcsy2009efficient, fushman2008controlled, sun2017direct}. The interactions have been mediated by atoms~\cite{birnbaum2005photon, bajcsy2009efficient}, quantum dots~\cite{fushman2008controlled}, semiconductor excitons~\cite{sun2017direct, tassone1999exciton}, or Rydberg-dressed atoms~\cite{jia2017strongly, firstenberg2016nonlinear}. Microwave experiments typically use resonating circuits and superconducting qubits~\cite{lang2011observation, roushan2016chiral}.

In addition to strong interactions, creating FQH states requires a magnetic field. Generating effective magnetic fields for photons is nontrivial. Nonetheless, by employing clever cavity designs to modify the photon dispersion~\cite{ozawa2018topological, lu2014topological}, experiments have created synthetic gauge fields in ``twisted'' optical cavities~\cite{schine2015synthetic, ningyuan2017photons}, microwave cavity arrays~\cite{roushan2016chiral, owens2017quarter}, radio-frequency circuits~\cite{ningyuan2015time}, and solid-state photonic devices~\cite{wang2009observation, hafezi2013imaging, rechtsman2013photonic, li2014photonic, rechtsman2013strain, tzuang2014non, mittal2014topologically}. These developments have set the stage to explore FQH physics in a single optical cavity~\cite{schine2015synthetic, jia2017strongly, ningyuan2017photons} or in a lattice of coupled microwave resonators~\cite{roushan2016chiral, owens2017quarter}.

\begin{figure}
\includegraphics[width=\columnwidth]{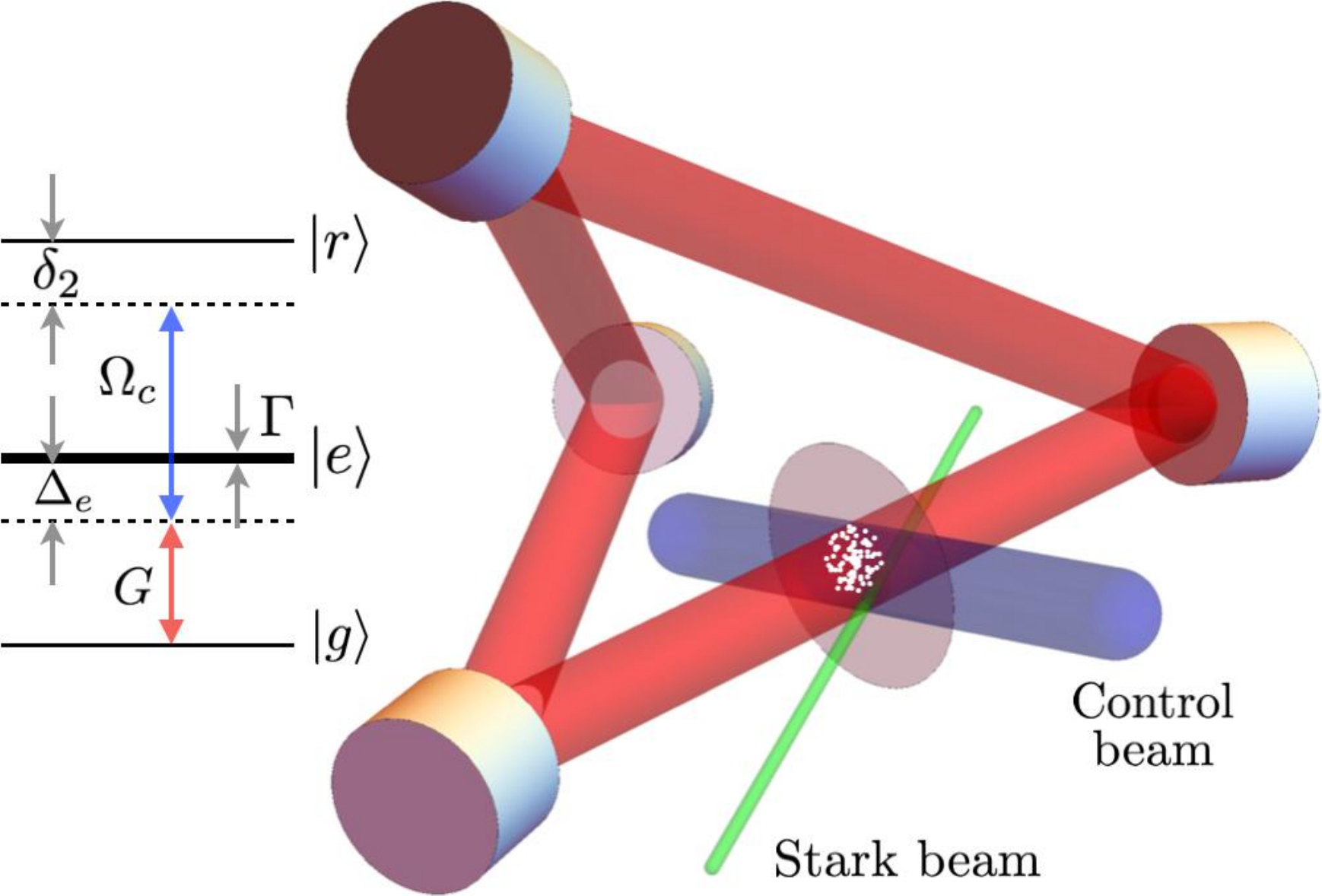}
\caption{\label{setup}Schematic of the proposed cavity experiment. The noncoplanar arrangement of the mirrors gives rise to an effective magnetic field for the transverse dynamics of the cavity photons (red beam). These photons are coupled to atoms (white dots) loaded into a transverse plane (transparent disk) of the cavity. As shown in the energy level diagram, the photons couple the ground state $|g\rangle$ and an excited state $|e\rangle$ of the atoms with a collective Rabi frequency $G$. The excited state has a single-photon detuning $\Delta_e$ and lifetime $1/\Gamma$ and is coupled to a metastable Rydberg level $|r\rangle$ by a control laser (blue beam) with Rabi frequency $\Omega_c$. When the two-photon detuning $\delta_2$ is smaller than a linewidth, long-lived Rydberg polaritons form which inherit the transverse photon dynamics and interact strongly with one another. In Sec.~\ref{preparation} we describe how one can drive these polaritons to form $\nu=1/2$ Laughlin states in the transverse plane. An additional laser (green beam) can be used to produce a localized potential for the polaritons via the ac Stark shift. As we show in Secs.~\ref{holegeneration} and \ref{braiding}, by moving such potentials relative to the polariton cloud, one can create and braid anyonic quasihole excitations in a Laughlin state.}
\end{figure}

As in Ref.~\cite{umucalilar2017generation}, we consider a near-degenerate cavity setup, similar to the one used to observe photonic Landau levels in Ref.~\cite{schine2015synthetic} and shown schematically in Fig.~\ref{setup}. Because of the noncoplanar mirror geometry in such a twisted cavity, the transverse light field obeys a 2D Schr\"{o}dinger equation with an effective magnetic field (see Sec.~\ref{system}). One can induce strong photon-photon interactions by loading $^{87}$Rb atoms into a transverse plane of the cavity and illuminating them with a control beam that resonantly couples the cavity photons to a long-lived highly excited atomic state~\cite{sommer2015quantum}. Experiments have demonstrated that the resulting Rydberg polaritons are both long lived~\cite{ningyuan2016observation} and strongly interacting~\cite{jia2017strongly}.

Initial theoretical proposals to construct FQH phases in single-cavity~\cite{umucalilar2013many, umucalilar2014probing} and coupled-cavity~\cite{umucalilar2012fractional, hafezi2013non} setups employed a monochromatic drive to excite Laughlin states via multiphoton resonances. These proposals produce states with very small overlap with the desired Laughlin state, and the overlaps fall off exponentially with the number of photons in the target state~\cite{umucalilar2014probing}. More sophisticated schemes have been proposed recently which use frequency-selective incoherent pumps~\cite{umucalilar2017generation, kapit2014induced, ma2017autonomous} or alternate flux insertions and coherent pumping~\cite{grusdt2014topological}. Unfortunately, even these\deleted[id=SD]{ very} complex approaches are lacking. For example, the scheme in Ref.~\cite{umucalilar2017generation} yields at best\deleted[id=SD]{ an 80\%}\added[id=SA]{ a 70\%} overlap with the $N=3$ Laughlin state. Using such a scheme to produce states with more particles seems impractical.

Here we describe a simpler and more effective protocol whereby one can reliably produce $N$-particle Laughlin states with high fidelity in a twisted optical cavity. As we explain in Sec.~\ref{preparation}, this is achieved by using rapid adiabatic passage ideas to sequentially transfer the state of the cavity from $n$- to $(n+1)$-particle Laughlin states [Fig.~\ref{rapanderror}(a)]. For adiabaticity, the duration of each transfer, $T$, must be large compared to the inverse of the many-body level splittings. These splittings are proportional to the two-particle interaction energy $V_0$, and, for the Laughlin states, the splittings are nearly independent of the particle number. Thus, one finds that the accumulated error scales as $N e^{-\xi}$, where $\xi \propto V_0 T$. For our protocol to be successful, the experiment must also be faster than the coherence time set by polariton loss from the cavity. Hence, the key technical impediment to implementing our scheme, which is also present in the earlier proposals~\cite{umucalilar2017generation, umucalilar2013many, umucalilar2014probing}, is engineering a sufficiently large ratio between the interaction strength and the single-polariton decay rate $\gamma$. In particular, for a high-fidelity generation of the $N$-particle Laughlin state, we require $V_0 / \gamma \gg 10 N^2 \ln N$. Since current experimental setups yield  $V_0 / \gamma \sim 50$~\footnote{See the Supplementary Information of Ref.~\cite{schine2015synthetic}
}, only the smallest $N$ states may be reliably produced. While nontrivial, it is reasonable to expect this figure of merit will increase in the next few years, enabling the creation of higher $N$ states.\added[id=SA]{ Despite this limitation, our protocol presents a significant step forward, as the existing protocols either have low fidelity ($\sim 0.01$ for $N=2$)~\cite{umucalilar2013many, umucalilar2014probing} or require prohibitively large values of $V_0 / \gamma$ ($\sim 3\times 10^4$ for $N=3$) \cite{umucalilar2017generation}.}

A central motivation for preparing a Laughlin state is to observe anyonic statistics by creating quasiparticles and braiding them~\cite{umucalilar2013many}. In Sec.~\ref{holegeneration} we show that one can generate quasiholes by adiabatically bringing in localized repulsive potentials through the edge of the cloud. These potentials can be engineered through the dynamical Stark shift from tightly focused lasers~\cite{hayat2012dynamic, li2013entanglement, cancellieri2014ultrafast} (see Fig.~\ref{setup}). The cavity geometry can be tuned to eliminate the excitation of surface modes (Fig.~\ref{hole-trapandU0}). Subsequently, one can drag the pinning potentials around one another to perform quasihole braiding~\cite{kapit2012non}, which we discuss in Sec.~\ref{braiding}. We find that both quasihole generation and braiding can be implemented with high fidelity over much shorter time scales than the preparation of the Laughlin state. In Sec.~\ref{statistics} we put forward an interferometric scheme to measure the braiding phase and extract the fractional exchange statistics (Fig.~\ref{statisticalphase}). We conclude with a summary and outlook in Sec.~\ref{summary}.

While our analysis is focused on Rydberg polaritons in optical cavities, nearly identical modeling applies to exciton-polaritons in semiconductors~\cite{carusotto2010feshbach, sun2017direct, tassone1999exciton, carusotto2013quantum, noh2016quantum, umucalilar2017generation}. Brief estimates of the energy scales suggest that our ideas are readily transferable to that domain.

\section{\label{system}The physical system}

\subsection{\label{transverse dynamics}Overview of polariton dynamics}
We envision the ``twisted'' cavity setup of Ref.~\cite{schine2015synthetic}, shown schematically in Fig.~\ref{setup}. The cavity is nearly degenerate; i.e., the transverse dynamics are much slower than the longitudinal dynamics. In this limit, an effective equation can be derived for the transverse field profile within the cavity. This equation is identical to the Schr\"{o}dinger equation for a 2D harmonically trapped charged particle in a uniform magnetic field. In Ref.~\cite{sommer2016engineering} they gave an intuitive derivation of this mapping by tracing the coordinates of a light ray as it repeatedly intersects a transverse plane within the cavity. One thereby constructs a dynamical map which describes the stroboscopic evolution of the transverse position and wave vector of a light ray. The latter plays the role of momentum. In the paraxial approximation, this map is linear and is equivalently generated by a 2D quadratic Hamiltonian. Quantizing this Hamiltonian yields the desired Schr\"{o}dinger equation. In the case of a planar cavity with flat mirrors, the dynamics map onto those of a free particle of mass $m_{\text{ph}} = \hbar \omega_0/c^2$ where $\omega_0$ is the frequency of the longitudinal mode and $c$ is the speed of light. The mirror curvature confines the light in the transverse direction, leading to a harmonic trapping potential. The deviation from a planar geometry rotates the light field about the longitudinal axis, which gives rise to Coriolis and centrifugal forces in the transverse plane. The former acts as a uniform magnetic field perpendicular to the plane. Thus, the twisted cavity realizes a Fock-Darwin Hamiltonian~\cite{darwin1927free, fock1928bemerkung} describing massive, trapped particles in two dimensions experiencing a uniform magnetic field. The effective photon mass, trap frequency, and magnetic field strength can be controlled independently by adjusting the cavity geometry.

Strong interactions can be introduced into the system by coupling the photons to an atomic ensemble in a Rydberg electromagnetically induced transparency (EIT) configuration~\cite{firstenberg2016nonlinear, fleischhauer2005electromagnetically}, as discussed in detail in Ref.~\cite{sommer2015quantum} and illustrated in Fig.~\ref{setup}. A thin layer of laser-cooled atoms is loaded into the cavity waist. The cavity photons couple the atomic ground state to an intermediate excited state $|e\rangle$ which is in turn coupled to a metastable Rydberg level $|r\rangle$ by a strong control laser. This light-matter coupling yields two ``bright'' and one ``dark'' polariton modes~\cite{fleischhauer2005electromagnetically, fleischhauer2000dark}. Near EIT resonance, the dark polariton mode has a long lifetime and represents a superposition of a collective Rydberg excitation and a cavity photon. The bright polariton modes, on the other hand, are short lived. For strong coupling, the splitting between the dark and bright modes is large compared to the energy scales of the transverse photon dynamics and Rydberg-Rydberg interactions. Then the problem reduces to describing the motion of dark polaritons in the cavity waist, which inherit the single-particle dynamics of photons and the interactions of Rydberg atoms.

\subsection{\label{single-particle hamiltonian}Single-particle Hamiltonian}
Projecting the 2D photon Hamiltonian onto the dark-polariton manifold renormalizes the photon mass and trap frequency, yielding the single-particle Hamiltonian
\begin{equation}
\label{H0}
\hat{H}_0 \hspace{-0.05cm}=\hspace{-0.05cm} \int\hspace{-0.05cm} d^2 r\hspace{0.05cm} \hat{\psi}^{\dagger} (\vec{r}) \bigg[\frac{(-i\hspace{0.02cm} \vec{\nabla} - M \omega_B r \hspace{0.02cm}\hat{\varphi})^2}{2M} + \frac{1}{2} M \omega_T^2 r^2 \bigg] \hat{\psi}(\vec{r})\hspace{0.05cm},
\end{equation}
where \smash{$\hat{\psi}(\vec{r})$} denotes the bosonic polariton field operator, $M$ and $\omega_T$ are the effective polariton mass and trap frequency, $\omega_B$ denotes half the cyclotron frequency, and \smash{$\hat{\varphi}$} is the unit vector in the azimuthal direction. Here we have explicitly used a symmetric-gauge vector potential to represent the uniform magnetic field and set $\hbar=1$. The cyclotron frequency $2\omega_B$ sets the energy gap between Landau levels and is typically a few GHz~\cite{schine2015synthetic}. This is much faster than the motion of polaritons, so the dynamics are confined to the lowest Landau level. The polariton mass $M$ is related to the photon mass $m_{\text{ph}}$ and the collective Rabi frequencies $G$ and $\Omega_c$ of the atomic transitions (see Fig.~\ref{setup}) via $M = m_{\text{ph}} / \cos^2\theta$, where $\theta \equiv \tan^{-1}(G/\Omega_c)$~\cite{sommer2015quantum}. For typical experimental parameters, $m_{\text{ph}} \sim 2 \times 10^{-5} m_e$~\cite{schine2015synthetic} and $\theta \approx 60^{\circ}$~\cite{ningyuan2016observation, jia2017strongly}, we get $M \sim 10^{-4} m_e$, where $m_e$ is the electron mass. Similarly, the trap frequency seen by polaritons is related to that seen by photons via $\omega_T = \omega_{T,\text{ph}} \cos^2\theta$, where $\omega_{T,\text{ph}}$ is calculated from the cavity geometry~\cite{Note1, sommer2016engineering}. This frequency was varied from zero to several tens of MHz in Ref.~\cite{schine2015synthetic} by changing the mirror separation. As we will see in Sec.~\ref{holegeneration}, one needs a finite $\omega_T$ in order to adiabatically produce quasiholes without exciting edge modes.

\subsection{\label{interaction hamiltonian}Interaction Hamiltonian}
Rydberg atoms interact through a strong dipole-dipole coupling of the form $V(r) = C_6/r^6$~\cite{beguin2013direct}. This leads to strong polariton-polariton interactions which are most simply modeled by a hard core of radius $r_b$, known as the ``blockade radius''~\cite{sommer2015quantum, bienias2014scattering}. For typical experimental conditions, $r_b$ is several micrometers and can be varied using the scaling $r_b \propto n^{11/6}$, where $n$ is the principal quantum number of the Rydberg state $|r\rangle$~\cite{firstenberg2016nonlinear, peyronel2012quantum}. For mean polariton separations larger than $r_b$, the interaction can be further approximated by a contact potential. In current experiments this regime can be reached for a few tens of polaritons by controlling the cavity waist radius $w \equiv \sqrt{2/(M \omega_B)}$ which sets the average polariton separation~\cite{Note1}. Under this approximation, we can write the interaction Hamiltonian
\begin{equation}
\label{Hint}
\hat{H}_{\text{int}} \hspace{-0.05cm}=g \hspace{-0.02cm} \int\hspace{-0.05cm} d^2 r\hspace{0.05cm} \hat{\psi}^{\dagger} (\vec{r}) \hat{\psi}^{\dagger} (\vec{r}) \hat{\psi}(\vec{r}) \hat{\psi}(\vec{r})\hspace{0.05cm},
\end{equation}
where $g$ is the effective interaction strength which depends on $C_6$ as well as the EIT parameters $\Omega_c$, $\Delta_e$, and $\Gamma$, where $\Omega_c$ denotes the Rabi frequency of the control laser, $\Delta_e$ is the detuning to the excited state $|e\rangle$, and $\Gamma$ is the decay rate of $|e\rangle$. The Rydberg-Rydberg interactions are generically inelastic, which can be modeled by taking $g$ complex. The imaginary part can, in principle, be made arbitrarily small by increasing both $\Omega_c$ and $\Delta_e$ while keeping the ratio $\Delta_e/\Omega_c\approx 0.25$~\cite{Note1}. Thus, we limit ourselves to real values of $g$ in this paper.

\subsection{\label{single-particle spectrum}Single-particle spectrum}
Combining Eqs.~\eqref{H0} and \eqref{Hint} we find the many-body Hamiltonian \smash{$\hat{H} = \hat{H}_0+\hat{H}_{\text{int}}$}. The single-particle spectrum in the absence of a trap consists of degenerate Landau levels separated by the cyclotron frequency $2 \omega_B$. The lowest Landau level (LLL) is spanned by angular momentum eigenstates $\phi_m(\vec{r}) \propto r^m e^{i m\varphi} \exp{(-r^2/w^2)}$ with $m=0,1,2,\dots$. The harmonic trap splits the energies of these states and rescales the wave functions, yielding new eigenstates $\phi_m(\vec{r}) \propto z^m \smash{e^{-|z|^2/2}}$ with energies $\epsilon_m = \omega_{\text{eff}}+ m\hspace{0.03cm} \varepsilon$, where $z \equiv r\hspace{0.02cm} e^{i \varphi}/l$, $l \equiv \smash{1/\sqrt{M \omega_{\text{eff}}}}$, $\omega_{\text{eff}} \equiv \smash{\sqrt{\omega_B^2 + \omega_T^2}}$, and $\varepsilon \equiv \omega_{\text{eff}} - \omega_B$. Note that $ | \phi_m (\vec{r}) |^2 $ is peaked at $ r = \sqrt{m}\hspace{0.05cm} l$ and has a width $\Delta r \sim l $. In the absence of interactions, the energy of a many-body state in the LLL depends only on the total particle number $N$ and total angular momentum $L$. A generic noninteracting eigenstate takes the form of a Gaussian times a symmetric polynomial in the coordinates $z_1$, $z_2$, $\dots$, $z_N$ representing the positions of the $N$ particles. Interactions split this degeneracy.

\subsection{\label{laughlin state}Laughlin states}
An exact $N$-particle eigenstate of the Hamiltonian is the $\nu=1/2$ Laughlin state~\cite{paredes20011}
\begin{equation}
\label{Laughlin}
\Phi_N (z_1,z_2,\dots,z_N) \propto \prod_{j<k} (z_j - z_k)^2\hspace{0.05cm} e^{-\sum_i |z_i|^2 /2}\hspace{0.05cm},
\end{equation}
which is composed of single-particle states in the LLL with $m = 0,1,\dots, 2(N-1)$. It has zero interaction energy since the wave function vanishes whenever two particles coincide. Further, it is an angular momentum eigenstate with $L = N(N-1)$ and has energy $E_N = N\omega_{\text{eff}} + N(N-1)\hspace{0.02cm}\varepsilon$. As we will see below, the Laughlin state is the lowest-energy $N$-particle state in the $L=N(N-1)$ manifold. Therefore, one way to excite $|\Phi_N\rangle$ is to pump on the single-particle mode with angular momentum $N-1$ and frequency $\omega_{\text{eff}} + (N-1)\hspace{0.02cm}\varepsilon$, which is the essence of the multiphoton resonance protocols proposed in Refs.~\cite{umucalilar2013many, umucalilar2014probing}. However, as discussed in Sec.~\ref{intro}, this approach produces an exponentially small overlap with $|\Phi_N\rangle$ due to the coupling with other many-body states. Here we circumvent this problem by employing a rapid adiabatic passage protocol which drives the system from $|\Phi_0\rangle \to |\Phi_1\rangle \to \dots \to |\Phi_N\rangle$ through a sequence of frequency sweeps. Physically, the transition from $|\Phi_n\rangle$ to $|\Phi_{n+1}\rangle$ is implemented by adding a particle with angular momentum $m=2n$ while maintaining the strong correlation in Eq.~\eqref{Laughlin}.

\subsection{\label{projection}Projection to lowest Landau level}
The efficiency of our drive mechanism is limited by the energy splittings between the Laughlin states and the neighboring many-body states. To quantify this efficiency, we assume that the dynamics are confined to the LLL, as in Refs.~\cite{umucalilar2013many, umucalilar2014probing, umucalilar2017generation}, and consider states within that manifold. To ensure that the LLL is spectrally well resolved from the higher Landau levels, we need to take $\varepsilon \ll 2\omega_B/m_{\text{max}}$, where the occupied single-particle states all have $m\leq m_{\text{max}}$. For Laughlin states $|\Phi_N\rangle$, $m_{\text{max}} = 2(N-1)$, so this requirement becomes $ N \ll \omega_B / \varepsilon \approx 2 (\omega_B / \omega_T)^2 $. For typical experiments, $(\omega_B/\omega_T)^2 \gtrsim 10^4$~\cite{schine2015synthetic}, so this requirement is not particularly limiting. Further, we assume that there is no Landau level mixing from interactions. Typical interaction energies between two particles in the LLL can be estimated from the zeroth Haldane pseudopotential \smash{$V_0 = g \langle \phi_0 | \delta(\hat{\vec{r}}) | \phi_0 \rangle = g/(\pi l^2)$}~\cite{umucalilar2014probing, haldane1983fractional}. Hence, our assumption is justified provided $V_0 \ll 2\omega_B$, which is indeed fulfilled in present-day experimental conditions, where $V_0$ is several MHz and $\omega_B \sim 1$ GHz~\cite{Note1}.

We project the dynamics onto the LLL by substituting $\hat{\psi}(\vec{r}) = \sum_{m=0}^{\infty} \phi_m (\vec{r}) \hspace{0.02cm}\hat{a}_m$ into Eqs.~\eqref{H0} and \eqref{Hint}, where $\hat{a}_m$ annihilates a particle in the state $|\phi_m\rangle$. Thus, we obtain the restricted Hamiltonian
\begin{equation}
\label{HLLL}
\hat{H}_{\text{LLL}} = \omega_{\text{eff}}\hspace{0.02cm} \hat{N} + \varepsilon\hspace{0.03cm} \hat{L} + V_0 \sum_{s=0}^{\infty} 2^{-(s+1)} \hat{A}_s^{\dagger} \hat{A}_s\hspace{0.05cm},
\end{equation}
where $\hat{N} \equiv \sum_{m=0}^{\infty} \hat{a}_m^{\dagger} \hat{a}_m$ and $\hat{L} \equiv \sum_{m=0}^{\infty} m \hspace{0.02cm}\hat{a}_m^{\dagger} \hat{a}_m$ measure the total particle number $N$ and total angular momentum $L$, respectively, and $\hat{A}_s \equiv \sum_{m=0}^{s} \sqrt{s!/[m! (s-m)!]}\hspace{0.05cm} \hat{a}_m \hat{a}_{s-m}$ annihilates two particles with net angular momentum $s$.

\subsection{\label{many-body spectrum}Many-body spectrum}

\begin{figure}
\includegraphics[width=\columnwidth]{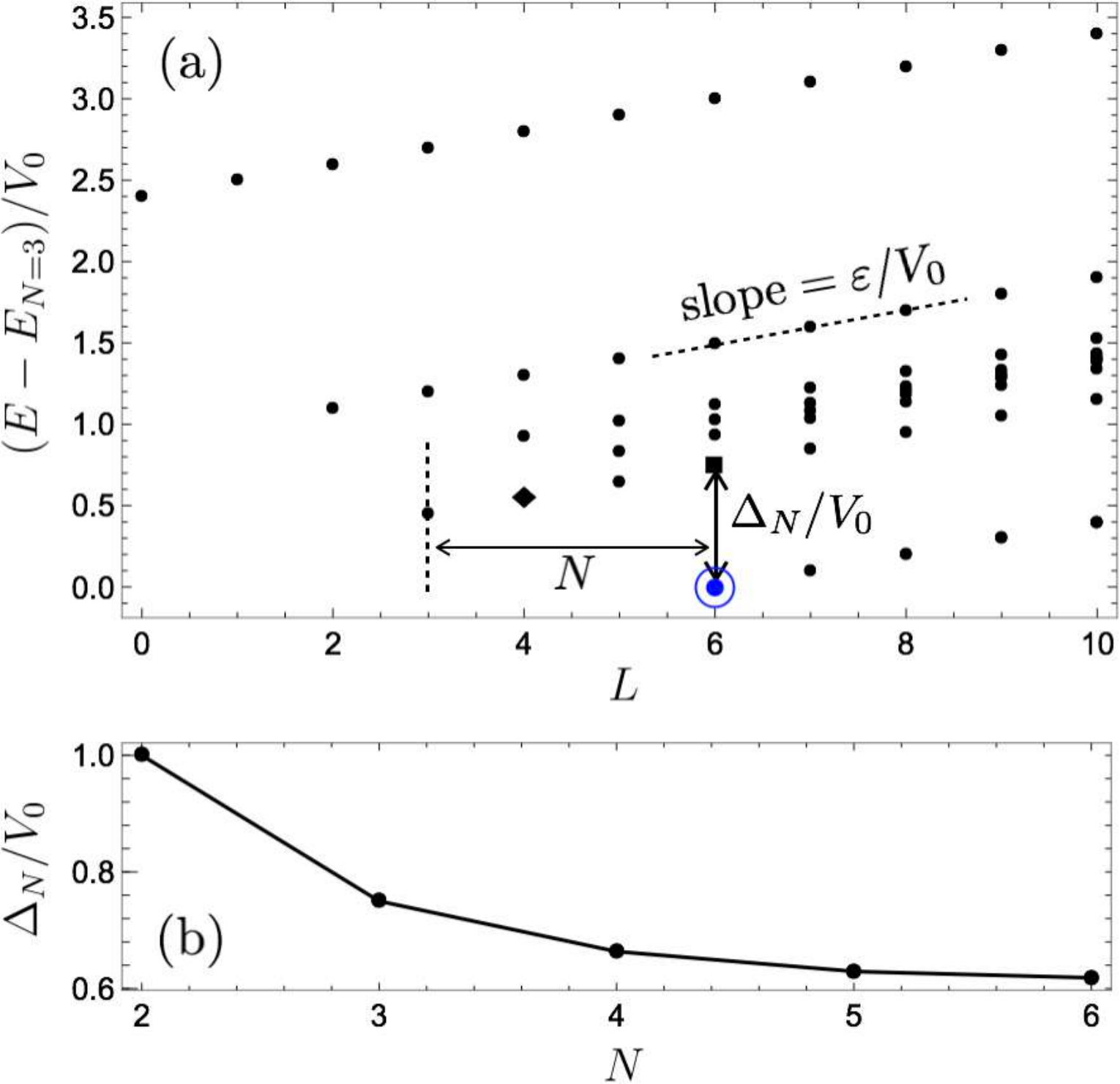}
\caption{\label{spectrumandgap}{\bf (a)} Spectrum of three polaritons in a twisted optical cavity, described by the Hamiltonian in Eq.~\eqref{HLLL}, which illustrates the general features of the $N$-body spectrum. The ground state, highlighted by the circled blue dot, has total angular momentum $L = N(N-1) = 6$ and corresponds to the $\nu=1/2$ Laughlin state $|\Phi_N\rangle$. At fixed $L$, the excitation gap from this state is $\Delta_N$, which arises from polariton-polariton interactions. The lowest-energy excitation with higher angular momentum has energy $\varepsilon$, whereas that with lower angular momentum has energy $\Delta_N - N \varepsilon$, where $\varepsilon$ is related to the harmonic confinement and the effective magnetic field. The square- and diamond-shaped dots represent the excited states $|\Phi_N^e\rangle$ and $|\Phi_N^g\rangle$ defined in Sec.~\ref{preparation}. {\bf (b)} Number dependence of the excitation gap $\Delta_N$. For $N\gtrsim 5$, it saturates at $0.6 V_0$, where $V_0$ is the interaction energy of two particles in the lowest Landau level with zero relative angular momentum (zeroth Haldane pseudopotential).}
\end{figure}

The eigenstates of $\hat{H}_{\text{LLL}}$ can be labeled by $N$ and $L$. Figure~\ref{spectrumandgap}(a) shows the spectrum in the $N=3$ manifold. The lowest-energy state with $L=N(N-1)$ is the Laughlin state $|\Phi_N\rangle$. The lowest-energy eigenstates with $L>N(N-1)$ represent quasihole and edge excitations of the Laughlin state and are degenerate with $|\Phi_N\rangle$ for $\varepsilon = 0$ (no trap)~\cite{umucalilar2017generation}. Each of these states is separated from the excited states with the same $L$ by an energy gap $\Delta_N \sim V_0$ (see Fig.~\ref{spectrumandgap}). As we will see in the next section, it is this gap which sets the maximum speed at which one can drive the system from $|\Phi_{n}\rangle$ to $|\Phi_{n+1}\rangle$. Any state with $L<N(N-1)$ also has an interaction energy $E_{\text{int}} \geq \Delta_N$. As the trap frequency is increased from zero, the eigenstate energies are simply increased by $\varepsilon L$. Consequently, there is a range of $\varepsilon$ for which the Laughlin state is the unique $N$-particle ground state and it costs energy to excite edge modes. As we describe in Sec.~\ref{holegeneration}, this energy cost will aid the adiabatic generation of quasiholes by suppressing unwanted edge excitations.

\subsection{\label{loss}Polariton loss}
A number of processes limit the polariton lifetime. First, the cavity has a finite finesse and a photon will eventually escape. Second, the atomic Rydberg states have finite lifetime, reflecting the fact that the atom can decay, emitting a photon into a noncavity mode. Third, as already discussed, the interactions between Rydberg atoms can have inelastic components and cause polariton loss. As we discussed in Sec.~\ref{interaction hamiltonian}, this latter process can be made negligible by carefully choosing parameters. In current experiments, the first two processes yield a net polariton decay rate $\gamma\sim 0.1$ MHz~\cite{Note1}.

Our protocol to create Laughlin states and braid quasiparticles relies on coherent evolution, and losing even a single polariton would be deleterious. Thus, the entire experiment must be conducted on microsecond time scales.

\section{\label{preparation}Laughlin state preparation}

\subsection{\label{laughlin overview}Overview}
Our protocol for creating the $N$-particle Laughlin state $|\Phi_N\rangle$ is based on a series of coherent optical drives which transfer the system from $|\Phi_{n}\rangle$ to $|\Phi_{n+1}\rangle$ via rapid adiabatic passage~\cite{malinovsky2001general}. Thus, the final state $ | \Phi_{N} \rangle $ is built up by sequentially injecting photons near the outer rim of the cloud. The idea of adding photons sequentially was also used in Ref.~\cite{grusdt2014topological}.

In Sec.~\ref{laughlin state} we explained that successive Laughlin states differ in their total angular momentum by $L_{n+1} - L_{n} = 2n$. Thus, in our protocol, we illuminate the cavity in state $|\Phi_{n}\rangle$ with a laser that couples strongly to the mode with $m=2n$ and sweep the detuning of the drive from negative to positive. If such a sweep is performed sufficiently slowly, the system will be adiabatically transferred to the state $|\Phi_{n+1}\rangle$. Our goal is to find the fastest possible sweep rate. We find that adiabaticity requires that the entire process take place over a time $T_L\gtrsim 40 \hspace{0.03cm} (N/V_0)\ln N$. In order to have negligible loss during this time, $T_L \ll 2/N\gamma$, where $\gamma$ is the single-polariton decay rate.

\subsection{\label{sweep protocol}Sweep protocol}
A coherent drive is expressed by the Hamiltonian
\begin{equation}
\label{Hdr}
\hspace{-0.1cm}\hat{H}_{\text{dr}} \hspace{-0.05cm}= \hspace{-0.05cm} \int \hspace{-0.05cm} d^2 r \hspace{0.03cm} \lambda(\vec{r},t) \hspace{0.02cm} \hat{\psi}^{\dagger}\hspace{-0.03cm} (\vec{r}) + \text{H.c.} \hspace{-0.02cm}=\hspace{-0.02cm} \sum_{m=0}^{\infty} \lambda_m (t)\hspace{0.02cm} \hat{a}^{\dagger}_m + \text{H.c.}\hspace{0.02cm},
\end{equation}
where $\lambda(\vec{r},t)$ denotes the optical drive field and $\lambda_m(t) \equiv \int \hspace{-0.03cm} d^2 r \hspace{0.02cm}\lambda(\vec{r},t) \hspace{0.02cm}\phi^*_m (\vec{r})$, whereby we have projected \smash{$\hat{H}_{\text{dr}}$} onto the LLL (see Sec.~\ref{single-particle spectrum}). Thus, $\lambda_m(t)$ represents the field component with a phase winding $e^{i m \varphi}$. The transition from $|\Phi_{n}\rangle$ to $|\Phi_{n+1}\rangle$ requires an optical drive with $\lambda_m \neq 0$ only for $m=2n$. Such helically phased laser beams are readily available~\cite{yao2011orbital, padgett2011tweezers}.

Thus, we consider a drive which couples $|\Phi_{n}\rangle$ to $|\Phi_{n+1}\rangle$,
\begin{equation}
\label{Hdrn}
\hat{H}^{(n)}_{\text{dr}}\hspace{-0.05cm} = \Lambda_n (t)\hspace{0.03cm} \exp\hspace{-0.05cm}\bigg[{-i\hspace{-0.03cm}\int^t \hspace{-0.1cm} d t^{\prime} [\omega_n + \delta_n (t^{\prime})]}\bigg] \hat{a}^{\dagger}_{2n} + \text{H.c.}\hspace{0.03cm},
\end{equation}
where $\omega_n$ is the resonant frequency, $\omega_n \equiv E_{n+1} - E_{n} =\omega_{\text{eff}} + 2 n \varepsilon$ (see Sec.~\ref{laughlin state}), $\delta_n (t)$ denotes the detuning which is swept from negative to positive values (or vice versa), and $\Lambda_n (t)$ is the amplitude which is controlled by the laser intensity and can be used to vary the Rabi frequency $\Omega_n(t) = \Lambda_n(t) \langle\Phi_{n+1}|\smash{\hat{a}^{\dagger}_{2n}}|\Phi_{n}\rangle$. This setup is similar to the two-state Landau-Zener problem~\cite{landau1932a, zener1932non}, where the amplitude is constant and the detuning is swept over a finite range $-\delta_{\text{max}}$ to $+\delta_{\text{max}}$ at a constant rate $\nu \equiv \partial_t \delta_n(t)$. In the Landau-Zener problem, the system will transition to the state $|\Phi_{n+1}\rangle$ provided $\delta_{\text{max}} \gg |\Omega_n| \gtrsim \sqrt{\nu}$~\cite{vitanov1996landau}. In our case, the transition probability will be modified because the coupling is not restricted to the two Laughlin states. In particular, the drive in Eq.~\eqref{Hdrn} couples any pair of states which differ in particle number by 1 and total angular momentum by $2n$. We also somewhat improve the transition probability by sculpting the profiles $\Lambda_n (t)$ and $\delta_n (t)$~\cite{shore2008coherent, vitanov2001laser, rangelov2010rapid, vitanov2015designer}.

\begin{figure}
\includegraphics[width=\columnwidth]{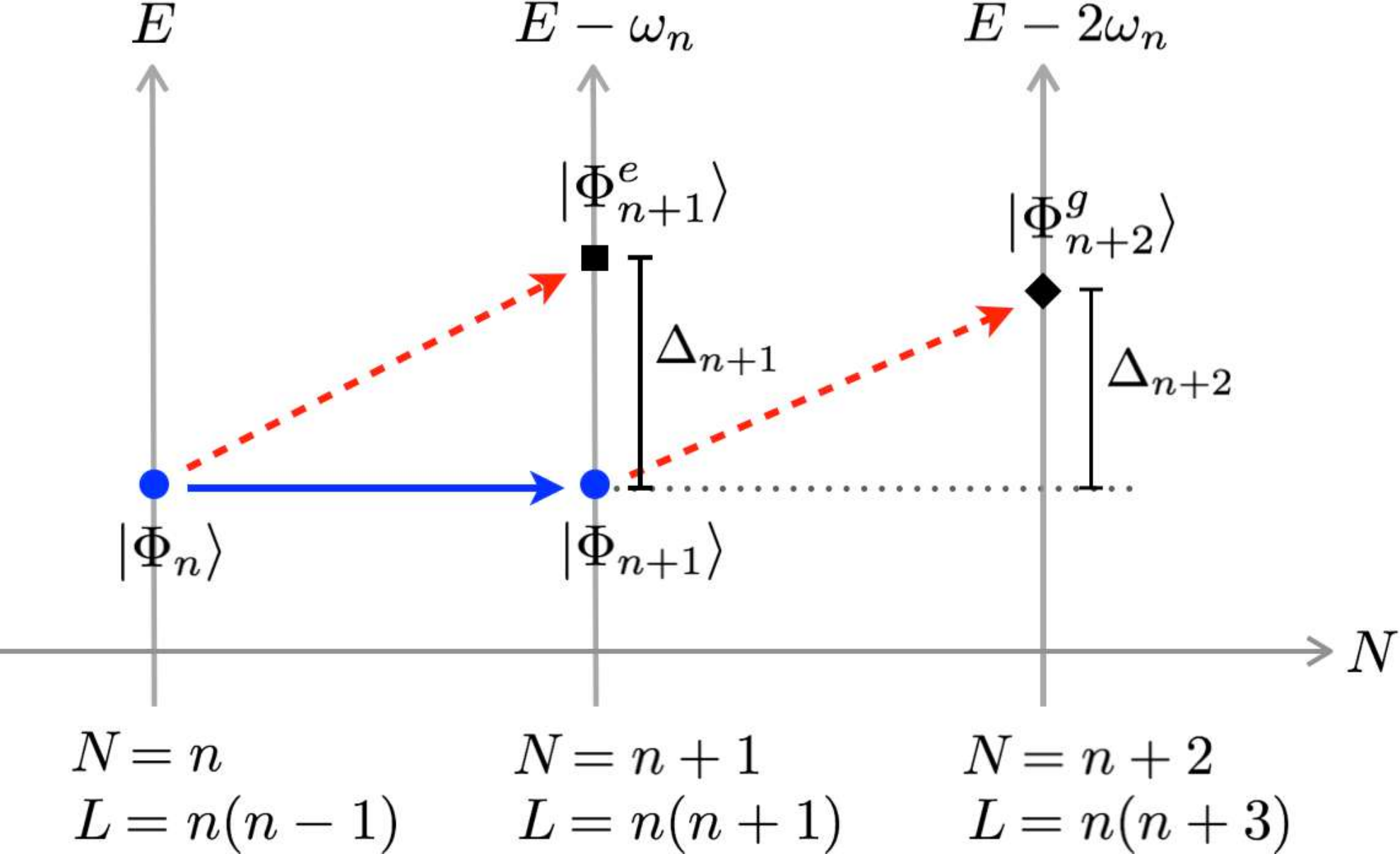}
\caption{\label{pathway}Spectrum of states coupled during our driving protocol. The solid blue arrow shows the desired transition from the $n$-particle Laughlin state $|\Phi_{n}\rangle$ to the $(n+1)$-particle Laughlin state $|\Phi_{n+1}\rangle$ with a resonant frequency $\omega_n$. Dashed red arrows show possible undesired transitions to the low-lying excited states $|\Phi_{n+1}^e\rangle$ and $|\Phi_{n+2}^g\rangle$. These transitions are off resonant by the many-body gaps $\Delta_{n+1}$ and $\Delta_{n+2}$.}
\end{figure}

The leading correction to the Landau-Zener problem comes from the in-coupled states which are closest to resonance. As sketched in Fig.~\ref{pathway}, these unwanted states are denoted by $|\Phi^{e}_{n+1}\rangle$ and $|\Phi^{g}_{n+2}\rangle$ which are the lowest-energy excited states with quantum numbers $N=n+1$, $L = n(n+1)$ and $N=n+2$, $L=n(n+3)$. The drive in Eq.~\eqref{Hdrn} couples $|\Phi_{n}\rangle$ to $|\Phi^{e}_{n+1}\rangle$ with Rabi frequency $\Omega^e_n = \Lambda_n \langle\Phi^e_{n+1}|\smash{\hat{a}^{\dagger}_{2n}}|\Phi_{n}\rangle$. Similarly, it couples $|\Phi_{n+1}\rangle$ to $|\Phi^{g}_{n+2}\rangle$ with Rabi frequency $\Omega^g_{n+1} = \Lambda_n \langle\Phi^g_{n+2}|\smash{\hat{a}^{\dagger}_{2n}}|\Phi_{n+1}\rangle$. The energy splittings of these transitions are $\omega^e_n = \omega_n + \Delta_{n+1}$ and $\omega^g_{n+1} = \omega_n + \Delta_{n+2}$, where $\Delta_n$ is the bulk excitation gap shown in Fig.~\ref{spectrumandgap}. To suppress these undesired excitations, we must have $\delta_{\text{max}}\lesssim \Delta_{n+1}, \Delta_{n+2} \sim V_0$.

\begin{figure}
\includegraphics[width=\columnwidth]{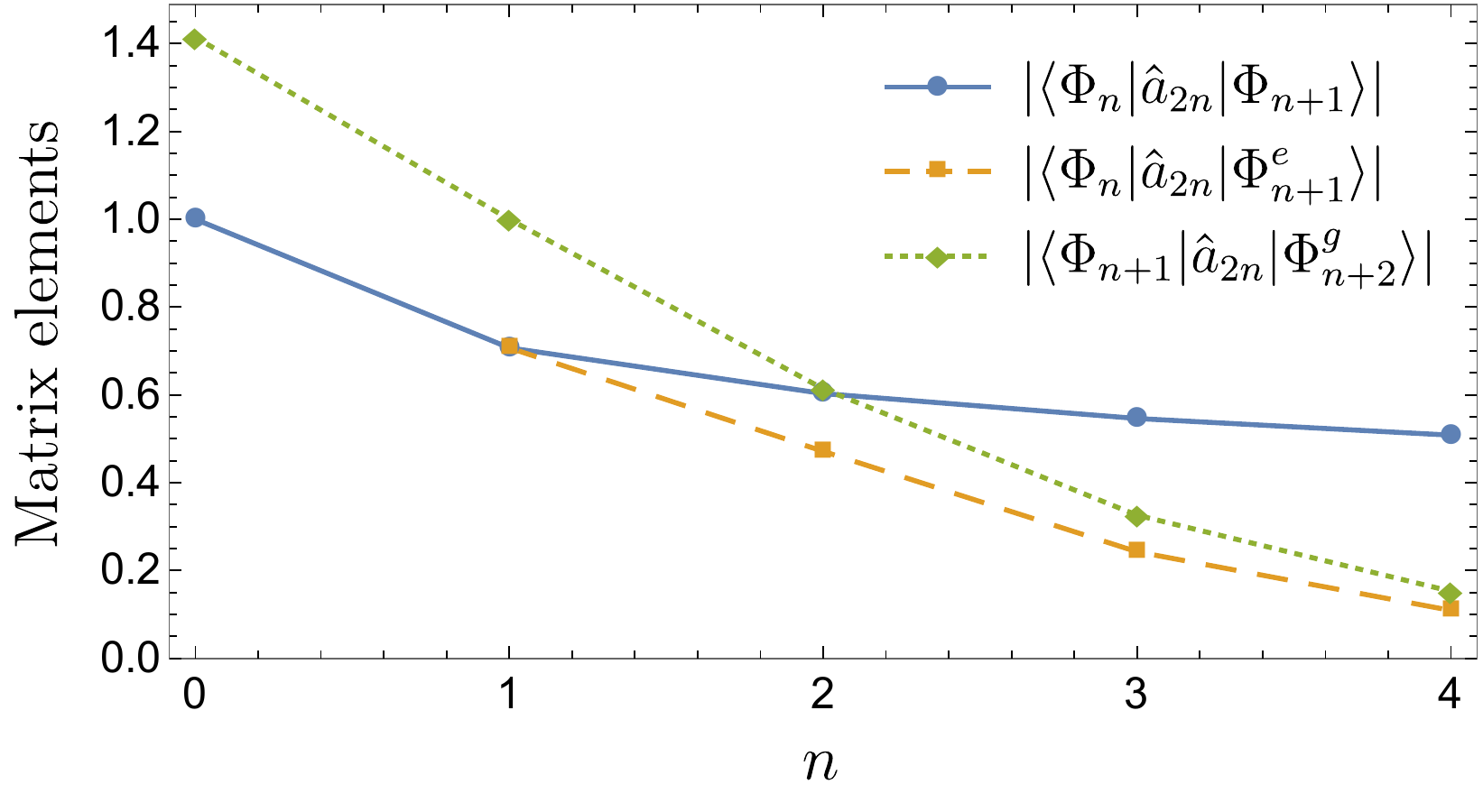}
\caption{\label{matrixelements}Matrix elements of the drive between the coupled many-body states. The operator $\hat{a}_{2n}$ annihilates a particle in the angular momentum mode $m=2n$.}
\end{figure}

As we discussed earlier, the desired transition from $|\Phi_{n}\rangle$ to $|\Phi_{n+1}\rangle$ occurs with near-unity probability only if $\delta_{\text{max}} \gg |\Omega_n| \gtrsim \sqrt{\nu}$. Thus, we have a bound on the sweep rate, $\nu \ll \smash{\Delta_{n+1}^2, \Delta_{n+2}^2}$. Figure~\ref{spectrumandgap}(b) shows that $\Delta_n$ varies weakly with $n$, saturating at $0.6 V_0$ for $n\gtrsim 5$. Hence, we can choose the same detuning range and sweep rate for each transfer. Further, as illustrated in Fig.~\ref{matrixelements}, $|\Omega_n| \propto |\langle\Phi_{n+1}|\smash{\hat{a}^{\dagger}_{2n}}|\Phi_{n}\rangle|$ is roughly independent of $n$ and therefore roughly the same laser intensity can be used for each transition. We also see that the undesired matrix elements fall off with $n$. Thus, we do not expect coupling to these states to be a problem even when $n$ is large.

As argued in Refs.~\cite{shore2008coherent, vitanov2001laser, rangelov2010rapid}, the adiabaticity requirements are somewhat relaxed if one takes smooth profiles for the laser intensity $\Lambda_n (t)$ and detuning $\delta_n (t)$. Thus, we take

\begin{align}
&\hspace{-0.2cm}\Lambda_n(t) = \frac{\Omega(t-(4n+2)\tau)}{|\langle\Phi_{n}|\hat{a}_{2n}|\Phi_{n+1}\rangle|} \hspace{0.05cm},\; \delta_n(t) = \delta(t-(4n+2)\tau)\hspace{0.03cm},
\label{sweepprofile1}\\
&\hspace{-0.2cm}\text{where}\quad \Omega(t) \equiv \frac{c_{\Omega}}{\tau} \hspace{0.03cm} e^{-\frac{t^2}{\tau^2}} \quad\text{and}\quad \delta(t) \equiv \frac{c_{\delta}}{\tau} \frac{t}{\tau} \hspace{0.03cm} e^{-\frac{t^2}{3\tau^2}}\hspace{0.05cm}.
\label{sweepprofile2}
\end{align}
These profiles are characterized by the parameters $c_{\Omega}$, $c_{\delta}$, and $\tau$. The first two parameters set the amplitudes of the Rabi frequency and the detuning, and $\tau$ sets the time scale of the frequency sweep. The Rabi frequency $\Omega_n$ is only significant in the interval $t=4 n \tau$ to $t = 4 (n+1) \tau$, during which the system is transferred from $|\Phi_{n}\rangle$ to $|\Phi_{n+1}\rangle$. The factor of 4 is chosen so that each sweep is well separated from the others. In the limit $c_{\Omega} \ll c_{\delta}$ and $\tau \gg c_{\delta}/V_0$, the sweep reduces to the original Landau-Zener problem with Rabi frequency $c_{\Omega}/\tau$ and sweep rate $\nu = c_{\delta}/\tau^2$. Then the transition probability is given by $P\approx 1 - \exp({-\pi c_{\Omega}^2/c_{\delta}})$ \cite{landau1932a, zener1932non}. Generically, we find $P\approx 1$ provided $c_{\Omega} \lesssim c_{\delta} \lesssim c_{\Omega}^2$ and $\tau\gg c_{\delta}/V_0$. Hence, the drive protocol is optimized by taking $c_{\Omega}$ and $c_{\delta}$ of order unity and $\tau$ sufficiently large compared to $1/V_0$.

\begin{figure}
\includegraphics[width=\columnwidth]{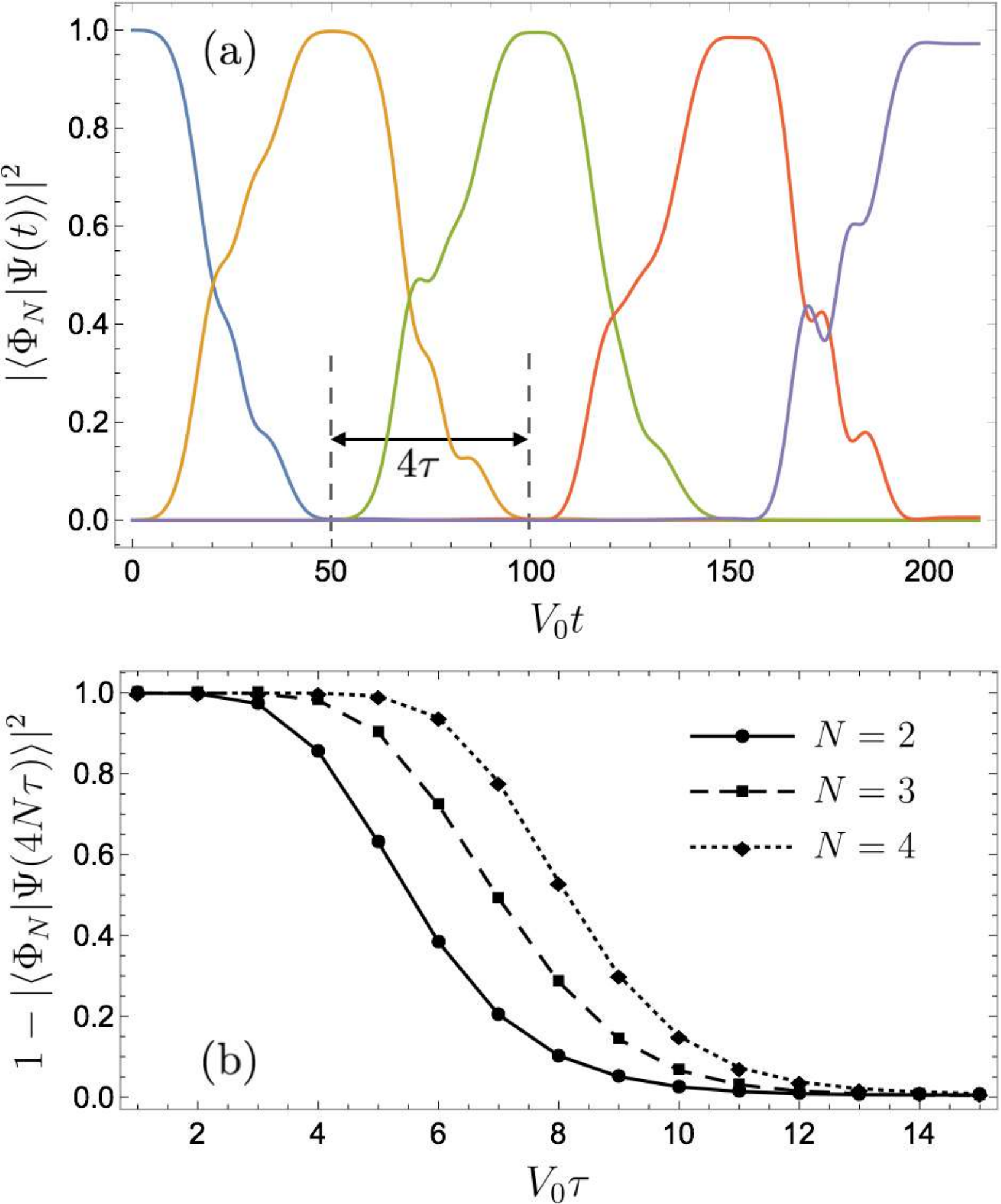}
\caption{\label{rapanderror}{\bf (a)} Overlap of the system wave function $|\Psi(t)\rangle$ with each $N$-particle Laughlin state $|\Phi_N\rangle$ as a function of time during our driving protocol, described by Eqs.~\eqref{Hdrn}--\eqref{sweepprofile2}. Each subsequent plateau corresponds to increasing $N$ by 1. The duration of each sweep is $4\tau$.\added[id=SA]{ As before, $V_0$ is the two-particle interaction energy}. {\bf (b)} Cumulative error in the final state preparation as a function of $\tau$ for three different particle numbers, with $c_{\Omega} = 4$, $c_{\delta} = 5.33$.}
\end{figure}

Figure~\ref{rapanderror}(a) shows the creation of the $N=4$ Laughlin state with $c_{\Omega} = 4$, $c_{\delta} = 5.33$, and $\tau = 12.5/V_0$. For $P\approx 1$, the error $1-P$ in a given sweep is roughly independent of $n$ and falls off exponentially as $V_0\tau$ is increased. Hence, the cumulative error after $N$ sweeps, for large $V_0 \tau$, scales as $N e^{-\xi}$ where $\xi \propto V_0 \tau$. This feature is apparent in Fig.~\ref{rapanderror}(b), where we plot the cumulative error as a function of $V_0\tau$ for different values of $N$. As a rough estimate, we find this error is less than 1\% for $\tau \gtrsim (10/V_0) \ln N$. Thus, one can prepare $|\Phi_{N}\rangle$ with such high fidelity in a total time $T_L\gtrsim 40 \hspace{0.03cm} (N/V_0) \ln N$.

In the Supplemental Material~\footnote{See Supplemental Material at \href{http://muellergroup.lassp.cornell.edu/twisted/}{\nolinkurl{http://muellergroup.lassp.cornell.edu/twisted/}} for polariton-density animations showing examples of adiabatic and nonadiabatic creation of Laughlin states, generation of quasiholes, and braiding of quasiholes.}, we show animations of the polariton density during our driving protocol. If the sweeps are adiabatic, the density is uniform and the radius of the Laughlin puddle grows as $\sqrt{n}$ as more photons are injected into the system. For nonadiabatic sweeps, we see the development of vortices arising from the coupling to other many-body states.

\subsection{\label{loss constraint}Constraint from polariton loss}
We require that $N_{\text{loss}}$, the expected number of polaritons lost during the preparation of $| \Phi_N\rangle$, be small compared to 1. We can estimate $N_{\text{loss}}$ by noting that the system approximately spends an interval $T_L/N$ in a state with $n$ polaritons, where $n$ varies from zero to $N-1$. For a single-polariton decay rate $\gamma$, the net loss rate from an $n$-polariton state is $n \gamma$. Hence,
\begin{equation}
N_{\text{loss}} \approx T_L/N \sum_{n=0}^{N-1} n \gamma \approx N\gamma \hspace{0.03cm} T_L/2\hspace{0.05cm}.
\end{equation}
Thus, our protocol can be used to prepare the $N$-particle Laughlin state provided $N_{\text{loss}} \ll 1$, or $V_0/\gamma \gg 20 N^2 \ln N$, where we have taken $T_L=40 \hspace{0.03cm} (N/V_0) \ln N$.

\section{\label{holegeneration}Quasihole generation}
\subsection{\label{quasihole generation overview}Overview}

\begin{figure}[b]
\includegraphics[width=\columnwidth]{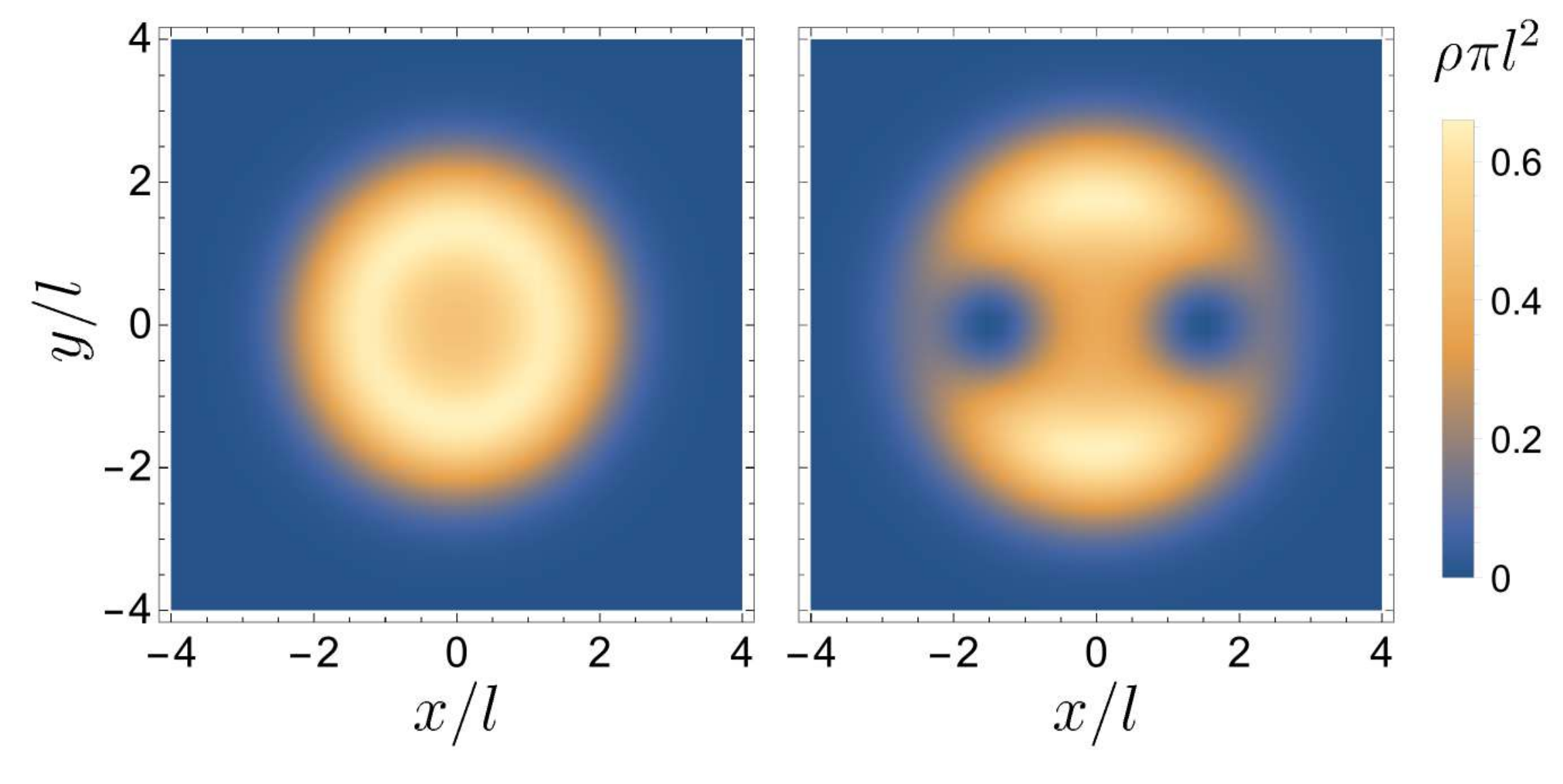}
\caption{\label{density}Polariton density $\rho$ in the Laughlin state $|\Phi_{N=3}\rangle$ (left) and a two-quasihole state $|\smash{\Phi^{\text{oo}}_{N=3}}\rangle$ (right).}
\end{figure}

A quasiparticle or quasihole is a collective excitation with particle like properties. For example, a quasihole at location $z_0$ in the Laughlin state $|\Phi_N\rangle$ is described by the wavefunction $\smash{\Phi^{\text{o}}_{N} (\{z_j\}) \propto \prod_{i=1}^N (z_i - z_0)\hspace{0.03cm} \Phi_N (\{z_j\})}$~\cite{arovas1984fractional}. This state has all the properties of the Laughlin state, except there is a density depletion near $z_0$. Integrating this depletion over space yields the surprising result that exactly half a particle has been removed from this region. The wave function $\smash{\Phi^{\text{o}}_{N}}(\{z_j\})$ is readily generalized to the case of multiple quasiholes. Thus, a state with two quasiholes at $\pm z_0$ is described by the wave function
\begin{equation}
\label{quasihole}
\Phi^{\text{oo}}_N (\{z_j\}) \propto \prod_{i=1}^N (z_i - z_0) (z_i + z_0)\hspace{0.03cm}\Phi_N (\{z_j\})\hspace{0.05cm}.
\end{equation}
As shown in Fig.~\ref{density}, the particle density in $|\smash{\Phi^{\text{oo}}_N}\rangle$ nearly vanishes within a circle of radius $\sim l$ centered at $\pm z_0$, thus forming holes in an otherwise uniform-density background of $|\Phi_N\rangle$. Past calculations have shown that exchanging the two defects yields a Berry phase of $\phi_s = \pi/2$ in the thermodynamic limit~\cite{arovas1984fractional, halperin1984statistics, stern2008anyons, paredes20011}. Thus, the quasiholes can be considered as quantum particles with fractional statistics. Here we show how to produce these defects by introducing additional laser potentials.

To produce a quasihole, we apply a localized repulsive potential just outside the Laughlin cloud and bring it radially inward through the edge. If the potential is strong enough and the radial sweep is sufficiently adiabatic, then we find that the final state will contain a quasihole bound to the potential. This procedure is more efficient than increasing the height of a potential barrier at a fixed location, as proposed in Ref.~\cite{paredes20011} for atomic systems. Experiments have demonstrated that such local potentials can be generated optically~\cite{hayat2012dynamic, li2013entanglement, amo2010light, sanvitto2011all}, e.g., by illuminating the atoms with a laser that Stark-shifts the intermediate state in the Rydberg transitions. This illumination can be tightly focused and moved spatially. By sweeping two such potentials through opposite sides of the Laughlin cloud, one can create a quasihole at each end, which can then be braided around one another.

\subsection{\label{quasihole generation modeling}Modeling}
We model the potentials by Dirac delta functions of strength $\alpha$ applied at positions $\pm \vec{r}_0 (t)$, where $\alpha$ is a constant and $\vec{r}_0 (t)$ is swept radially inward along the $x$ axis. This model is good as long as the spatial extent of the actual potential is smaller than the scaled magnetic length $l$. The potential energy is then described by the Hamiltonian
\begin{equation}
\hat{U}(t) = \alpha\hspace{-0.02cm} \int\hspace{-0.05cm} d^2 r\hspace{0.03cm} [\delta(\vec{r}\hspace{-0.03cm} -\hspace{-0.03cm}  \vec{r}_0(t)) \hspace{-0.03cm} +\hspace{-0.03cm}  \delta(\vec{r} \hspace{-0.03cm} +\hspace{-0.03cm}  \vec{r}_0(t))]\hspace{0.03cm} \hat{\psi}^{\dagger}\hspace{-0.02cm} (\vec{r}) \hat{\psi}(\vec{r})\hspace{0.05cm}.\label{potential}
\end{equation}
Projecting into the LLL, we find
\begin{equation}
\hat{U}_{\text{LLL}}(t) = U_0\hspace{0.03cm} e^{-(z_0(t))^2} \sum_{s=0}^{\infty} [z_0(t)]^{2 s}\hspace{0.02cm} \hat{Q}_{2s}\hspace{0.05cm},\label{potentialLLL}
\end{equation}
where $z_0 \equiv  r_0/l$, $\smash{\hat{Q}_s} \equiv\smash{\sum_{m=0}^{s} \hat{a}^{\dagger}_{s-m} \hat{a}_{m} / \sqrt{m! (s-m)!}}$, and $U_0 \equiv 2\alpha/(\pi l^2)$. Hence, the Hamiltonian conserves the particle number $N$ but changes the total angular momentum $L$ through the operator \smash{$\hat{Q}_s$}.

The potentials must be strong enough to fully deplete the density at $\pm z_0$. If the sweep is adiabatic, the system will always be in an eigenstate of $\smash{\hat{H}_{\text{LLL}}} + \smash{\hat{U}_{\text{LLL}}}(t)$, where $\smash{\hat{H}_{\text{LLL}}}$ is the unperturbed Hamiltonian given by Eq.~\eqref{HLLL}. For $U_0$ sufficiently large, the ground state belongs to the null space of $\smash{\hat{U}_{\text{LLL}}}$. This space is heavily degenerate and spanned by wavefunctions of the form $\smash{\prod_{i=1}^N} (z_i - z_0) (z_i + z_0) f(\{z_j\})$, where $f$ is a symmetric polynomial times a Gaussian. The two-quasihole state $|\smash{\Phi^{\text{oo}}_N}\rangle$ in Eq.~\eqref{quasihole} is the lowest-energy eigenstate of this form in the absence of a trap ($\varepsilon=0$). However, for $\varepsilon=0$, the ground-state manifold is degenerate, consisting of all $f(\{z_j\}) = \Phi_N(\{z_j\}) \chi(\{z_j\})$ for arbitrary symmetric polynomials $\chi$. The harmonic trap splits the energies of different angular momentum states, thus lifting the degeneracy. For small $\varepsilon$, the ground state $|\Psi_g\rangle$ can be found by applying degenerate perturbation theory, which yields $1-|\langle\smash{\Phi^{\text{oo}}_N}|\Psi_g\rangle|^2 \sim \varepsilon^2/\Delta_N^2$, where $\Delta_N$ is the many-body interaction splitting  shown in Fig.~\ref{spectrumandgap}. Thus, $|\smash{\Phi^{\text{oo}}_N}\rangle$ represents the approximate ground state. We find numerically that the overlap $|\langle\Phi^{\text{oo}}_N|\Psi_g\rangle|^2$ remains near unity as long as $\varepsilon \ll U_0$ and $\varepsilon \lesssim \Delta_N / N$.

Thus, we consider a sweep where the instantaneous ground state of the system evolves from $|\Phi_N\rangle$ when the potentials are outside the cloud to approximately $|\Phi^{\text{oo}}_N\rangle$ when they are fully inside. To produce quasiholes, the sweep must be sufficiently slow that the system resides in the instantaneous ground state at all times. Similar to the analysis in Sec.~\ref{preparation}, we numerically integrate the time-dependent Schr\"{o}dinger equation to evaluate the fidelity of this process. Owing to the presence of edge modes, we find that the most sensitive part of the process is when the potential moves through the edge of the cloud at $R \sim 2\sqrt{N-1}\hspace{0.05cm}l$~\cite{paredes20011}. In particular, if the motion from $r_0 = R + l$ to $r_0 = R - l$ is adiabatic, then the entire sweep is adiabatic. For simplicity, we consider linear sweeps in which $r_0$ is reduced at a constant rate.

The maximum allowed sweep rate $|\partial_t(r_0/l)|$ can be estimated by requiring that the rate be smaller than the energy gap $\Delta E$ between the ground state and the first excited state. When the potentials are near the edge of the cloud, the system is largely unperturbed; then $\Delta E$ is roughly the minimum of $\varepsilon$ and $\Delta_N - N \varepsilon$ (see Fig.~\ref{spectrumandgap}). The former corresponds to the lowest-energy surface waves which increase the total angular momentum by 1 unit but do not result in any density increases. The latter corresponds to bulk excitations which increase the density and decrease the total angular momentum. To prevent exciting these modes, one must have $|\partial_t(r_0/l)| \lesssim \varepsilon,\Delta_N - N \varepsilon$. Thus, we need a small but finite trap frequency such that $0<\varepsilon < \Delta_N/N$. This energy gap is maximized for $\varepsilon = \Delta_N / (N+1)$. However, the potentials modify the excitation spectrum as they enter the cloud. We numerically find that adiabaticity throughout the sweep requires $|\partial_t(r_0/l)| \lesssim \varepsilon\lesssim \Delta_N / (2 N)$.

\begin{figure}
\includegraphics[width=\columnwidth]{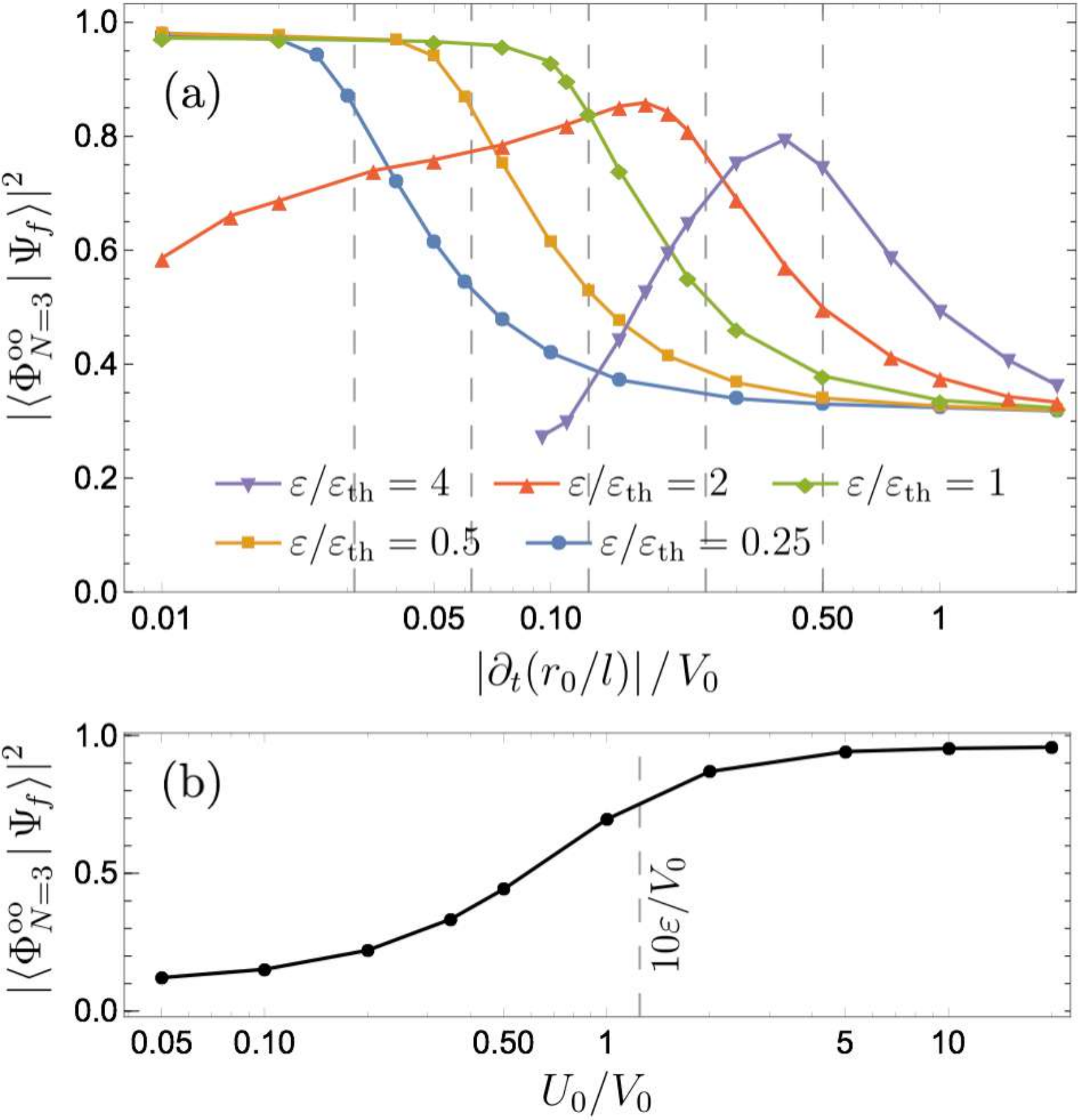}
\caption{\label{hole-trapandU0}{\bf (a)} Fidelity of the two-quasihole state preparation with a strong impurity potential ($U_0 = 20 V_0$) as a function of the sweep rate $|\partial_t (r_0/l)|$ for different trap frequencies parametrized by $\varepsilon$.\added[id=SA]{ Here, $r_0$ is the radial distance of each pinning potential from the center and $l$ is the scaled magnetic length.} Dashed vertical lines show where $|\partial_t (r_0/l)| = \varepsilon$ for each curve. The\added[id=SA]{ final} overlap approaches unity for $|\partial_t(r_0/l)| \lesssim \varepsilon$ provided $\varepsilon \lesssim \varepsilon_{\text{th}} \equiv \Delta_N / (2 N) = V_0/8$ (for $N=3$)\added[id=SA]{, where $\Delta_N$ is the many-body interaction splitting shown in Fig.~\ref{spectrumandgap}}. To avoid visual distraction, the curve for $\varepsilon/\varepsilon_{\text{th}} = 4$ is only shown for $|\partial_t (r_0/l)| > 0.1 \hspace{0.03cm} V_0$. {\bf (b)} Fidelity of the two-quasihole state preparation as a function of the strength $U_0$ of the applied potential. Here, $|\partial_t(r_0/l)| = \varepsilon/2 = \varepsilon_{\text{th}}/2$ (adiabatic sweep).}
\end{figure}

As a measure of adiabaticity, we plot the final overlap $|\langle \Phi^{\text{oo}}_N | \Psi_{\hspace{-0.05cm}f} \rangle|^2$ for $N=3$ in Fig.~\ref{hole-trapandU0}(a) as a function of the sweep rate for different values of $\varepsilon$, with $U_0 \gg \varepsilon$. We see that the overlap approaches 1 for $|\partial_t(r_0/l)| \lesssim \varepsilon \lesssim \Delta_N / (2 N)$. We show animations of the polariton density during the sweep in the Supplemental Material~\cite{Note2}. For nonadiabatic sweeps, the potentials excite surface modes and shape deformations in the density profile.

For smaller $U_0$, the ground state is not well approximated by $|\Phi_N^{\text{oo}}\rangle$. This feature is illustrated in Fig.~\ref{hole-trapandU0}(b), which shows the overlap following an adiabatic evolution. As expected, the overlap is near unity if $U_0 \gg \varepsilon$.

We note that a strong attractive potential ($U_0<0$) will also produce quasiholes. This is because the total energy is conserved and for $|U_0| \gg \varepsilon, V_0$, the dynamics get projected onto the zero-energy subspace of the applied potential, regardless of the sign of $U_0$.

We can calculate the time required to generate the two-quasihole state, $T_h$, by noting that $r_0$ is being swept over a distance $d \gtrsim 2 l$ at a rate $|\partial_t(r_0/l)| \lesssim \Delta_N / (2 N)$. Hence, $T_h \gtrsim 4 N / \Delta_N$. We found earlier that $\Delta_N$ saturates at $3 V_0 /5$ for $N\gtrsim 5$ and $\Delta_2 = V_0$, where $V_0$ is the zeroth Haldane pseudopotential [Fig.~\ref{spectrumandgap}(b)]. Thus, the minimum quasihole preparation time will vary from $4N/V_0$ for small $N$  to $(20/3) N/V_0$ for $N\gtrsim 5$. This bound is much smaller than the time required to prepare the $N$-particle Laughlin state, $T_L \gtrsim 40\hspace{0.03cm} (N/V_0) \ln N$ (see Sec.~\ref{preparation}).

\section{\label{braiding}Quasihole braiding}
\subsection{\label{braiding overview}Overview}
In the previous section, we showed how one can create a pair of quasiholes at opposite ends of a Laughlin state, each bound to a local external potential. The same potentials can be dragged around one another to braid the two quasiholes~\cite{kapit2012non}. One must move the potentials slowly enough to ensure that the quasiholes remain bound to the potentials throughout the process. The adiabaticity condition also differs for clockwise and counterclockwise motion, as the effective magnetic field breaks time-reversal symmetry. Below we investigate the conditions for an adiabatic braiding.

As we explained in the last section, the ground state $|\Psi_g\rangle$ in the presence of the applied potentials approximates the desired two-quasihole state $|\Phi_N^{\text{oo}}\rangle$ in Eq.~\eqref{quasihole}. We consider braiding these quasiholes by rotating the two potentials on a circle by an angle $\pi$. This rotation can be modeled by taking $\vec{r}_0(t) = r_0 (\cos\varphi_0(t), \sin\varphi_0(t))$ in Eq.~\eqref{potential}, where $\varphi_0(t)$ varies from zero to $\pm\pi$. For an infinitely slow braiding, the system follows the instantaneous ground state $|\Psi_g(t)\rangle$, which is simply the rotated version of the initial state $|\Psi_g\rangle$. Hence, in this case, the two quasiholes move with the potentials. However, for a finite rotation speed, the overlap with the ground state is no longer unity. Then the ``braiding error'' can be calculated as $\eta \equiv 1 - |\langle\Psi_g|\Psi_f\rangle|^2$, where $|\Psi_f\rangle$ is the final state of the system. Since polaritons are lost in the experiment at a finite rate, our goal is to minimize the braiding duration $T_b$ while keeping $\eta$ below a cutoff $\eta_c$.

\subsection{\label{braiding modeling}Modeling}
For simplicity, we only consider rotations where $\varphi_0(t)$ changes at a constant rate $\omega_b$. Then we can transform to the corotating frame where the system evolves (within the LLL) under a time-independent Hamiltonian $\smash{\hat{H}_{\text{rot}}} = \smash{\hat{H}_{\text{LLL}}  + \hat{U}_{\text{LLL}}(0) - \omega_b \hat{L}}$, where $\smash{\hat{H}_{\text{LLL}}}$ and $\smash{\hat{U}_{\text{LLL}}}$ are defined in Eqs.~\eqref{HLLL} and \eqref{potentialLLL}. Hence, the braiding is equivalent to introducing a perturbation $\delta \smash{\hat{H}} = -\omega_b \smash{\hat{L}}$ for a duration $T_b = \pi/|\omega_b|$. The error $\eta$ is set by the dimensionless parameters $\omega_b / V_0$, $\varepsilon / V_0$, $r_0/l$, and $U_0 / V_0$.

\begin{figure}[b]
\includegraphics[width=\columnwidth]{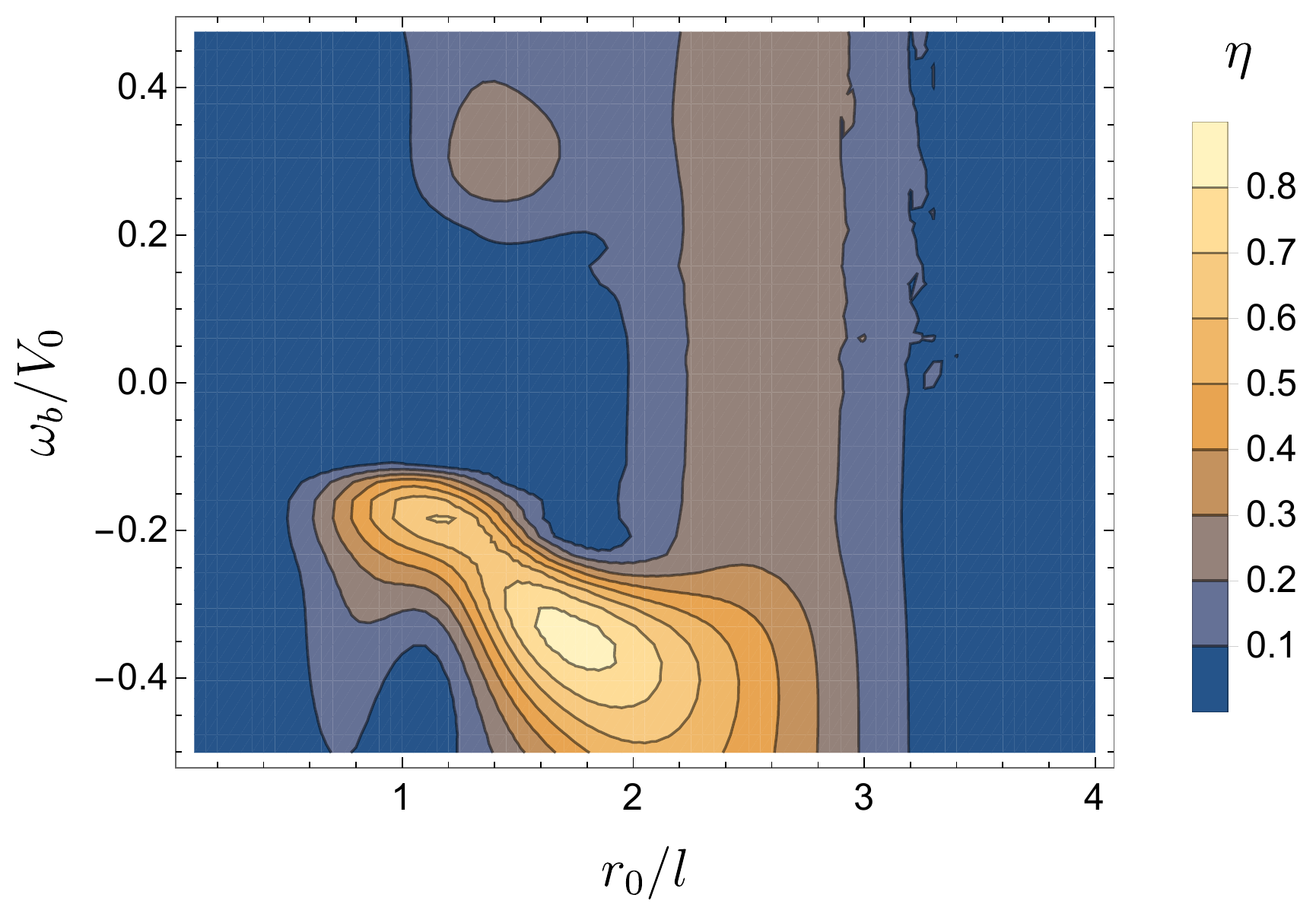}
\caption{\label{contourplot}Contour plot showing the braiding error $\eta$ when two strong impurity potentials ($U_0 = 100 V_0$), each binding a quasihole at $\pm r_0$, are rotated by $\pi$ at a constant angular speed $\omega_b$. Here, $N=3$ and $\varepsilon=0$ (no trap). The vertical band centered around $r_0/l \approx 2.5$ corresponds to edge excitations. Other peaks correspond to bulk resonances.\added[id=SA]{ As before, $l$ denotes the scaled magnetic length.}}
\end{figure}

\begin{figure}
\includegraphics[width=\columnwidth]{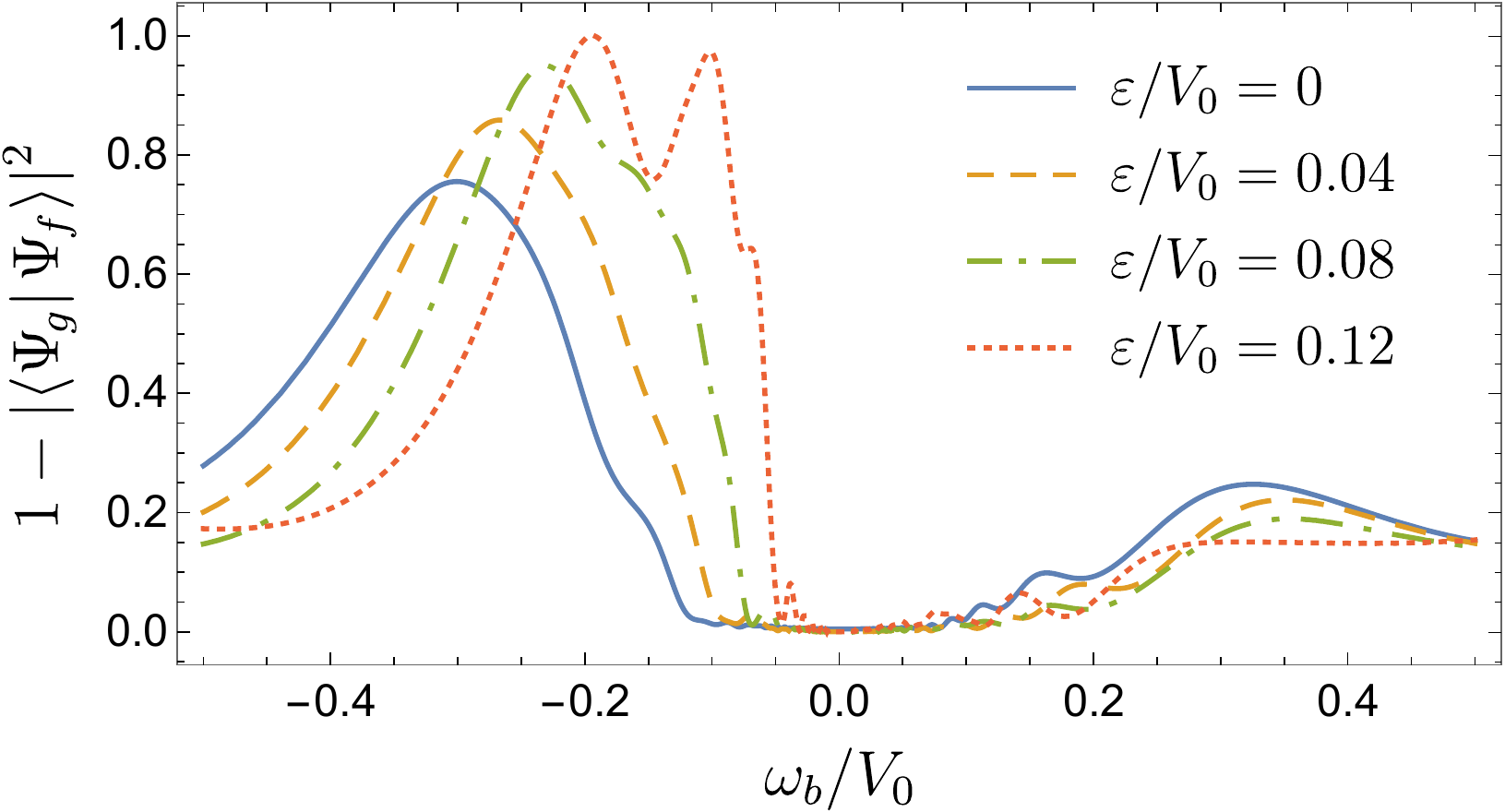}
\caption{\label{braid-trapandU0}Braiding error $\eta$ vs rotation rate $\omega_b$ at different trap frequencies for $N=3$, $U_0/V_0 = 100$, and $r_0/l = 1.5$. Note that for adiabatic quasihole generation, one must have $\varepsilon \lesssim \varepsilon_{\text{th}} \equiv \Delta_N / (2N) = 0.125 V_0$ [Figs.~\ref{hole-trapandU0} and \ref{spectrumandgap}(b)].}
\end{figure}

Figure~\ref{contourplot} shows the error as a function of $\omega_b$ and $r_0$ for $\varepsilon=0$ and $U_0/V_0 \gg 1$. If $r_0$ is near the edge of the cloud, the braiding can excite surface modes, resulting in braiding error. Similarly, there appear to be bulk resonances at particular radii and rotation frequencies. As more clearly illustrated by the line cuts in Fig.~\ref{braid-trapandU0}, this structure results in a threshold behavior, where $\eta \approx 0$ when $|\omega_b| / V_0$ is sufficiently small. The threshold for positive $\omega_b$ (counterclockwise rotations) is roughly independent of $\varepsilon$, while that for negative $\omega_b$ drops, and becomes sharper, as $\varepsilon$ grows. The thresholds also move to lower values as one decreases $r_0$. Generally, the braiding is more adiabatic for rotations in the direction of the Lorentz force, which is counterclockwise in our case. In the Supplemental Material~\cite{Note2}, we show videos of the excitations created in the nonadiabatic regime. For $\smash{|\omega_b| / V_0 \gg 1}$, the quasiholes do not have time to move, so the system remains in the initial state and $\eta \to 0$. This limit is clearly not suitable for quasihole braiding.

The threshold frequency for $\omega_b >0$ and $N=3$ is approximately $0.1 V_0$. Thus, one can perform an adiabatic braiding of two quasiholes in a three-particle Laughlin state in a time $T_b \gtrsim 10\pi/V_0$ with vanishingly small error. This duration is much smaller than the $N=3$ Laughlin state preparation time $T_L\gtrsim 130/V_0$ but comparable to the quasihole generation time $T_h \gtrsim 16/V_0$. One can further reduce $T_b$ by moving the potentials in a more smooth manner~\cite{knapp2016nature}.

\section{\label{statistics}Measuring anyonic statistics}
\subsection{\label{statistics overview}Overview}
During an adiabatic braiding of two quasiholes, the many-body wave function picks up a geometric (or Berry) phase $\phi_g$, in addition to a dynamical phase $\phi_d$ associated with the time evolution. The geometric phase can be further decomposed into two pieces, $\phi_g = 2 \phi_1 + \phi_s$, where $\phi_1$ corresponds to the phase which would be acquired if one had a single quasihole and moved it through the same path. One can interpret $\phi_1$ as the Aharonov-Bohm phase resulting from an effective magnetic field felt by a quasihole. The remainder, $\phi_s$, is interpreted as a statistical phase which originates from exchanging the two quasiholes. Equivalently, $\phi_s$ can be understood as encoding how the presence of one quasihole influences the magnetic field which the other experiences. Past theoretical studies have shown that $\phi_s=\pm\pi/2$ in the thermodynamic limit (depending on the direction of rotation)~\cite{arovas1984fractional, halperin1984statistics, stern2008anyons, paredes20011}. Here we examine how these ``anyonic'' statistics manifest for finite particle numbers and show how one can measure $\phi_s$ in experiments.

\subsection{\label{extracting statistical phase}Extracting statistical phase}
\begin{figure}
\includegraphics[width=\columnwidth]{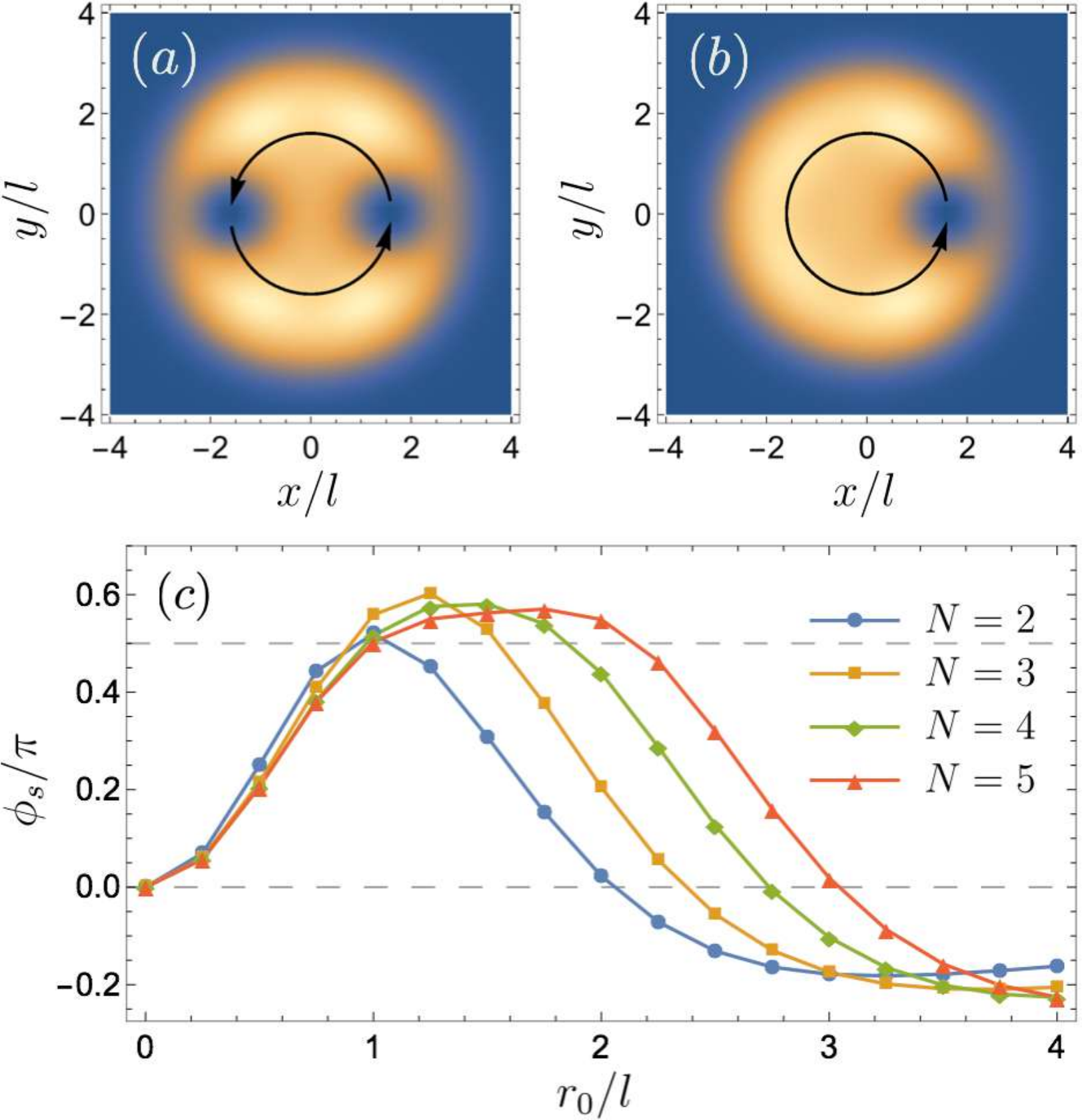}
\caption{\label{statisticalphase}{\bf (a,b)} Polariton density in the $x$--$y$ plane for the two experiments needed to extract the statistical phase $\phi_s$ associated with exchanging two quasiholes. As in Fig.~\ref{density}, brighter colors represent higher density and the dark disks correspond to quasiholes bound to potentials. Here $N=4$. In (a) two quasiholes are exchanged while in (b) one quasihole is moved in a circle. {\bf (c)} Statistical phase inferred from subtracting the geometric phases that would be found in these two experiments. In the thermodynamic limit, with well-separated quasiholes, one expects $\phi_s = \pi/2$.\added[id=SA]{ Note that $r_0$ denotes the radial distance of each quasihole from the center and $l$ is the scaled magnetic length.}}
\end{figure}

In the next section we describe an interferometric protocol for measuring the total many-body phase $\phi = \phi_g + \phi_d$. Given such a protocol, it is straightforward to extract $\phi_s$: First, by repeating the experiment multiple times with different sweep rates, one can distinguish between $\phi_d$ and $\phi_g$. Second, $\phi_s$ can be extracted from $\phi_g$ by comparing two experiments. In the first experiment, illustrated in Fig.~\ref{statisticalphase}(a), one rotates two quasiholes by $\pi$. In the second experiment, illustrated in Fig.~\ref{statisticalphase}(b), a single quasihole is rotated by $2\pi$. The latter yields the same Aharonov-Bohm phase $2\phi_1$ but no statistical phase. This approach is similar to the ones suggested in Refs.~\cite{paredes20011, umucalilar2013many}. Figure~\ref{statisticalphase}(c) shows the value of $\phi_s$ which would thereby be extracted.

If the two quasiholes are too close together, they interact and it is not appropriate to interpret $\phi_s$ as being due to statistics. Similarly, if the quasiholes are moved outside of the bulk region, their properties are modified. Thus, in the small clouds we study, one only expects $\phi_s = \pi/2$ over a finite range of $r_0$ (radial position of each quasihole). As $N$ increases, so should the bulk region. This trend is clear in Fig.~\ref{statisticalphase}(c). To calculate the curves in this figure, we took advantage of a\added[id=SA]{ simple} relationship\deleted[id=SD]{ derived in~\footnote{See the Supplementary Material of Ref.~\cite{umucalilar2013many}
}} between the geometric phase and the total angular momentum\added[id=SA]{~\cite{NoteX, umucalilar2017observing}}, which yields $\phi_s/\pi = N(N-1) + \langle \smash{\hat{L}}\rangle_{\text{oo}} - 2 \langle \smash{\hat{L}}\rangle_{\text{o}}$. Here $\langle \smash{\hat{L}}\rangle_{\text{o}}$ and $\langle \smash{\hat{L}}\rangle_{\text{oo}}$ are the expectation of $\smash{\hat{L}}$ in the single-quasihole and two-quasihole states, respectively.

\subsection{\label{measuring total phase}Measuring total braiding phase}
Our approach to measuring the total many-body phase $\phi$ requires being able to create a reference state $|R^{\prime}\rangle$ which is unaffected by the sweep protocols that we use to fill the cavity with polaritons. That is, if we put the system in state $|R^{\prime}\rangle$, then apply the manipulations in Secs.~\ref{preparation}, \ref{holegeneration}, and \ref{braiding}, it will still be in state $|R^{\prime}\rangle$. One way to generate this reference is to drive an atom into a Rydberg state $|r^{\prime}\rangle$ with a large blockade radius. Then  $|R^{\prime}\rangle$ will represent a collective Rydberg excitation. Clearly, $|r^{\prime}\rangle$ should be distinct from the state $|r\rangle$ used to produce polaritons. Blockade physics will then prevent any further excitations during our protocol~\cite{jia2017strongly, firstenberg2016nonlinear}.

To measure $\phi$, one first uses a $\pi/2$ pulse to prepare the system in the superposition $|0\rangle + |R^{\prime}\rangle$, where $|0\rangle$ denotes all atoms being in the ground state. One then follows the procedures in Secs.~\ref{preparation} to \ref{braiding} to create the desired Laughlin state, generate quasiholes, and braid them. Then the process is repeated backwards, removing the quasiholes and coherently converting the Laughlin state to the vacuum. During this cycle, $|R^{\prime}\rangle$ is unaffected and $|0\rangle$ gains a total phase $\phi=\phi_d + \phi_g$, i.e., $|0\rangle + |R^{\prime}\rangle \to e^{i\phi}|0\rangle + |R^{\prime}\rangle$. Finally, a second $\pi/2$ pulse is applied to recombine the states $|0\rangle$ and $|R^{\prime}\rangle$, and the phase $\phi$ is read out by measuring the ground-state occupation. This approach is related to the one proposed in Ref.~\cite{grusdt2016interferometric} for measuring topological invariants and is similar to quantum computing protocols for measuring expectation values~\cite{quantum2000nielsen}.

In order to maintain coherence, the entire experiment must be performed on a time scale that is short compared to the polariton lifetime and the lifetime of the Rydberg state $|r^{\prime}\rangle$, which are typically a few tens of microseconds~\cite{Note1}. 

\section{\label{summary}Summary and outlook}
The rapidly growing field of many-body cavity quantum electrodynamics presents new opportunities to realize novel quantum states in a driven dissipative environment. In particular, with strong light-matter coupling and synthetic gauge fields, experiments now have the necessary ingredients to prepare fractional quantum Hall states of polaritons~\cite{schine2015synthetic, jia2017strongly, ningyuan2017photons, roushan2016chiral, owens2017quarter}. Here we have developed a protocol by which one can create the simplest of such states, the $\nu=1/2$ Laughlin states, in a twisted optical cavity (Fig.~\ref{setup}). We further explained how to generate quasihole excitations and directly measure their fractional exchange statistics.

In our protocol, one sequentially drives the system between the $n$- and $(n+1)$-particle Laughlin states, $|\Phi_n\rangle \to |\Phi_{n+1}\rangle$. This transition requires injecting a single photon with angular momentum $2n$. We showed how the transition can be achieved by illuminating the cavity mirrors with an appropriately tuned laser and sweeping its frequency. We find that one can create a very high-fidelity $N$-particle Laughlin state in a time $T\propto N \ln N$ (Fig.~\ref{rapanderror}). This can be contrasted with previous proposals for which the fidelity was exponentially small in $N$~\cite{umucalilar2013many, umucalilar2014probing, umucalilar2012fractional, hafezi2013non}.

We have also shown how one can adiabatically produce and braid quasiholes in $|\Phi_N\rangle$ by moving local pinning potentials (Figs.~\ref{hole-trapandU0}--\ref{braid-trapandU0}) and extract their anyonic statistics via interferometry (Fig.~\ref{statisticalphase}). Our results will facilitate ongoing experiments aiming to observe fractional quantum Hall physics in photonic systems~\cite{schine2015synthetic, jia2017strongly, ningyuan2017photons}.

High-fidelity preparation of Laughlin states requires a separation of energy scales between the two-particle interaction energy $V_0$ and the single-polariton decay rate $\gamma$. In our protocol, this condition arises from the need to maintain both adiabaticity and coherence, and takes the form $V_0/\gamma \gg 10 N^2 \ln N$. While this condition is very demanding in current experiments, where $V_0/\gamma \approx 50$, this figure of merit is continually improving.\added[id=SA]{ Note that the bound $10 N^2 \ln N$ is still much smaller than in Ref.~\cite{umucalilar2017generation}, where one needs $V_0/\gamma \gtrsim 3\times 10^4$ for $N=3$.}

Directly measuring the exchange statistics of two quasiholes in the bosonic $\nu=1/2$ Laughlin state would be extremely impactful and would be a step towards more complicated braiding protocols. For example, at $\nu=1$, bosons in the lowest Landau level form a paired Pfaffian state in which the quasiholes behave like Majorana fermions. Exchanging two of them rotates the system among a set of degenerate levels. At $\nu=3/2$, the exchange statistics are sufficiently rich that one can perform arbitrary unitary gates by braiding the particles~\cite{nayak2008non, kapit2012non, cooper2001quantum}.

One fascinating feature of using optical cavities as a platform for many-body physics is that the underlying system is coupled to a highly controllable environment, which can be used to manipulate the system~\cite{umucalilar2017generation, kapit2014induced, ma2017autonomous}. For example, one can implement a feedback stabilization mechanism where the photons emitted from the cavity are filtered by their angular momenta~\cite{umucalilar2017generation, mair2001entanglement} and the lost angular momentum is replenished by an appropriate drive. Despite such obvious potential, it is not yet clear how to best utilize the environment. Future studies can look deeper into this resource.

\section{\label{acknowledgments}Acknowledgments}We thank Jon Simon for several illuminating discussions. This material is based upon work supported by the National Science Foundation under Grant No. PHY-1508300 and the ARO-MURI Non-equilibrium Many-body Dynamics Grant No. W9111NF-14-1-0003.


\begin{thebibliography}{122}%
\makeatletter
\providecommand \@ifxundefined [1]{%
 \@ifx{#1\undefined}
}%
\providecommand \@ifnum [1]{%
 \ifnum #1\expandafter \@firstoftwo
 \else \expandafter \@secondoftwo
 \fi
}%
\providecommand \@ifx [1]{%
 \ifx #1\expandafter \@firstoftwo
 \else \expandafter \@secondoftwo
 \fi
}%
\providecommand \natexlab [1]{#1}%
\providecommand \enquote  [1]{``#1''}%
\providecommand \bibnamefont  [1]{#1}%
\providecommand \bibfnamefont [1]{#1}%
\providecommand \citenamefont [1]{#1}%
\providecommand \href@noop [0]{\@secondoftwo}%
\providecommand \href [0]{\begingroup \@sanitize@url \@href}%
\providecommand \@href[1]{\@@startlink{#1}\@@href}%
\providecommand \@@href[1]{\endgroup#1\@@endlink}%
\providecommand \@sanitize@url [0]{\catcode `\\12\catcode `\$12\catcode
  `\&12\catcode `\#12\catcode `\^12\catcode `\_12\catcode `\%12\relax}%
\providecommand \@@startlink[1]{}%
\providecommand \@@endlink[0]{}%
\providecommand \url  [0]{\begingroup\@sanitize@url \@url }%
\providecommand \@url [1]{\endgroup\@href {#1}{\urlprefix }}%
\providecommand \urlprefix  [0]{URL }%
\providecommand \Eprint [0]{\href }%
\providecommand \doibase [0]{http://dx.doi.org/}%
\providecommand \selectlanguage [0]{\@gobble}%
\providecommand \bibinfo  [0]{\@secondoftwo}%
\providecommand \bibfield  [0]{\@secondoftwo}%
\providecommand \translation [1]{[#1]}%
\providecommand \BibitemOpen [0]{}%
\providecommand \bibitemStop [0]{}%
\providecommand \bibitemNoStop [0]{.\EOS\space}%
\providecommand \EOS [0]{\spacefactor3000\relax}%
\providecommand \BibitemShut  [1]{\csname bibitem#1\endcsname}%
\let\auto@bib@innerbib\@empty
\bibitem [{\citenamefont {Stormer}\ \emph {et~al.}(1999)\citenamefont
  {Stormer}, \citenamefont {Tsui},\ and\ \citenamefont
  {Gossard}}]{stormer1999fractional}%
  \BibitemOpen
  \bibfield  {author} {\bibinfo {author} {\bibfnamefont {Horst~L.}\
  \bibnamefont {Stormer}}, \bibinfo {author} {\bibfnamefont {Daniel~C.}\
  \bibnamefont {Tsui}}, \ and\ \bibinfo {author} {\bibfnamefont {Arthur~C.}\
  \bibnamefont {Gossard}},\ }\bibfield  {title} {\enquote {\bibinfo {title}
  {The fractional quantum {Hall} effect},}\ }\href {\doibase
  10.1103/RevModPhys.71.S298} {\bibfield  {journal} {\bibinfo  {journal} {Rev.
  Mod. Phys.}\ }\textbf {\bibinfo {volume} {71}},\ \bibinfo {pages} {S298}
  (\bibinfo {year} {1999})}\BibitemShut {NoStop}%
\bibitem [{\citenamefont {Stormer}(1999)}]{stormer1999nobel}%
  \BibitemOpen
  \bibfield  {author} {\bibinfo {author} {\bibfnamefont {Horst~L.}\
  \bibnamefont {Stormer}},\ }\bibfield  {title} {\enquote {\bibinfo {title}
  {Nobel {Lecture}: The fractional quantum {Hall} effect},}\ }\href {\doibase
  10.1103/RevModPhys.71.875} {\bibfield  {journal} {\bibinfo  {journal} {Rev.
  Mod. Phys.}\ }\textbf {\bibinfo {volume} {71}},\ \bibinfo {pages} {875}
  (\bibinfo {year} {1999})}\BibitemShut {NoStop}%
\bibitem [{\citenamefont {Wen}(1995)}]{wen1995topological}%
  \BibitemOpen
  \bibfield  {author} {\bibinfo {author} {\bibfnamefont {Xiao-Gang}\
  \bibnamefont {Wen}},\ }\bibfield  {title} {\enquote {\bibinfo {title}
  {Topological orders and edge excitations in fractional quantum {Hall}
  states},}\ }\href {\doibase 10.1080/00018739500101566} {\bibfield  {journal}
  {\bibinfo  {journal} {Adv. Phys.}\ }\textbf {\bibinfo {volume} {44}},\
  \bibinfo {pages} {405} (\bibinfo {year} {1995})}\BibitemShut {NoStop}%
\bibitem [{\citenamefont {Arovas}\ \emph {et~al.}(1984)\citenamefont {Arovas},
  \citenamefont {Schrieffer},\ and\ \citenamefont
  {Wilczek}}]{arovas1984fractional}%
  \BibitemOpen
  \bibfield  {author} {\bibinfo {author} {\bibfnamefont {Daniel}\ \bibnamefont
  {Arovas}}, \bibinfo {author} {\bibfnamefont {John~R.}\ \bibnamefont
  {Schrieffer}}, \ and\ \bibinfo {author} {\bibfnamefont {Frank}\ \bibnamefont
  {Wilczek}},\ }\bibfield  {title} {\enquote {\bibinfo {title} {Fractional
  statistics and the quantum {Hall} effect},}\ }\href {\doibase
  10.1103/PhysRevLett.53.722} {\bibfield  {journal} {\bibinfo  {journal} {Phys.
  Rev. Lett.}\ }\textbf {\bibinfo {volume} {53}},\ \bibinfo {pages} {722}
  (\bibinfo {year} {1984})}\BibitemShut {NoStop}%
\bibitem [{\citenamefont {Halperin}(1984)}]{halperin1984statistics}%
  \BibitemOpen
  \bibfield  {author} {\bibinfo {author} {\bibfnamefont {Bertrand~I.}\
  \bibnamefont {Halperin}},\ }\bibfield  {title} {\enquote {\bibinfo {title}
  {Statistics of quasiparticles and the hierarchy of fractional quantized
  {Hall} states},}\ }\href {\doibase 10.1103/PhysRevLett.52.1583} {\bibfield
  {journal} {\bibinfo  {journal} {Phys. Rev. Lett.}\ }\textbf {\bibinfo
  {volume} {52}},\ \bibinfo {pages} {1583} (\bibinfo {year}
  {1984})}\BibitemShut {NoStop}%
\bibitem [{\citenamefont {Stern}(2008)}]{stern2008anyons}%
  \BibitemOpen
  \bibfield  {author} {\bibinfo {author} {\bibfnamefont {Ady}\ \bibnamefont
  {Stern}},\ }\bibfield  {title} {\enquote {\bibinfo {title} {Anyons and the
  quantum {Hall} effect---{A} pedagogical review},}\ }\href {\doibase
  10.1016/j.aop.2007.10.008} {\bibfield  {journal} {\bibinfo  {journal} {Ann.
  Phys.}\ }\textbf {\bibinfo {volume} {323}},\ \bibinfo {pages} {204} (\bibinfo
  {year} {2008})}\BibitemShut {NoStop}%
\bibitem [{\citenamefont {Camino}\ \emph {et~al.}(2005)\citenamefont {Camino},
  \citenamefont {Zhou},\ and\ \citenamefont {Goldman}}]{camino2005realization}%
  \BibitemOpen
  \bibfield  {author} {\bibinfo {author} {\bibfnamefont {F.~E.}\ \bibnamefont
  {Camino}}, \bibinfo {author} {\bibfnamefont {Wei}\ \bibnamefont {Zhou}}, \
  and\ \bibinfo {author} {\bibfnamefont {V.~J.}\ \bibnamefont {Goldman}},\
  }\bibfield  {title} {\enquote {\bibinfo {title} {Realization of a {Laughlin}
  quasiparticle interferometer: Observation of fractional statistics},}\ }\href
  {\doibase 10.1103/PhysRevB.72.075342} {\bibfield  {journal} {\bibinfo
  {journal} {Phys. Rev. B}\ }\textbf {\bibinfo {volume} {72}},\ \bibinfo
  {pages} {075342} (\bibinfo {year} {2005})}\BibitemShut {NoStop}%
\bibitem [{\citenamefont {Willett}\ \emph {et~al.}(2010)\citenamefont
  {Willett}, \citenamefont {Pfeiffer},\ and\ \citenamefont
  {West}}]{willett2010alternation}%
  \BibitemOpen
  \bibfield  {author} {\bibinfo {author} {\bibfnamefont {Robert~L.}\
  \bibnamefont {Willett}}, \bibinfo {author} {\bibfnamefont {Loren~N.}\
  \bibnamefont {Pfeiffer}}, \ and\ \bibinfo {author} {\bibfnamefont {K.~W.}\
  \bibnamefont {West}},\ }\bibfield  {title} {\enquote {\bibinfo {title}
  {Alternation and interchange of e/4 and e/2 period interference oscillations
  consistent with filling factor 5/2 non-{Abelian} quasiparticles},}\ }\href
  {\doibase 10.1103/PhysRevB.82.205301} {\bibfield  {journal} {\bibinfo
  {journal} {Phys. Rev. B}\ }\textbf {\bibinfo {volume} {82}},\ \bibinfo
  {pages} {205301} (\bibinfo {year} {2010})}\BibitemShut {NoStop}%
\bibitem [{\citenamefont {An}\ \emph {et~al.}()\citenamefont {An},
  \citenamefont {Jiang}, \citenamefont {Choi}, \citenamefont {Kang},
  \citenamefont {Simon}, \citenamefont {Pfeiffer}, \citenamefont {West},\ and\
  \citenamefont {Baldwin}}]{an2011braiding}%
  \BibitemOpen
  \bibfield  {author} {\bibinfo {author} {\bibfnamefont {Sanghun}\ \bibnamefont
  {An}}, \bibinfo {author} {\bibfnamefont {P.}~\bibnamefont {Jiang}}, \bibinfo
  {author} {\bibfnamefont {H.}~\bibnamefont {Choi}}, \bibinfo {author}
  {\bibfnamefont {W.}~\bibnamefont {Kang}}, \bibinfo {author} {\bibfnamefont
  {S.~H.}\ \bibnamefont {Simon}}, \bibinfo {author} {\bibfnamefont {L.~N.}\
  \bibnamefont {Pfeiffer}}, \bibinfo {author} {\bibfnamefont {K.~W.}\
  \bibnamefont {West}}, \ and\ \bibinfo {author} {\bibfnamefont {K.~W.}\
  \bibnamefont {Baldwin}},\ }\bibfield  {title} {\enquote {\bibinfo {title}
  {Braiding of {Abelian} and non-{Abelian} anyons in the fractional quantum
  {Hall} effect},}\ }\href@noop {} {\ }\Eprint
  {http://arxiv.org/abs/arXiv:1112.3400} {arXiv:1112.3400} \BibitemShut
  {NoStop}%
\bibitem [{\citenamefont {Kitaev}(2003)}]{kitaev2003fault}%
  \BibitemOpen
  \bibfield  {author} {\bibinfo {author} {\bibfnamefont {A.~Yu}\ \bibnamefont
  {Kitaev}},\ }\bibfield  {title} {\enquote {\bibinfo {title} {Fault-tolerant
  quantum computation by anyons},}\ }\href {\doibase
  10.1016/S0003-4916(02)00018-0} {\bibfield  {journal} {\bibinfo  {journal}
  {Ann. Phys.}\ }\textbf {\bibinfo {volume} {303}},\ \bibinfo {pages} {2}
  (\bibinfo {year} {2003})}\BibitemShut {NoStop}%
\bibitem [{\citenamefont {Nayak}\ \emph {et~al.}(2008)\citenamefont {Nayak},
  \citenamefont {Simon}, \citenamefont {Stern}, \citenamefont {Freedman},\ and\
  \citenamefont {Sarma}}]{nayak2008non}%
  \BibitemOpen
  \bibfield  {author} {\bibinfo {author} {\bibfnamefont {Chetan}\ \bibnamefont
  {Nayak}}, \bibinfo {author} {\bibfnamefont {Steven~H.}\ \bibnamefont
  {Simon}}, \bibinfo {author} {\bibfnamefont {Ady}\ \bibnamefont {Stern}},
  \bibinfo {author} {\bibfnamefont {Michael}\ \bibnamefont {Freedman}}, \ and\
  \bibinfo {author} {\bibfnamefont {Sankar~Das}\ \bibnamefont {Sarma}},\
  }\bibfield  {title} {\enquote {\bibinfo {title} {Non-{Abelian} anyons and
  topological quantum computation},}\ }\href {\doibase
  10.1103/RevModPhys.80.1083} {\bibfield  {journal} {\bibinfo  {journal} {Rev.
  Mod. Phys.}\ }\textbf {\bibinfo {volume} {80}},\ \bibinfo {pages} {1083}
  (\bibinfo {year} {2008})}\BibitemShut {NoStop}%
\bibitem [{\citenamefont {Georgescu}\ \emph {et~al.}(2014)\citenamefont
  {Georgescu}, \citenamefont {Ashhab},\ and\ \citenamefont
  {Nori}}]{georgescu2014quantum}%
  \BibitemOpen
  \bibfield  {author} {\bibinfo {author} {\bibfnamefont {I.~M.}\ \bibnamefont
  {Georgescu}}, \bibinfo {author} {\bibfnamefont {Sahel}\ \bibnamefont
  {Ashhab}}, \ and\ \bibinfo {author} {\bibfnamefont {Franco}\ \bibnamefont
  {Nori}},\ }\bibfield  {title} {\enquote {\bibinfo {title} {Quantum
  simulation},}\ }\href {\doibase 10.1103/RevModPhys.86.153} {\bibfield
  {journal} {\bibinfo  {journal} {Rev. Mod. Phys.}\ }\textbf {\bibinfo {volume}
  {86}},\ \bibinfo {pages} {153} (\bibinfo {year} {2014})}\BibitemShut
  {NoStop}%
\bibitem [{\citenamefont {Bloch}\ \emph {et~al.}(2008)\citenamefont {Bloch},
  \citenamefont {Dalibard},\ and\ \citenamefont {Zwerger}}]{bloch2008many}%
  \BibitemOpen
  \bibfield  {author} {\bibinfo {author} {\bibfnamefont {Immanuel}\
  \bibnamefont {Bloch}}, \bibinfo {author} {\bibfnamefont {Jean}\ \bibnamefont
  {Dalibard}}, \ and\ \bibinfo {author} {\bibfnamefont {Wilhelm}\ \bibnamefont
  {Zwerger}},\ }\bibfield  {title} {\enquote {\bibinfo {title} {Many-body
  physics with ultracold gases},}\ }\href {\doibase 10.1103/RevModPhys.80.885}
  {\bibfield  {journal} {\bibinfo  {journal} {Rev. Mod. Phys.}\ }\textbf
  {\bibinfo {volume} {80}},\ \bibinfo {pages} {885} (\bibinfo {year}
  {2008})}\BibitemShut {NoStop}%
\bibitem [{\citenamefont {Lewenstein}\ \emph {et~al.}(2007)\citenamefont
  {Lewenstein}, \citenamefont {Sanpera}, \citenamefont {Ahufinger},
  \citenamefont {Damski}, \citenamefont {Sen},\ and\ \citenamefont
  {Sen}}]{lewenstein2007ultracold}%
  \BibitemOpen
  \bibfield  {author} {\bibinfo {author} {\bibfnamefont {Maciej}\ \bibnamefont
  {Lewenstein}}, \bibinfo {author} {\bibfnamefont {Anna}\ \bibnamefont
  {Sanpera}}, \bibinfo {author} {\bibfnamefont {Veronica}\ \bibnamefont
  {Ahufinger}}, \bibinfo {author} {\bibfnamefont {Bogdan}\ \bibnamefont
  {Damski}}, \bibinfo {author} {\bibfnamefont {Aditi}\ \bibnamefont {Sen}}, \
  and\ \bibinfo {author} {\bibfnamefont {Ujjwal}\ \bibnamefont {Sen}},\
  }\bibfield  {title} {\enquote {\bibinfo {title} {Ultracold atomic gases in
  optical lattices: mimicking condensed matter physics and beyond},}\ }\href
  {\doibase 10.1080/00018730701223200} {\bibfield  {journal} {\bibinfo
  {journal} {Adv. Phys.}\ }\textbf {\bibinfo {volume} {56}},\ \bibinfo {pages}
  {243} (\bibinfo {year} {2007})}\BibitemShut {NoStop}%
\bibitem [{\citenamefont {Bloch}\ \emph {et~al.}(2012)\citenamefont {Bloch},
  \citenamefont {Dalibard},\ and\ \citenamefont
  {Nascimbene}}]{bloch2012quantum}%
  \BibitemOpen
  \bibfield  {author} {\bibinfo {author} {\bibfnamefont {Immanuel}\
  \bibnamefont {Bloch}}, \bibinfo {author} {\bibfnamefont {Jean}\ \bibnamefont
  {Dalibard}}, \ and\ \bibinfo {author} {\bibfnamefont {Sylvain}\ \bibnamefont
  {Nascimbene}},\ }\bibfield  {title} {\enquote {\bibinfo {title} {Quantum
  simulations with ultracold quantum gases},}\ }\href {\doibase
  10.1038/nphys2259} {\bibfield  {journal} {\bibinfo  {journal} {Nat. Phys.}\
  }\textbf {\bibinfo {volume} {8}},\ \bibinfo {pages} {267} (\bibinfo {year}
  {2012})}\BibitemShut {NoStop}%
\bibitem [{\citenamefont {Carusotto}\ and\ \citenamefont
  {Ciuti}(2013)}]{carusotto2013quantum}%
  \BibitemOpen
  \bibfield  {author} {\bibinfo {author} {\bibfnamefont {Iacopo}\ \bibnamefont
  {Carusotto}}\ and\ \bibinfo {author} {\bibfnamefont {Cristiano}\ \bibnamefont
  {Ciuti}},\ }\bibfield  {title} {\enquote {\bibinfo {title} {Quantum fluids of
  light},}\ }\href {\doibase 10.1103/RevModPhys.85.299} {\bibfield  {journal}
  {\bibinfo  {journal} {Rev. Mod. Phys.}\ }\textbf {\bibinfo {volume} {85}},\
  \bibinfo {pages} {299} (\bibinfo {year} {2013})}\BibitemShut {NoStop}%
\bibitem [{\citenamefont {Noh}\ and\ \citenamefont
  {Angelakis}(2016)}]{noh2016quantum}%
  \BibitemOpen
  \bibfield  {author} {\bibinfo {author} {\bibfnamefont {Changsuk}\
  \bibnamefont {Noh}}\ and\ \bibinfo {author} {\bibfnamefont {Dimitris~G.}\
  \bibnamefont {Angelakis}},\ }\bibfield  {title} {\enquote {\bibinfo {title}
  {Quantum simulations and many-body physics with light},}\ }\href {\doibase
  10.1088/0034-4885/80/1/016401} {\bibfield  {journal} {\bibinfo  {journal}
  {Rep. Prog. Phys.}\ }\textbf {\bibinfo {volume} {80}},\ \bibinfo {pages}
  {016401} (\bibinfo {year} {2016})}\BibitemShut {NoStop}%
\bibitem [{\citenamefont {Hartmann}(2016)}]{hartmann2016quantum}%
  \BibitemOpen
  \bibfield  {author} {\bibinfo {author} {\bibfnamefont {Michael~J.}\
  \bibnamefont {Hartmann}},\ }\bibfield  {title} {\enquote {\bibinfo {title}
  {Quantum simulation with interacting photons},}\ }\href {\doibase
  10.1088/2040-8978/18/10/104005} {\bibfield  {journal} {\bibinfo  {journal}
  {J. Opt.}\ }\textbf {\bibinfo {volume} {18}},\ \bibinfo {pages} {104005}
  (\bibinfo {year} {2016})}\BibitemShut {NoStop}%
\bibitem [{\citenamefont {Houck}\ \emph {et~al.}(2012)\citenamefont {Houck},
  \citenamefont {T{\"u}reci},\ and\ \citenamefont {Koch}}]{houck2012chip}%
  \BibitemOpen
  \bibfield  {author} {\bibinfo {author} {\bibfnamefont {Andrew~A.}\
  \bibnamefont {Houck}}, \bibinfo {author} {\bibfnamefont {Hakan~E.}\
  \bibnamefont {T{\"u}reci}}, \ and\ \bibinfo {author} {\bibfnamefont {Jens}\
  \bibnamefont {Koch}},\ }\bibfield  {title} {\enquote {\bibinfo {title}
  {On-chip quantum simulation with superconducting circuits},}\ }\href
  {\doibase 10.1038/nphys2251} {\bibfield  {journal} {\bibinfo  {journal} {Nat.
  Phys.}\ }\textbf {\bibinfo {volume} {8}},\ \bibinfo {pages} {292} (\bibinfo
  {year} {2012})}\BibitemShut {NoStop}%
\bibitem [{\citenamefont {Laughlin}(1983)}]{laughlin1983anomalous}%
  \BibitemOpen
  \bibfield  {author} {\bibinfo {author} {\bibfnamefont {Robert~B.}\
  \bibnamefont {Laughlin}},\ }\bibfield  {title} {\enquote {\bibinfo {title}
  {Anomalous quantum {Hall} effect: An incompressible quantum fluid with
  fractionally charged excitations},}\ }\href {\doibase
  10.1103/PhysRevLett.50.1395} {\bibfield  {journal} {\bibinfo  {journal}
  {Phys. Rev. Lett.}\ }\textbf {\bibinfo {volume} {50}},\ \bibinfo {pages}
  {1395} (\bibinfo {year} {1983})}\BibitemShut {NoStop}%
\bibitem [{\citenamefont {Wilkin}\ and\ \citenamefont
  {Gunn}(2000)}]{wilkin2000condensation}%
  \BibitemOpen
  \bibfield  {author} {\bibinfo {author} {\bibfnamefont {N.~K.}\ \bibnamefont
  {Wilkin}}\ and\ \bibinfo {author} {\bibfnamefont {J.~M.~F.}\ \bibnamefont
  {Gunn}},\ }\bibfield  {title} {\enquote {\bibinfo {title} {Condensation of
  ``composite bosons'' in a rotating {BEC}},}\ }\href {\doibase
  10.1103/PhysRevLett.84.6} {\bibfield  {journal} {\bibinfo  {journal} {Phys.
  Rev. Lett.}\ }\textbf {\bibinfo {volume} {84}},\ \bibinfo {pages} {6}
  (\bibinfo {year} {2000})}\BibitemShut {NoStop}%
\bibitem [{\citenamefont {Paredes}\ \emph {et~al.}(2001)\citenamefont
  {Paredes}, \citenamefont {Fedichev}, \citenamefont {Cirac},\ and\
  \citenamefont {Zoller}}]{paredes20011}%
  \BibitemOpen
  \bibfield  {author} {\bibinfo {author} {\bibfnamefont {B.}~\bibnamefont
  {Paredes}}, \bibinfo {author} {\bibfnamefont {P.}~\bibnamefont {Fedichev}},
  \bibinfo {author} {\bibfnamefont {J.~I.}\ \bibnamefont {Cirac}}, \ and\
  \bibinfo {author} {\bibfnamefont {P.}~\bibnamefont {Zoller}},\ }\bibfield
  {title} {\enquote {\bibinfo {title} {1/2-anyons in small atomic
  {Bose}-{Einstein} condensates},}\ }\href {\doibase
  10.1103/PhysRevLett.87.010402} {\bibfield  {journal} {\bibinfo  {journal}
  {Phys. Rev. Lett.}\ }\textbf {\bibinfo {volume} {87}},\ \bibinfo {pages}
  {010402} (\bibinfo {year} {2001})}\BibitemShut {NoStop}%
\bibitem [{\citenamefont {Kapit}\ \emph {et~al.}(2012)\citenamefont {Kapit},
  \citenamefont {Ginsparg},\ and\ \citenamefont {Mueller}}]{kapit2012non}%
  \BibitemOpen
  \bibfield  {author} {\bibinfo {author} {\bibfnamefont {Eliot}\ \bibnamefont
  {Kapit}}, \bibinfo {author} {\bibfnamefont {Paul}\ \bibnamefont {Ginsparg}},
  \ and\ \bibinfo {author} {\bibfnamefont {Erich}\ \bibnamefont {Mueller}},\
  }\bibfield  {title} {\enquote {\bibinfo {title} {Non-{Abelian} braiding of
  lattice bosons},}\ }\href {\doibase 10.1103/PhysRevLett.108.066802}
  {\bibfield  {journal} {\bibinfo  {journal} {Phys. Rev. Lett.}\ }\textbf
  {\bibinfo {volume} {108}},\ \bibinfo {pages} {066802} (\bibinfo {year}
  {2012})}\BibitemShut {NoStop}%
\bibitem [{\citenamefont {Cooper}\ \emph {et~al.}(2001)\citenamefont {Cooper},
  \citenamefont {Wilkin},\ and\ \citenamefont {Gunn}}]{cooper2001quantum}%
  \BibitemOpen
  \bibfield  {author} {\bibinfo {author} {\bibfnamefont {Nigel~R.}\
  \bibnamefont {Cooper}}, \bibinfo {author} {\bibfnamefont {Nicola~K.}\
  \bibnamefont {Wilkin}}, \ and\ \bibinfo {author} {\bibfnamefont {J.~M.~F.}\
  \bibnamefont {Gunn}},\ }\bibfield  {title} {\enquote {\bibinfo {title}
  {Quantum phases of vortices in rotating {Bose}-{Einstein} condensates},}\
  }\href {\doibase 10.1103/PhysRevLett.87.120405} {\bibfield  {journal}
  {\bibinfo  {journal} {Phys. Rev. Lett.}\ }\textbf {\bibinfo {volume} {87}},\
  \bibinfo {pages} {120405} (\bibinfo {year} {2001})}\BibitemShut {NoStop}%
\bibitem [{\citenamefont {Paredes}\ \emph {et~al.}(2003)\citenamefont
  {Paredes}, \citenamefont {Zoller},\ and\ \citenamefont
  {Cirac}}]{paredes2003fractional}%
  \BibitemOpen
  \bibfield  {author} {\bibinfo {author} {\bibfnamefont {Belen}\ \bibnamefont
  {Paredes}}, \bibinfo {author} {\bibfnamefont {P.}~\bibnamefont {Zoller}}, \
  and\ \bibinfo {author} {\bibfnamefont {J.~Ignacio}\ \bibnamefont {Cirac}},\
  }\bibfield  {title} {\enquote {\bibinfo {title} {Fractional quantum {Hall}
  regime of a gas of ultracold atoms},}\ }\href {\doibase
  10.1016/S0038-1098(03)00314-4} {\bibfield  {journal} {\bibinfo  {journal}
  {Solid State Commun.}\ }\textbf {\bibinfo {volume} {127}},\ \bibinfo {pages}
  {155} (\bibinfo {year} {2003})}\BibitemShut {NoStop}%
\bibitem [{\citenamefont {Popp}\ \emph {et~al.}(2004)\citenamefont {Popp},
  \citenamefont {Paredes},\ and\ \citenamefont {Cirac}}]{popp2004adiabatic}%
  \BibitemOpen
  \bibfield  {author} {\bibinfo {author} {\bibfnamefont {Markus}\ \bibnamefont
  {Popp}}, \bibinfo {author} {\bibfnamefont {Belen}\ \bibnamefont {Paredes}}, \
  and\ \bibinfo {author} {\bibfnamefont {J.~Ignacio}\ \bibnamefont {Cirac}},\
  }\bibfield  {title} {\enquote {\bibinfo {title} {Adiabatic path to fractional
  quantum {Hall} states of a few bosonic atoms},}\ }\href {\doibase
  10.1103/PhysRevA.70.053612} {\bibfield  {journal} {\bibinfo  {journal} {Phys.
  Rev. A}\ }\textbf {\bibinfo {volume} {70}},\ \bibinfo {pages} {053612}
  (\bibinfo {year} {2004})}\BibitemShut {NoStop}%
\bibitem [{\citenamefont {Chang}\ \emph {et~al.}(2005)\citenamefont {Chang},
  \citenamefont {Regnault}, \citenamefont {Jolicoeur},\ and\ \citenamefont
  {Jain}}]{chang2005composite}%
  \BibitemOpen
  \bibfield  {author} {\bibinfo {author} {\bibfnamefont {Chia-Chen}\
  \bibnamefont {Chang}}, \bibinfo {author} {\bibfnamefont {Nicolas}\
  \bibnamefont {Regnault}}, \bibinfo {author} {\bibfnamefont {Thierry}\
  \bibnamefont {Jolicoeur}}, \ and\ \bibinfo {author} {\bibfnamefont
  {Jainendra~K.}\ \bibnamefont {Jain}},\ }\bibfield  {title} {\enquote
  {\bibinfo {title} {Composite fermionization of bosons in rapidly rotating
  atomic traps},}\ }\href {\doibase 10.1103/PhysRevA.72.013611} {\bibfield
  {journal} {\bibinfo  {journal} {Phys. Rev. A}\ }\textbf {\bibinfo {volume}
  {72}},\ \bibinfo {pages} {013611} (\bibinfo {year} {2005})}\BibitemShut
  {NoStop}%
\bibitem [{\citenamefont {Baranov}\ \emph {et~al.}(2005)\citenamefont
  {Baranov}, \citenamefont {Osterloh},\ and\ \citenamefont
  {Lewenstein}}]{baranov2005fractional}%
  \BibitemOpen
  \bibfield  {author} {\bibinfo {author} {\bibfnamefont {M.~A.}\ \bibnamefont
  {Baranov}}, \bibinfo {author} {\bibfnamefont {Klaus}\ \bibnamefont
  {Osterloh}}, \ and\ \bibinfo {author} {\bibfnamefont {M.}~\bibnamefont
  {Lewenstein}},\ }\bibfield  {title} {\enquote {\bibinfo {title} {Fractional
  quantum {Hall} states in ultracold rapidly rotating dipolar {Fermi} gases},}\
  }\href {\doibase 10.1103/PhysRevLett.94.070404} {\bibfield  {journal}
  {\bibinfo  {journal} {Phys. Rev. Lett.}\ }\textbf {\bibinfo {volume} {94}},\
  \bibinfo {pages} {070404} (\bibinfo {year} {2005})}\BibitemShut {NoStop}%
\bibitem [{\citenamefont {S{\o}rensen}\ \emph {et~al.}(2005)\citenamefont
  {S{\o}rensen}, \citenamefont {Demler},\ and\ \citenamefont
  {Lukin}}]{sorensen2005fractional}%
  \BibitemOpen
  \bibfield  {author} {\bibinfo {author} {\bibfnamefont {Anders~S.}\
  \bibnamefont {S{\o}rensen}}, \bibinfo {author} {\bibfnamefont {Eugene}\
  \bibnamefont {Demler}}, \ and\ \bibinfo {author} {\bibfnamefont {Mikhail~D.}\
  \bibnamefont {Lukin}},\ }\bibfield  {title} {\enquote {\bibinfo {title}
  {Fractional quantum {Hall} states of atoms in optical lattices},}\ }\href
  {\doibase 10.1103/PhysRevLett.94.086803} {\bibfield  {journal} {\bibinfo
  {journal} {Phys. Rev. Lett.}\ }\textbf {\bibinfo {volume} {94}},\ \bibinfo
  {pages} {086803} (\bibinfo {year} {2005})}\BibitemShut {NoStop}%
\bibitem [{\citenamefont {Palmer}\ and\ \citenamefont
  {Jaksch}(2006)}]{palmer2006high}%
  \BibitemOpen
  \bibfield  {author} {\bibinfo {author} {\bibfnamefont {R.~N.}\ \bibnamefont
  {Palmer}}\ and\ \bibinfo {author} {\bibfnamefont {D.}~\bibnamefont
  {Jaksch}},\ }\bibfield  {title} {\enquote {\bibinfo {title} {High-field
  fractional quantum {Hall} effect in optical lattices},}\ }\href {\doibase
  10.1103/PhysRevLett.96.180407} {\bibfield  {journal} {\bibinfo  {journal}
  {Phys. Rev. Lett.}\ }\textbf {\bibinfo {volume} {96}},\ \bibinfo {pages}
  {180407} (\bibinfo {year} {2006})}\BibitemShut {NoStop}%
\bibitem [{\citenamefont {Hafezi}\ \emph {et~al.}(2007)\citenamefont {Hafezi},
  \citenamefont {S{\o}rensen}, \citenamefont {Demler},\ and\ \citenamefont
  {Lukin}}]{hafezi2007fractional}%
  \BibitemOpen
  \bibfield  {author} {\bibinfo {author} {\bibfnamefont {Mohammad}\
  \bibnamefont {Hafezi}}, \bibinfo {author} {\bibfnamefont
  {Anders~S{\o}ndberg}\ \bibnamefont {S{\o}rensen}}, \bibinfo {author}
  {\bibfnamefont {Eugene}\ \bibnamefont {Demler}}, \ and\ \bibinfo {author}
  {\bibfnamefont {Mikhail~D.}\ \bibnamefont {Lukin}},\ }\bibfield  {title}
  {\enquote {\bibinfo {title} {Fractional quantum {Hall} effect in optical
  lattices},}\ }\href {\doibase 10.1103/PhysRevA.76.023613} {\bibfield
  {journal} {\bibinfo  {journal} {Phys. Rev. A}\ }\textbf {\bibinfo {volume}
  {76}},\ \bibinfo {pages} {023613} (\bibinfo {year} {2007})}\BibitemShut
  {NoStop}%
\bibitem [{\citenamefont {Bhat}\ \emph {et~al.}(2007)\citenamefont {Bhat},
  \citenamefont {Kr{\"a}mer}, \citenamefont {Cooper},\ and\ \citenamefont
  {Holland}}]{bhat2007hall}%
  \BibitemOpen
  \bibfield  {author} {\bibinfo {author} {\bibfnamefont {Rajiv}\ \bibnamefont
  {Bhat}}, \bibinfo {author} {\bibfnamefont {M.}~\bibnamefont {Kr{\"a}mer}},
  \bibinfo {author} {\bibfnamefont {J.}~\bibnamefont {Cooper}}, \ and\ \bibinfo
  {author} {\bibfnamefont {M.~J.}\ \bibnamefont {Holland}},\ }\bibfield
  {title} {\enquote {\bibinfo {title} {Hall effects in {Bose}-{Einstein}
  condensates in a rotating optical lattice},}\ }\href {\doibase
  10.1103/PhysRevA.76.043601} {\bibfield  {journal} {\bibinfo  {journal} {Phys.
  Rev. A}\ }\textbf {\bibinfo {volume} {76}},\ \bibinfo {pages} {043601}
  (\bibinfo {year} {2007})}\BibitemShut {NoStop}%
\bibitem [{\citenamefont {Baur}\ \emph {et~al.}(2008)\citenamefont {Baur},
  \citenamefont {Hazzard},\ and\ \citenamefont {Mueller}}]{baur2008stirring}%
  \BibitemOpen
  \bibfield  {author} {\bibinfo {author} {\bibfnamefont {Stefan~K.}\
  \bibnamefont {Baur}}, \bibinfo {author} {\bibfnamefont {Kaden R.~A.}\
  \bibnamefont {Hazzard}}, \ and\ \bibinfo {author} {\bibfnamefont {Erich~J.}\
  \bibnamefont {Mueller}},\ }\bibfield  {title} {\enquote {\bibinfo {title}
  {Stirring trapped atoms into fractional quantum {Hall} puddles},}\ }\href
  {\doibase 10.1103/PhysRevA.78.061608} {\bibfield  {journal} {\bibinfo
  {journal} {Phys. Rev. A}\ }\textbf {\bibinfo {volume} {78}},\ \bibinfo
  {pages} {061608} (\bibinfo {year} {2008})}\BibitemShut {NoStop}%
\bibitem [{\citenamefont {M{\"o}ller}\ and\ \citenamefont
  {Cooper}(2009)}]{moller2009composite}%
  \BibitemOpen
  \bibfield  {author} {\bibinfo {author} {\bibfnamefont {Gunnar}\ \bibnamefont
  {M{\"o}ller}}\ and\ \bibinfo {author} {\bibfnamefont {Nigel~R.}\ \bibnamefont
  {Cooper}},\ }\bibfield  {title} {\enquote {\bibinfo {title} {Composite
  fermion theory for bosonic quantum {Hall} states on lattices},}\ }\href
  {\doibase 10.1103/PhysRevLett.103.105303} {\bibfield  {journal} {\bibinfo
  {journal} {Phys. Rev. Lett.}\ }\textbf {\bibinfo {volume} {103}},\ \bibinfo
  {pages} {105303} (\bibinfo {year} {2009})}\BibitemShut {NoStop}%
\bibitem [{\citenamefont {Gemelke}\ \emph {et~al.}()\citenamefont {Gemelke},
  \citenamefont {Sarajlic},\ and\ \citenamefont {Chu}}]{gemelke2010rotating}%
  \BibitemOpen
  \bibfield  {author} {\bibinfo {author} {\bibfnamefont {Nathan}\ \bibnamefont
  {Gemelke}}, \bibinfo {author} {\bibfnamefont {Edina}\ \bibnamefont
  {Sarajlic}}, \ and\ \bibinfo {author} {\bibfnamefont {Steven}\ \bibnamefont
  {Chu}},\ }\bibfield  {title} {\enquote {\bibinfo {title} {Rotating few-body
  atomic systems in the fractional quantum {Hall} regime},}\ }\href@noop {} {\
  }\Eprint {http://arxiv.org/abs/arXiv:1007.2677} {arXiv:1007.2677}
  \BibitemShut {NoStop}%
\bibitem [{\citenamefont {Kapit}\ and\ \citenamefont
  {Mueller}(2010)}]{kapit2010exact}%
  \BibitemOpen
  \bibfield  {author} {\bibinfo {author} {\bibfnamefont {Eliot}\ \bibnamefont
  {Kapit}}\ and\ \bibinfo {author} {\bibfnamefont {Erich}\ \bibnamefont
  {Mueller}},\ }\bibfield  {title} {\enquote {\bibinfo {title} {Exact parent
  {Hamiltonian} for the quantum {Hall} states in a lattice},}\ }\href {\doibase
  10.1103/PhysRevLett.105.215303} {\bibfield  {journal} {\bibinfo  {journal}
  {Phys. Rev. Lett.}\ }\textbf {\bibinfo {volume} {105}},\ \bibinfo {pages}
  {215303} (\bibinfo {year} {2010})}\BibitemShut {NoStop}%
\bibitem [{\citenamefont {Roncaglia}\ \emph {et~al.}(2011)\citenamefont
  {Roncaglia}, \citenamefont {Rizzi},\ and\ \citenamefont
  {Dalibard}}]{roncaglia2011rotating}%
  \BibitemOpen
  \bibfield  {author} {\bibinfo {author} {\bibfnamefont {Marco}\ \bibnamefont
  {Roncaglia}}, \bibinfo {author} {\bibfnamefont {Matteo}\ \bibnamefont
  {Rizzi}}, \ and\ \bibinfo {author} {\bibfnamefont {Jean}\ \bibnamefont
  {Dalibard}},\ }\bibfield  {title} {\enquote {\bibinfo {title} {From rotating
  atomic rings to quantum {Hall} states},}\ }\href {\doibase 10.1038/srep00043}
  {\bibfield  {journal} {\bibinfo  {journal} {Sci. Rep.}\ }\textbf {\bibinfo
  {volume} {1}},\ \bibinfo {pages} {43} (\bibinfo {year} {2011})}\BibitemShut
  {NoStop}%
\bibitem [{\citenamefont {Juli{\'a}-D{\'\i}az}\ \emph
  {et~al.}(2012)\citenamefont {Juli{\'a}-D{\'\i}az}, \citenamefont {Gra{\ss}},
  \citenamefont {Barber{\'a}n},\ and\ \citenamefont
  {Lewenstein}}]{julia2012fractional}%
  \BibitemOpen
  \bibfield  {author} {\bibinfo {author} {\bibfnamefont {B.}~\bibnamefont
  {Juli{\'a}-D{\'\i}az}}, \bibinfo {author} {\bibfnamefont {T.}~\bibnamefont
  {Gra{\ss}}}, \bibinfo {author} {\bibfnamefont {N.}~\bibnamefont
  {Barber{\'a}n}}, \ and\ \bibinfo {author} {\bibfnamefont {M.}~\bibnamefont
  {Lewenstein}},\ }\bibfield  {title} {\enquote {\bibinfo {title} {Fractional
  quantum {Hall} states of a few bosonic atoms in geometric gauge fields},}\
  }\href {\doibase 10.1088/1367-2630/14/5/055003} {\bibfield  {journal}
  {\bibinfo  {journal} {New J. Phys.}\ }\textbf {\bibinfo {volume} {14}},\
  \bibinfo {pages} {055003} (\bibinfo {year} {2012})}\BibitemShut {NoStop}%
\bibitem [{\citenamefont {Nielsen}\ \emph {et~al.}(2013)\citenamefont
  {Nielsen}, \citenamefont {Sierra},\ and\ \citenamefont
  {Cirac}}]{nielsen2013local}%
  \BibitemOpen
  \bibfield  {author} {\bibinfo {author} {\bibfnamefont {Anne E.~B.}\
  \bibnamefont {Nielsen}}, \bibinfo {author} {\bibfnamefont {Germ{\'a}n}\
  \bibnamefont {Sierra}}, \ and\ \bibinfo {author} {\bibfnamefont {J.~Ignacio}\
  \bibnamefont {Cirac}},\ }\bibfield  {title} {\enquote {\bibinfo {title}
  {Local models of fractional quantum {Hall} states in lattices and physical
  implementation},}\ }\href {\doibase 10.1038/ncomms3864} {\bibfield  {journal}
  {\bibinfo  {journal} {Nat. Commun.}\ }\textbf {\bibinfo {volume} {4}},\
  \bibinfo {pages} {2864} (\bibinfo {year} {2013})}\BibitemShut {NoStop}%
\bibitem [{\citenamefont {Cooper}\ and\ \citenamefont
  {Dalibard}(2013)}]{cooper2013reaching}%
  \BibitemOpen
  \bibfield  {author} {\bibinfo {author} {\bibfnamefont {Nigel~R.}\
  \bibnamefont {Cooper}}\ and\ \bibinfo {author} {\bibfnamefont {Jean}\
  \bibnamefont {Dalibard}},\ }\bibfield  {title} {\enquote {\bibinfo {title}
  {Reaching fractional quantum {Hall} states with optical flux lattices},}\
  }\href {\doibase 10.1103/PhysRevLett.110.185301} {\bibfield  {journal}
  {\bibinfo  {journal} {Phys. Rev. Lett.}\ }\textbf {\bibinfo {volume} {110}},\
  \bibinfo {pages} {185301} (\bibinfo {year} {2013})}\BibitemShut {NoStop}%
\bibitem [{\citenamefont {Zhang}\ \emph {et~al.}(2016)\citenamefont {Zhang},
  \citenamefont {Beugnon},\ and\ \citenamefont
  {Nascimbene}}]{zhang2016creating}%
  \BibitemOpen
  \bibfield  {author} {\bibinfo {author} {\bibfnamefont {Junyi}\ \bibnamefont
  {Zhang}}, \bibinfo {author} {\bibfnamefont {J{\'e}r{\^o}me}\ \bibnamefont
  {Beugnon}}, \ and\ \bibinfo {author} {\bibfnamefont {Sylvain}\ \bibnamefont
  {Nascimbene}},\ }\bibfield  {title} {\enquote {\bibinfo {title} {Creating
  fractional quantum {Hall} states with atomic clusters using light-assisted
  insertion of angular momentum},}\ }\href {\doibase
  10.1103/PhysRevA.94.043610} {\bibfield  {journal} {\bibinfo  {journal} {Phys.
  Rev. A}\ }\textbf {\bibinfo {volume} {94}},\ \bibinfo {pages} {043610}
  (\bibinfo {year} {2016})}\BibitemShut {NoStop}%
\bibitem [{\citenamefont {He}\ \emph {et~al.}(2017)\citenamefont {He},
  \citenamefont {Grusdt}, \citenamefont {Kaufman}, \citenamefont {Greiner},\
  and\ \citenamefont {Vishwanath}}]{he2017realizing}%
  \BibitemOpen
  \bibfield  {author} {\bibinfo {author} {\bibfnamefont {Yin-Chen}\
  \bibnamefont {He}}, \bibinfo {author} {\bibfnamefont {Fabian}\ \bibnamefont
  {Grusdt}}, \bibinfo {author} {\bibfnamefont {Adam}\ \bibnamefont {Kaufman}},
  \bibinfo {author} {\bibfnamefont {Markus}\ \bibnamefont {Greiner}}, \ and\
  \bibinfo {author} {\bibfnamefont {Ashvin}\ \bibnamefont {Vishwanath}},\
  }\bibfield  {title} {\enquote {\bibinfo {title} {Realizing and adiabatically
  preparing bosonic integer and fractional quantum {Hall} states in optical
  lattices},}\ }\href {\doibase 10.1103/PhysRevB.96.201103} {\bibfield
  {journal} {\bibinfo  {journal} {Phys. Rev. B}\ }\textbf {\bibinfo {volume}
  {96}},\ \bibinfo {pages} {201103(R)} (\bibinfo {year} {2017})}\BibitemShut
  {NoStop}%
\bibitem [{\citenamefont {Cho}\ \emph {et~al.}(2008)\citenamefont {Cho},
  \citenamefont {Angelakis},\ and\ \citenamefont {Bose}}]{cho2008fractional}%
  \BibitemOpen
  \bibfield  {author} {\bibinfo {author} {\bibfnamefont {Jaeyoon}\ \bibnamefont
  {Cho}}, \bibinfo {author} {\bibfnamefont {Dimitris~G.}\ \bibnamefont
  {Angelakis}}, \ and\ \bibinfo {author} {\bibfnamefont {Sougato}\ \bibnamefont
  {Bose}},\ }\bibfield  {title} {\enquote {\bibinfo {title} {Fractional quantum
  {Hall} state in coupled cavities},}\ }\href {\doibase
  10.1103/PhysRevLett.101.246809} {\bibfield  {journal} {\bibinfo  {journal}
  {Phys. Rev. Lett.}\ }\textbf {\bibinfo {volume} {101}},\ \bibinfo {pages}
  {246809} (\bibinfo {year} {2008})}\BibitemShut {NoStop}%
\bibitem [{\citenamefont {Hayward}\ \emph {et~al.}(2012)\citenamefont
  {Hayward}, \citenamefont {Martin},\ and\ \citenamefont
  {Greentree}}]{hayward2012fractional}%
  \BibitemOpen
  \bibfield  {author} {\bibinfo {author} {\bibfnamefont {Andrew L.~C.}\
  \bibnamefont {Hayward}}, \bibinfo {author} {\bibfnamefont {Andrew~M.}\
  \bibnamefont {Martin}}, \ and\ \bibinfo {author} {\bibfnamefont {Andrew~D.}\
  \bibnamefont {Greentree}},\ }\bibfield  {title} {\enquote {\bibinfo {title}
  {Fractional quantum {Hall} physics in {Jaynes}-{Cummings}-{Hubbard}
  lattices},}\ }\href {\doibase 10.1103/PhysRevLett.108.223602} {\bibfield
  {journal} {\bibinfo  {journal} {Phys. Rev. Lett.}\ }\textbf {\bibinfo
  {volume} {108}},\ \bibinfo {pages} {223602} (\bibinfo {year}
  {2012})}\BibitemShut {NoStop}%
\bibitem [{\citenamefont {Maghrebi}\ \emph {et~al.}(2015)\citenamefont
  {Maghrebi}, \citenamefont {Yao}, \citenamefont {Hafezi}, \citenamefont
  {Pohl}, \citenamefont {Firstenberg},\ and\ \citenamefont
  {Gorshkov}}]{maghrebi2015fractional}%
  \BibitemOpen
  \bibfield  {author} {\bibinfo {author} {\bibfnamefont {Mohammad~F.}\
  \bibnamefont {Maghrebi}}, \bibinfo {author} {\bibfnamefont {Norman~Y.}\
  \bibnamefont {Yao}}, \bibinfo {author} {\bibfnamefont {Mohammad}\
  \bibnamefont {Hafezi}}, \bibinfo {author} {\bibfnamefont {Thomas}\
  \bibnamefont {Pohl}}, \bibinfo {author} {\bibfnamefont {Ofer}\ \bibnamefont
  {Firstenberg}}, \ and\ \bibinfo {author} {\bibfnamefont {Alexey~V.}\
  \bibnamefont {Gorshkov}},\ }\bibfield  {title} {\enquote {\bibinfo {title}
  {Fractional quantum {Hall} states of {Rydberg} polaritons},}\ }\href
  {\doibase 10.1103/PhysRevA.91.033838} {\bibfield  {journal} {\bibinfo
  {journal} {Phys. Rev. A}\ }\textbf {\bibinfo {volume} {91}},\ \bibinfo
  {pages} {033838} (\bibinfo {year} {2015})}\BibitemShut {NoStop}%
\bibitem [{\citenamefont {Anderson}\ \emph {et~al.}(2016)\citenamefont
  {Anderson}, \citenamefont {Ma}, \citenamefont {Owens}, \citenamefont
  {Schuster},\ and\ \citenamefont {Simon}}]{anderson2016engineering}%
  \BibitemOpen
  \bibfield  {author} {\bibinfo {author} {\bibfnamefont {Brandon~M.}\
  \bibnamefont {Anderson}}, \bibinfo {author} {\bibfnamefont {Ruichao}\
  \bibnamefont {Ma}}, \bibinfo {author} {\bibfnamefont {Clai}\ \bibnamefont
  {Owens}}, \bibinfo {author} {\bibfnamefont {David~I.}\ \bibnamefont
  {Schuster}}, \ and\ \bibinfo {author} {\bibfnamefont {Jonathan}\ \bibnamefont
  {Simon}},\ }\bibfield  {title} {\enquote {\bibinfo {title} {Engineering
  topological many-body materials in microwave cavity arrays},}\ }\href
  {\doibase 10.1103/PhysRevX.6.041043} {\bibfield  {journal} {\bibinfo
  {journal} {Phys. Rev. X}\ }\textbf {\bibinfo {volume} {6}},\ \bibinfo {pages}
  {041043} (\bibinfo {year} {2016})}\BibitemShut {NoStop}%
\bibitem [{\citenamefont {Umucal{\i}lar}\ and\ \citenamefont
  {Carusotto}(2013)}]{umucalilar2013many}%
  \BibitemOpen
  \bibfield  {author} {\bibinfo {author} {\bibfnamefont {R.~O.}\ \bibnamefont
  {Umucal{\i}lar}}\ and\ \bibinfo {author} {\bibfnamefont {I.}~\bibnamefont
  {Carusotto}},\ }\bibfield  {title} {\enquote {\bibinfo {title} {Many-body
  braiding phases in a rotating strongly correlated photon gas},}\ }\href
  {\doibase 10.1016/j.physleta.2013.06.011} {\bibfield  {journal} {\bibinfo
  {journal} {Phys. Lett. A}\ }\textbf {\bibinfo {volume} {377}},\ \bibinfo
  {pages} {2074} (\bibinfo {year} {2013})}\BibitemShut {NoStop}%
\bibitem [{\citenamefont {Umucal{\i}lar}\ \emph {et~al.}(2014)\citenamefont
  {Umucal{\i}lar}, \citenamefont {Wouters},\ and\ \citenamefont
  {Carusotto}}]{umucalilar2014probing}%
  \BibitemOpen
  \bibfield  {author} {\bibinfo {author} {\bibfnamefont {R.~O.}\ \bibnamefont
  {Umucal{\i}lar}}, \bibinfo {author} {\bibfnamefont {M.}~\bibnamefont
  {Wouters}}, \ and\ \bibinfo {author} {\bibfnamefont {I.}~\bibnamefont
  {Carusotto}},\ }\bibfield  {title} {\enquote {\bibinfo {title} {Probing
  few-particle {Laughlin} states of photons via correlation measurements},}\
  }\href {\doibase 10.1103/PhysRevA.89.023803} {\bibfield  {journal} {\bibinfo
  {journal} {Phys. Rev. A}\ }\textbf {\bibinfo {volume} {89}},\ \bibinfo
  {pages} {023803} (\bibinfo {year} {2014})}\BibitemShut {NoStop}%
\bibitem [{\citenamefont {Umucal{\i}lar}\ and\ \citenamefont
  {Carusotto}(2012)}]{umucalilar2012fractional}%
  \BibitemOpen
  \bibfield  {author} {\bibinfo {author} {\bibfnamefont {R.~O.}\ \bibnamefont
  {Umucal{\i}lar}}\ and\ \bibinfo {author} {\bibfnamefont {I.}~\bibnamefont
  {Carusotto}},\ }\bibfield  {title} {\enquote {\bibinfo {title} {Fractional
  quantum {Hall} states of photons in an array of dissipative coupled
  cavities},}\ }\href {\doibase 10.1103/PhysRevLett.108.206809} {\bibfield
  {journal} {\bibinfo  {journal} {Phys. Rev. Lett.}\ }\textbf {\bibinfo
  {volume} {108}},\ \bibinfo {pages} {206809} (\bibinfo {year}
  {2012})}\BibitemShut {NoStop}%
\bibitem [{\citenamefont {Hafezi}\ \emph
  {et~al.}(2013{\natexlab{a}})\citenamefont {Hafezi}, \citenamefont {Lukin},\
  and\ \citenamefont {Taylor}}]{hafezi2013non}%
  \BibitemOpen
  \bibfield  {author} {\bibinfo {author} {\bibfnamefont {Mohammad}\
  \bibnamefont {Hafezi}}, \bibinfo {author} {\bibfnamefont {Mikhail~D.}\
  \bibnamefont {Lukin}}, \ and\ \bibinfo {author} {\bibfnamefont {Jacob~M.}\
  \bibnamefont {Taylor}},\ }\bibfield  {title} {\enquote {\bibinfo {title}
  {Non-equilibrium fractional quantum {Hall} state of light},}\ }\href
  {\doibase 10.1088/1367-2630/15/6/063001} {\bibfield  {journal} {\bibinfo
  {journal} {New J. Phys.}\ }\textbf {\bibinfo {volume} {15}},\ \bibinfo
  {pages} {063001} (\bibinfo {year} {2013}{\natexlab{a}})}\BibitemShut
  {NoStop}%
\bibitem [{\citenamefont {Umucalilar}\ and\ \citenamefont
  {Carusotto}(2017)}]{umucalilar2017generation}%
  \BibitemOpen
  \bibfield  {author} {\bibinfo {author} {\bibfnamefont {R.~O.}\ \bibnamefont
  {Umucalilar}}\ and\ \bibinfo {author} {\bibfnamefont {I.}~\bibnamefont
  {Carusotto}},\ }\bibfield  {title} {\enquote {\bibinfo {title} {Generation
  and spectroscopic signatures of a fractional quantum {Hall} liquid of photons
  in an incoherently pumped optical cavity},}\ }\href {\doibase
  10.1103/PhysRevA.96.053808} {\bibfield  {journal} {\bibinfo  {journal} {Phys.
  Rev. A}\ }\textbf {\bibinfo {volume} {96}},\ \bibinfo {pages} {053808}
  (\bibinfo {year} {2017})}\BibitemShut {NoStop}%
\bibitem [{\citenamefont {Kapit}\ \emph {et~al.}(2014)\citenamefont {Kapit},
  \citenamefont {Hafezi},\ and\ \citenamefont {Simon}}]{kapit2014induced}%
  \BibitemOpen
  \bibfield  {author} {\bibinfo {author} {\bibfnamefont {Eliot}\ \bibnamefont
  {Kapit}}, \bibinfo {author} {\bibfnamefont {Mohammad}\ \bibnamefont
  {Hafezi}}, \ and\ \bibinfo {author} {\bibfnamefont {Steven~H.}\ \bibnamefont
  {Simon}},\ }\bibfield  {title} {\enquote {\bibinfo {title} {Induced
  self-stabilization in fractional quantum {Hall} states of light},}\ }\href
  {\doibase 10.1103/PhysRevX.4.031039} {\bibfield  {journal} {\bibinfo
  {journal} {Phys. Rev. X}\ }\textbf {\bibinfo {volume} {4}},\ \bibinfo {pages}
  {031039} (\bibinfo {year} {2014})}\BibitemShut {NoStop}%
\bibitem [{\citenamefont {Grusdt}\ \emph {et~al.}(2014)\citenamefont {Grusdt},
  \citenamefont {Letscher}, \citenamefont {Hafezi},\ and\ \citenamefont
  {Fleischhauer}}]{grusdt2014topological}%
  \BibitemOpen
  \bibfield  {author} {\bibinfo {author} {\bibfnamefont {Fabian}\ \bibnamefont
  {Grusdt}}, \bibinfo {author} {\bibfnamefont {Fabian}\ \bibnamefont
  {Letscher}}, \bibinfo {author} {\bibfnamefont {Mohammad}\ \bibnamefont
  {Hafezi}}, \ and\ \bibinfo {author} {\bibfnamefont {Michael}\ \bibnamefont
  {Fleischhauer}},\ }\bibfield  {title} {\enquote {\bibinfo {title}
  {Topological growing of {Laughlin} states in synthetic gauge fields},}\
  }\href {\doibase 10.1103/PhysRevLett.113.155301} {\bibfield  {journal}
  {\bibinfo  {journal} {Phys. Rev. Lett.}\ }\textbf {\bibinfo {volume} {113}},\
  \bibinfo {pages} {155301} (\bibinfo {year} {2014})}\BibitemShut {NoStop}%
\bibitem [{\citenamefont {Ozawa}\ \emph {et~al.}()\citenamefont {Ozawa},
  \citenamefont {Price}, \citenamefont {Amo}, \citenamefont {Goldman},
  \citenamefont {Hafezi}, \citenamefont {Lu}, \citenamefont {Rechtsman},
  \citenamefont {Schuster}, \citenamefont {Simon}, \citenamefont {Zilberberg}
  \emph {et~al.}}]{ozawa2018topological}%
  \BibitemOpen
  \bibfield  {author} {\bibinfo {author} {\bibfnamefont {Tomoki}\ \bibnamefont
  {Ozawa}}, \bibinfo {author} {\bibfnamefont {Hannah~M.}\ \bibnamefont
  {Price}}, \bibinfo {author} {\bibfnamefont {Alberto}\ \bibnamefont {Amo}},
  \bibinfo {author} {\bibfnamefont {Nathan}\ \bibnamefont {Goldman}}, \bibinfo
  {author} {\bibfnamefont {Mohammad}\ \bibnamefont {Hafezi}}, \bibinfo {author}
  {\bibfnamefont {Ling}\ \bibnamefont {Lu}}, \bibinfo {author} {\bibfnamefont
  {Mikael}\ \bibnamefont {Rechtsman}}, \bibinfo {author} {\bibfnamefont
  {David}\ \bibnamefont {Schuster}}, \bibinfo {author} {\bibfnamefont
  {Jonathan}\ \bibnamefont {Simon}}, \bibinfo {author} {\bibfnamefont {Oded}\
  \bibnamefont {Zilberberg}},  \emph {et~al.},\ }\bibfield  {title} {\enquote
  {\bibinfo {title} {Topological photonics},}\ }\href@noop {} {\ }\Eprint
  {http://arxiv.org/abs/arXiv:1802.04173} {arXiv:1802.04173} \BibitemShut
  {NoStop}%
\bibitem [{\citenamefont {Lu}\ \emph {et~al.}(2014)\citenamefont {Lu},
  \citenamefont {Joannopoulos},\ and\ \citenamefont
  {Solja{\v{c}}i{\'c}}}]{lu2014topological}%
  \BibitemOpen
  \bibfield  {author} {\bibinfo {author} {\bibfnamefont {Ling}\ \bibnamefont
  {Lu}}, \bibinfo {author} {\bibfnamefont {John~D.}\ \bibnamefont
  {Joannopoulos}}, \ and\ \bibinfo {author} {\bibfnamefont {Marin}\
  \bibnamefont {Solja{\v{c}}i{\'c}}},\ }\bibfield  {title} {\enquote {\bibinfo
  {title} {Topological photonics},}\ }\href {\doibase 10.1038/nphoton.2014.248}
  {\bibfield  {journal} {\bibinfo  {journal} {Nat. Photon.}\ }\textbf {\bibinfo
  {volume} {8}},\ \bibinfo {pages} {821} (\bibinfo {year} {2014})}\BibitemShut
  {NoStop}%
\bibitem [{\citenamefont {Chang}\ \emph {et~al.}(2014)\citenamefont {Chang},
  \citenamefont {Vuleti{\'c}},\ and\ \citenamefont {Lukin}}]{chang2014quantum}%
  \BibitemOpen
  \bibfield  {author} {\bibinfo {author} {\bibfnamefont {Darrick~E.}\
  \bibnamefont {Chang}}, \bibinfo {author} {\bibfnamefont {Vladan}\
  \bibnamefont {Vuleti{\'c}}}, \ and\ \bibinfo {author} {\bibfnamefont
  {Mikhail~D.}\ \bibnamefont {Lukin}},\ }\bibfield  {title} {\enquote {\bibinfo
  {title} {Quantum nonlinear optics --- photon by photon},}\ }\href {\doibase
  10.1038/nphoton.2014.192} {\bibfield  {journal} {\bibinfo  {journal} {Nat.
  Photon.}\ }\textbf {\bibinfo {volume} {8}},\ \bibinfo {pages} {685} (\bibinfo
  {year} {2014})}\BibitemShut {NoStop}%
\bibitem [{\citenamefont {Schine}\ \emph {et~al.}(2016)\citenamefont {Schine},
  \citenamefont {Ryou}, \citenamefont {Gromov}, \citenamefont {Sommer},\ and\
  \citenamefont {Simon}}]{schine2015synthetic}%
  \BibitemOpen
  \bibfield  {author} {\bibinfo {author} {\bibfnamefont {Nathan}\ \bibnamefont
  {Schine}}, \bibinfo {author} {\bibfnamefont {Albert}\ \bibnamefont {Ryou}},
  \bibinfo {author} {\bibfnamefont {Andrey}\ \bibnamefont {Gromov}}, \bibinfo
  {author} {\bibfnamefont {Ariel}\ \bibnamefont {Sommer}}, \ and\ \bibinfo
  {author} {\bibfnamefont {Jonathan}\ \bibnamefont {Simon}},\ }\bibfield
  {title} {\enquote {\bibinfo {title} {Synthetic {Landau} levels for
  photons},}\ }\href {\doibase 10.1038/nature17943} {\bibfield  {journal}
  {\bibinfo  {journal} {Nature (London)}\ }\textbf {\bibinfo {volume} {534}},\
  \bibinfo {pages} {671} (\bibinfo {year} {2016})}\BibitemShut {NoStop}%
\bibitem [{\citenamefont {Jia}\ \emph {et~al.}()\citenamefont {Jia},
  \citenamefont {Schine}, \citenamefont {Georgakopoulos}, \citenamefont {Ryou},
  \citenamefont {Sommer},\ and\ \citenamefont {Simon}}]{jia2017strongly}%
  \BibitemOpen
  \bibfield  {author} {\bibinfo {author} {\bibfnamefont {Ningyuan}\
  \bibnamefont {Jia}}, \bibinfo {author} {\bibfnamefont {Nathan}\ \bibnamefont
  {Schine}}, \bibinfo {author} {\bibfnamefont {Alexandros}\ \bibnamefont
  {Georgakopoulos}}, \bibinfo {author} {\bibfnamefont {Albert}\ \bibnamefont
  {Ryou}}, \bibinfo {author} {\bibfnamefont {Ariel}\ \bibnamefont {Sommer}}, \
  and\ \bibinfo {author} {\bibfnamefont {Jonathan}\ \bibnamefont {Simon}},\
  }\bibfield  {title} {\enquote {\bibinfo {title} {A strongly interacting
  polaritonic quantum dot},}\ }\href@noop {} {\ }\Eprint
  {http://arxiv.org/abs/arXiv:1705.07475} {arXiv:1705.07475} \BibitemShut
  {NoStop}%
\bibitem [{\citenamefont {Ningyuan}\ \emph {et~al.}(2018)\citenamefont
  {Ningyuan}, \citenamefont {Schine}, \citenamefont {Georgakopoulos},
  \citenamefont {Ryou}, \citenamefont {Sommer},\ and\ \citenamefont
  {Simon}}]{ningyuan2017photons}%
  \BibitemOpen
  \bibfield  {author} {\bibinfo {author} {\bibfnamefont {Jia}\ \bibnamefont
  {Ningyuan}}, \bibinfo {author} {\bibfnamefont {Nathan}\ \bibnamefont
  {Schine}}, \bibinfo {author} {\bibfnamefont {Alexandros}\ \bibnamefont
  {Georgakopoulos}}, \bibinfo {author} {\bibfnamefont {Albert}\ \bibnamefont
  {Ryou}}, \bibinfo {author} {\bibfnamefont {Ariel}\ \bibnamefont {Sommer}}, \
  and\ \bibinfo {author} {\bibfnamefont {Jonathan}\ \bibnamefont {Simon}},\
  }\bibfield  {title} {\enquote {\bibinfo {title} {Photons and polaritons in a
  broken-time-reversal non-planar resonator},}\ }\href {\doibase
  10.1103/PhysRevA.97.013802} {\bibfield  {journal} {\bibinfo  {journal} {Phys.
  Rev. A}\ }\textbf {\bibinfo {volume} {97}},\ \bibinfo {pages} {013802}
  (\bibinfo {year} {2018})}\BibitemShut {NoStop}%
\bibitem [{\citenamefont {Grusdt}\ \emph {et~al.}(2016)\citenamefont {Grusdt},
  \citenamefont {Yao}, \citenamefont {Abanin}, \citenamefont {Fleischhauer},\
  and\ \citenamefont {Demler}}]{grusdt2016interferometric}%
  \BibitemOpen
  \bibfield  {author} {\bibinfo {author} {\bibfnamefont {Fabian}\ \bibnamefont
  {Grusdt}}, \bibinfo {author} {\bibfnamefont {Norman~Y.}\ \bibnamefont {Yao}},
  \bibinfo {author} {\bibfnamefont {D.}~\bibnamefont {Abanin}}, \bibinfo
  {author} {\bibfnamefont {Michael}\ \bibnamefont {Fleischhauer}}, \ and\
  \bibinfo {author} {\bibfnamefont {E.}~\bibnamefont {Demler}},\ }\bibfield
  {title} {\enquote {\bibinfo {title} {Interferometric measurements of
  many-body topological invariants using mobile impurities},}\ }\href {\doibase
  10.1038/ncomms11994} {\bibfield  {journal} {\bibinfo  {journal} {Nat.
  Commun.}\ }\textbf {\bibinfo {volume} {7}},\ \bibinfo {pages} {11994}
  (\bibinfo {year} {2016})}\BibitemShut {NoStop}%
\bibitem [{\citenamefont {Northup}\ and\ \citenamefont
  {Blatt}(2014)}]{northup2014quantum}%
  \BibitemOpen
  \bibfield  {author} {\bibinfo {author} {\bibfnamefont {T.~E.}\ \bibnamefont
  {Northup}}\ and\ \bibinfo {author} {\bibfnamefont {R.}~\bibnamefont
  {Blatt}},\ }\bibfield  {title} {\enquote {\bibinfo {title} {Quantum
  information transfer using photons},}\ }\href {\doibase
  10.1038/nphoton.2014.53} {\bibfield  {journal} {\bibinfo  {journal} {Nat.
  Photon.}\ }\textbf {\bibinfo {volume} {8}},\ \bibinfo {pages} {356} (\bibinfo
  {year} {2014})}\BibitemShut {NoStop}%
\bibitem [{\citenamefont {Kimble}(2008)}]{kimble2008the}%
  \BibitemOpen
  \bibfield  {author} {\bibinfo {author} {\bibfnamefont {H.~J.}\ \bibnamefont
  {Kimble}},\ }\bibfield  {title} {\enquote {\bibinfo {title} {The quantum
  internet},}\ }\href {\doibase 10.1038/nature07127} {\bibfield  {journal}
  {\bibinfo  {journal} {Nature (London)}\ }\textbf {\bibinfo {volume} {453}},\
  \bibinfo {pages} {1023} (\bibinfo {year} {2008})}\BibitemShut {NoStop}%
\bibitem [{\citenamefont {Kok}\ \emph {et~al.}(2007)\citenamefont {Kok},
  \citenamefont {Munro}, \citenamefont {Nemoto}, \citenamefont {Ralph},
  \citenamefont {Dowling},\ and\ \citenamefont {Milburn}}]{kok2007linear}%
  \BibitemOpen
  \bibfield  {author} {\bibinfo {author} {\bibfnamefont {Pieter}\ \bibnamefont
  {Kok}}, \bibinfo {author} {\bibfnamefont {William~J.}\ \bibnamefont {Munro}},
  \bibinfo {author} {\bibfnamefont {Kae}\ \bibnamefont {Nemoto}}, \bibinfo
  {author} {\bibfnamefont {Timothy~C.}\ \bibnamefont {Ralph}}, \bibinfo
  {author} {\bibfnamefont {Jonathan~P.}\ \bibnamefont {Dowling}}, \ and\
  \bibinfo {author} {\bibfnamefont {Gerard~J.}\ \bibnamefont {Milburn}},\
  }\bibfield  {title} {\enquote {\bibinfo {title} {Linear optical quantum
  computing with photonic qubits},}\ }\href {\doibase
  10.1103/RevModPhys.79.135} {\bibfield  {journal} {\bibinfo  {journal} {Rev.
  Mod. Phys.}\ }\textbf {\bibinfo {volume} {79}},\ \bibinfo {pages} {135}
  (\bibinfo {year} {2007})}\BibitemShut {NoStop}%
\bibitem [{\citenamefont {Monroe}(2002)}]{monroe2002quantum}%
  \BibitemOpen
  \bibfield  {author} {\bibinfo {author} {\bibfnamefont {Chris}\ \bibnamefont
  {Monroe}},\ }\bibfield  {title} {\enquote {\bibinfo {title} {Quantum
  information processing with atoms and photons},}\ }\href {\doibase
  10.1038/416238a} {\bibfield  {journal} {\bibinfo  {journal} {Nature
  (London)}\ }\textbf {\bibinfo {volume} {416}},\ \bibinfo {pages} {238}
  (\bibinfo {year} {2002})}\BibitemShut {NoStop}%
\bibitem [{\citenamefont {Klaers}\ \emph {et~al.}(2010)\citenamefont {Klaers},
  \citenamefont {Schmitt}, \citenamefont {Vewinger},\ and\ \citenamefont
  {Weitz}}]{klaers2010bose}%
  \BibitemOpen
  \bibfield  {author} {\bibinfo {author} {\bibfnamefont {Jan}\ \bibnamefont
  {Klaers}}, \bibinfo {author} {\bibfnamefont {Julian}\ \bibnamefont
  {Schmitt}}, \bibinfo {author} {\bibfnamefont {Frank}\ \bibnamefont
  {Vewinger}}, \ and\ \bibinfo {author} {\bibfnamefont {Martin}\ \bibnamefont
  {Weitz}},\ }\bibfield  {title} {\enquote {\bibinfo {title} {Bose-{Einstein}
  condensation of photons in an optical microcavity},}\ }\href {\doibase
  10.1038/nature09567} {\bibfield  {journal} {\bibinfo  {journal} {Nature
  (London)}\ }\textbf {\bibinfo {volume} {468}},\ \bibinfo {pages} {545}
  (\bibinfo {year} {2010})}\BibitemShut {NoStop}%
\bibitem [{\citenamefont {Marelic}\ and\ \citenamefont
  {Nyman}(2015)}]{marelic2015experimental}%
  \BibitemOpen
  \bibfield  {author} {\bibinfo {author} {\bibfnamefont {Jakov}\ \bibnamefont
  {Marelic}}\ and\ \bibinfo {author} {\bibfnamefont {R.~A.}\ \bibnamefont
  {Nyman}},\ }\bibfield  {title} {\enquote {\bibinfo {title} {Experimental
  evidence for inhomogeneous pumping and energy-dependent effects in photon
  {Bose}-{Einstein} condensation},}\ }\href {\doibase
  10.1103/PhysRevA.91.033813} {\bibfield  {journal} {\bibinfo  {journal} {Phys.
  Rev. A}\ }\textbf {\bibinfo {volume} {91}},\ \bibinfo {pages} {033813}
  (\bibinfo {year} {2015})}\BibitemShut {NoStop}%
\bibitem [{\citenamefont {Damm}\ \emph {et~al.}(2016)\citenamefont {Damm},
  \citenamefont {Schmitt}, \citenamefont {Liang}, \citenamefont {Dung},
  \citenamefont {Vewinger}, \citenamefont {Weitz},\ and\ \citenamefont
  {Klaers}}]{damm2016calorimetry}%
  \BibitemOpen
  \bibfield  {author} {\bibinfo {author} {\bibfnamefont {Tobias}\ \bibnamefont
  {Damm}}, \bibinfo {author} {\bibfnamefont {Julian}\ \bibnamefont {Schmitt}},
  \bibinfo {author} {\bibfnamefont {Qi}~\bibnamefont {Liang}}, \bibinfo
  {author} {\bibfnamefont {David}\ \bibnamefont {Dung}}, \bibinfo {author}
  {\bibfnamefont {Frank}\ \bibnamefont {Vewinger}}, \bibinfo {author}
  {\bibfnamefont {Martin}\ \bibnamefont {Weitz}}, \ and\ \bibinfo {author}
  {\bibfnamefont {Jan}\ \bibnamefont {Klaers}},\ }\bibfield  {title} {\enquote
  {\bibinfo {title} {Calorimetry of a {Bose}--{Einstein}-condensed photon
  gas},}\ }\href {\doibase 10.1038/ncomms11340} {\bibfield  {journal} {\bibinfo
   {journal} {Nat. Commun.}\ }\textbf {\bibinfo {volume} {7}},\ \bibinfo
  {pages} {11340} (\bibinfo {year} {2016})}\BibitemShut {NoStop}%
\bibitem [{\citenamefont {Marelic}\ \emph {et~al.}(2016)\citenamefont
  {Marelic}, \citenamefont {Zajiczek}, \citenamefont {Hesten}, \citenamefont
  {Leung}, \citenamefont {Ong}, \citenamefont {Mintert},\ and\ \citenamefont
  {Nyman}}]{marelic2016spatiotemporal}%
  \BibitemOpen
  \bibfield  {author} {\bibinfo {author} {\bibfnamefont {Jakov}\ \bibnamefont
  {Marelic}}, \bibinfo {author} {\bibfnamefont {Lydia~F.}\ \bibnamefont
  {Zajiczek}}, \bibinfo {author} {\bibfnamefont {Henry~J.}\ \bibnamefont
  {Hesten}}, \bibinfo {author} {\bibfnamefont {Kon~H.}\ \bibnamefont {Leung}},
  \bibinfo {author} {\bibfnamefont {Edward Y.~X.}\ \bibnamefont {Ong}},
  \bibinfo {author} {\bibfnamefont {Florian}\ \bibnamefont {Mintert}}, \ and\
  \bibinfo {author} {\bibfnamefont {Robert~A.}\ \bibnamefont {Nyman}},\
  }\bibfield  {title} {\enquote {\bibinfo {title} {Spatiotemporal coherence of
  non-equilibrium multimode photon condensates},}\ }\href {\doibase
  10.1088/1367-2630/18/10/103012} {\bibfield  {journal} {\bibinfo  {journal}
  {New J. Phys.}\ }\textbf {\bibinfo {volume} {18}},\ \bibinfo {pages} {103012}
  (\bibinfo {year} {2016})}\BibitemShut {NoStop}%
\bibitem [{\citenamefont {Damm}\ \emph {et~al.}(2017)\citenamefont {Damm},
  \citenamefont {Dung}, \citenamefont {Vewinger}, \citenamefont {Weitz},\ and\
  \citenamefont {Schmitt}}]{damm2017first}%
  \BibitemOpen
  \bibfield  {author} {\bibinfo {author} {\bibfnamefont {Tobias}\ \bibnamefont
  {Damm}}, \bibinfo {author} {\bibfnamefont {David}\ \bibnamefont {Dung}},
  \bibinfo {author} {\bibfnamefont {Frank}\ \bibnamefont {Vewinger}}, \bibinfo
  {author} {\bibfnamefont {Martin}\ \bibnamefont {Weitz}}, \ and\ \bibinfo
  {author} {\bibfnamefont {Julian}\ \bibnamefont {Schmitt}},\ }\bibfield
  {title} {\enquote {\bibinfo {title} {First-order spatial coherence
  measurements in a thermalized two-dimensional photonic quantum gas},}\ }\href
  {\doibase 10.1038/s41467-017-00270-8} {\bibfield  {journal} {\bibinfo
  {journal} {Nat. Commun.}\ }\textbf {\bibinfo {volume} {8}},\ \bibinfo {pages}
  {158} (\bibinfo {year} {2017})}\BibitemShut {NoStop}%
\bibitem [{\citenamefont {Lukin}(2003)}]{lukin2003colloquium}%
  \BibitemOpen
  \bibfield  {author} {\bibinfo {author} {\bibfnamefont {M.~D.}\ \bibnamefont
  {Lukin}},\ }\bibfield  {title} {\enquote {\bibinfo {title} {Colloquium:
  {Trapping} and manipulating photon states in atomic ensembles},}\ }\href
  {\doibase 10.1103/RevModPhys.75.457} {\bibfield  {journal} {\bibinfo
  {journal} {Rev. Mod. Phys.}\ }\textbf {\bibinfo {volume} {75}},\ \bibinfo
  {pages} {457} (\bibinfo {year} {2003})}\BibitemShut {NoStop}%
\bibitem [{\citenamefont {Birnbaum}\ \emph {et~al.}(2005)\citenamefont
  {Birnbaum}, \citenamefont {Boca}, \citenamefont {Miller}, \citenamefont
  {Boozer}, \citenamefont {Northup},\ and\ \citenamefont
  {Kimble}}]{birnbaum2005photon}%
  \BibitemOpen
  \bibfield  {author} {\bibinfo {author} {\bibfnamefont {Kevin~M.}\
  \bibnamefont {Birnbaum}}, \bibinfo {author} {\bibfnamefont {Andreea}\
  \bibnamefont {Boca}}, \bibinfo {author} {\bibfnamefont {Russell}\
  \bibnamefont {Miller}}, \bibinfo {author} {\bibfnamefont {Allen~D.}\
  \bibnamefont {Boozer}}, \bibinfo {author} {\bibfnamefont {Tracy~E.}\
  \bibnamefont {Northup}}, \ and\ \bibinfo {author} {\bibfnamefont {H.~Jeff}\
  \bibnamefont {Kimble}},\ }\bibfield  {title} {\enquote {\bibinfo {title}
  {Photon blockade in an optical cavity with one trapped atom},}\ }\href
  {\doibase 10.1038/nature03804} {\bibfield  {journal} {\bibinfo  {journal}
  {Nature (London)}\ }\textbf {\bibinfo {volume} {436}},\ \bibinfo {pages} {87}
  (\bibinfo {year} {2005})}\BibitemShut {NoStop}%
\bibitem [{\citenamefont {Bajcsy}\ \emph {et~al.}(2009)\citenamefont {Bajcsy},
  \citenamefont {Hofferberth}, \citenamefont {Balic}, \citenamefont {Peyronel},
  \citenamefont {Hafezi}, \citenamefont {Zibrov}, \citenamefont {Vuletic},\
  and\ \citenamefont {Lukin}}]{bajcsy2009efficient}%
  \BibitemOpen
  \bibfield  {author} {\bibinfo {author} {\bibfnamefont {Michal}\ \bibnamefont
  {Bajcsy}}, \bibinfo {author} {\bibfnamefont {Sebastian}\ \bibnamefont
  {Hofferberth}}, \bibinfo {author} {\bibfnamefont {Vlatko}\ \bibnamefont
  {Balic}}, \bibinfo {author} {\bibfnamefont {Thibault}\ \bibnamefont
  {Peyronel}}, \bibinfo {author} {\bibfnamefont {Mohammad}\ \bibnamefont
  {Hafezi}}, \bibinfo {author} {\bibfnamefont {Alexander~S.}\ \bibnamefont
  {Zibrov}}, \bibinfo {author} {\bibfnamefont {Vladan}\ \bibnamefont
  {Vuletic}}, \ and\ \bibinfo {author} {\bibfnamefont {Mikhail~D.}\
  \bibnamefont {Lukin}},\ }\bibfield  {title} {\enquote {\bibinfo {title}
  {Efficient all-optical switching using slow light within a hollow fiber},}\
  }\href {\doibase 10.1103/PhysRevLett.102.203902} {\bibfield  {journal}
  {\bibinfo  {journal} {Phys. Rev. Lett.}\ }\textbf {\bibinfo {volume} {102}},\
  \bibinfo {pages} {203902} (\bibinfo {year} {2009})}\BibitemShut {NoStop}%
\bibitem [{\citenamefont {Fushman}\ \emph {et~al.}(2008)\citenamefont
  {Fushman}, \citenamefont {Englund}, \citenamefont {Faraon}, \citenamefont
  {Stoltz}, \citenamefont {Petroff},\ and\ \citenamefont
  {Vu{\v{c}}kovi{\'c}}}]{fushman2008controlled}%
  \BibitemOpen
  \bibfield  {author} {\bibinfo {author} {\bibfnamefont {Ilya}\ \bibnamefont
  {Fushman}}, \bibinfo {author} {\bibfnamefont {Dirk}\ \bibnamefont {Englund}},
  \bibinfo {author} {\bibfnamefont {Andrei}\ \bibnamefont {Faraon}}, \bibinfo
  {author} {\bibfnamefont {Nick}\ \bibnamefont {Stoltz}}, \bibinfo {author}
  {\bibfnamefont {Pierre}\ \bibnamefont {Petroff}}, \ and\ \bibinfo {author}
  {\bibfnamefont {Jelena}\ \bibnamefont {Vu{\v{c}}kovi{\'c}}},\ }\bibfield
  {title} {\enquote {\bibinfo {title} {Controlled phase shifts with a single
  quantum dot},}\ }\href {\doibase 10.1126/science.1154643} {\bibfield
  {journal} {\bibinfo  {journal} {Science}\ }\textbf {\bibinfo {volume}
  {320}},\ \bibinfo {pages} {769} (\bibinfo {year} {2008})}\BibitemShut
  {NoStop}%
\bibitem [{\citenamefont {Sun}\ \emph {et~al.}(2017)\citenamefont {Sun},
  \citenamefont {Yoon}, \citenamefont {Steger}, \citenamefont {Liu},
  \citenamefont {Pfeiffer}, \citenamefont {West}, \citenamefont {Snoke},\ and\
  \citenamefont {Nelson}}]{sun2017direct}%
  \BibitemOpen
  \bibfield  {author} {\bibinfo {author} {\bibfnamefont {Yongbao}\ \bibnamefont
  {Sun}}, \bibinfo {author} {\bibfnamefont {Yoseob}\ \bibnamefont {Yoon}},
  \bibinfo {author} {\bibfnamefont {Mark}\ \bibnamefont {Steger}}, \bibinfo
  {author} {\bibfnamefont {Gangqiang}\ \bibnamefont {Liu}}, \bibinfo {author}
  {\bibfnamefont {Loren~N.}\ \bibnamefont {Pfeiffer}}, \bibinfo {author}
  {\bibfnamefont {Ken}\ \bibnamefont {West}}, \bibinfo {author} {\bibfnamefont
  {David~W.}\ \bibnamefont {Snoke}}, \ and\ \bibinfo {author} {\bibfnamefont
  {Keith~A.}\ \bibnamefont {Nelson}},\ }\bibfield  {title} {\enquote {\bibinfo
  {title} {Direct measurement of polariton-polariton interaction strength},}\
  }\href {\doibase 10.1038/nphys4148} {\bibfield  {journal} {\bibinfo
  {journal} {Nat. Phys.}\ }\textbf {\bibinfo {volume} {13}},\ \bibinfo {pages}
  {870} (\bibinfo {year} {2017})}\BibitemShut {NoStop}%
\bibitem [{\citenamefont {Tassone}\ and\ \citenamefont
  {Yamamoto}(1999)}]{tassone1999exciton}%
  \BibitemOpen
  \bibfield  {author} {\bibinfo {author} {\bibfnamefont {F.}~\bibnamefont
  {Tassone}}\ and\ \bibinfo {author} {\bibfnamefont {Y.}~\bibnamefont
  {Yamamoto}},\ }\bibfield  {title} {\enquote {\bibinfo {title}
  {Exciton-exciton scattering dynamics in a semiconductor microcavity and
  stimulated scattering into polaritons},}\ }\href {\doibase
  10.1103/PhysRevB.59.10830} {\bibfield  {journal} {\bibinfo  {journal} {Phys.
  Rev. B}\ }\textbf {\bibinfo {volume} {59}},\ \bibinfo {pages} {10830}
  (\bibinfo {year} {1999})}\BibitemShut {NoStop}%
\bibitem [{\citenamefont {Firstenberg}\ \emph {et~al.}(2016)\citenamefont
  {Firstenberg}, \citenamefont {Adams},\ and\ \citenamefont
  {Hofferberth}}]{firstenberg2016nonlinear}%
  \BibitemOpen
  \bibfield  {author} {\bibinfo {author} {\bibfnamefont {Ofer}\ \bibnamefont
  {Firstenberg}}, \bibinfo {author} {\bibfnamefont {Charles~S.}\ \bibnamefont
  {Adams}}, \ and\ \bibinfo {author} {\bibfnamefont {Sebastian}\ \bibnamefont
  {Hofferberth}},\ }\bibfield  {title} {\enquote {\bibinfo {title} {Nonlinear
  quantum optics mediated by {Rydberg} interactions},}\ }\href {\doibase
  10.1088/0953-4075/49/15/152003} {\bibfield  {journal} {\bibinfo  {journal}
  {J. Phys. B}\ }\textbf {\bibinfo {volume} {49}},\ \bibinfo {pages} {152003}
  (\bibinfo {year} {2016})}\BibitemShut {NoStop}%
\bibitem [{\citenamefont {Lang}\ \emph {et~al.}(2011)\citenamefont {Lang},
  \citenamefont {Bozyigit}, \citenamefont {Eichler}, \citenamefont {Steffen},
  \citenamefont {Fink}, \citenamefont {Abdumalikov~Jr.}, \citenamefont {Baur},
  \citenamefont {Filipp}, \citenamefont {da~Silva}, \citenamefont {Blais},\
  and\ \citenamefont {Wallraff}}]{lang2011observation}%
  \BibitemOpen
  \bibfield  {author} {\bibinfo {author} {\bibfnamefont {C.}~\bibnamefont
  {Lang}}, \bibinfo {author} {\bibfnamefont {D.}~\bibnamefont {Bozyigit}},
  \bibinfo {author} {\bibfnamefont {C.}~\bibnamefont {Eichler}}, \bibinfo
  {author} {\bibfnamefont {L.}~\bibnamefont {Steffen}}, \bibinfo {author}
  {\bibfnamefont {J.~M.}\ \bibnamefont {Fink}}, \bibinfo {author}
  {\bibfnamefont {A.~A.}\ \bibnamefont {Abdumalikov~Jr.}}, \bibinfo {author}
  {\bibfnamefont {M.}~\bibnamefont {Baur}}, \bibinfo {author} {\bibfnamefont
  {S.}~\bibnamefont {Filipp}}, \bibinfo {author} {\bibfnamefont {M.~P.}\
  \bibnamefont {da~Silva}}, \bibinfo {author} {\bibfnamefont {Alexandre}\
  \bibnamefont {Blais}}, \ and\ \bibinfo {author} {\bibfnamefont
  {A.}~\bibnamefont {Wallraff}},\ }\bibfield  {title} {\enquote {\bibinfo
  {title} {Observation of resonant photon blockade at microwave frequencies
  using correlation function measurements},}\ }\href {\doibase
  10.1103/PhysRevLett.106.243601} {\bibfield  {journal} {\bibinfo  {journal}
  {Phys. Rev. Lett.}\ }\textbf {\bibinfo {volume} {106}},\ \bibinfo {pages}
  {243601} (\bibinfo {year} {2011})}\BibitemShut {NoStop}%
\bibitem [{\citenamefont {Roushan}\ \emph {et~al.}(2017)\citenamefont
  {Roushan}, \citenamefont {Neill}, \citenamefont {Megrant}, \citenamefont
  {Chen}, \citenamefont {Babbush}, \citenamefont {Barends}, \citenamefont
  {Campbell}, \citenamefont {Chen}, \citenamefont {Chiaro}, \citenamefont
  {Dunsworth} \emph {et~al.}}]{roushan2016chiral}%
  \BibitemOpen
  \bibfield  {author} {\bibinfo {author} {\bibfnamefont {Pedram}\ \bibnamefont
  {Roushan}}, \bibinfo {author} {\bibfnamefont {Charles}\ \bibnamefont
  {Neill}}, \bibinfo {author} {\bibfnamefont {Anthony}\ \bibnamefont
  {Megrant}}, \bibinfo {author} {\bibfnamefont {Yu}~\bibnamefont {Chen}},
  \bibinfo {author} {\bibfnamefont {Ryan}\ \bibnamefont {Babbush}}, \bibinfo
  {author} {\bibfnamefont {Rami}\ \bibnamefont {Barends}}, \bibinfo {author}
  {\bibfnamefont {Brooks}\ \bibnamefont {Campbell}}, \bibinfo {author}
  {\bibfnamefont {Zijun}\ \bibnamefont {Chen}}, \bibinfo {author}
  {\bibfnamefont {Ben}\ \bibnamefont {Chiaro}}, \bibinfo {author}
  {\bibfnamefont {Andrew}\ \bibnamefont {Dunsworth}},  \emph {et~al.},\
  }\bibfield  {title} {\enquote {\bibinfo {title} {Chiral groundstate currents
  of interacting photons in a synthetic magnetic field},}\ }\href {\doibase
  10.1038/nphys3930} {\bibfield  {journal} {\bibinfo  {journal} {Nat. Phys.}\
  }\textbf {\bibinfo {volume} {13}},\ \bibinfo {pages} {146} (\bibinfo {year}
  {2017})}\BibitemShut {NoStop}%
\bibitem [{\citenamefont {Owens}\ \emph {et~al.}()\citenamefont {Owens},
  \citenamefont {LaChapelle}, \citenamefont {Saxberg}, \citenamefont
  {Anderson}, \citenamefont {Ma}, \citenamefont {Simon},\ and\ \citenamefont
  {Schuster}}]{owens2017quarter}%
  \BibitemOpen
  \bibfield  {author} {\bibinfo {author} {\bibfnamefont {Clai}\ \bibnamefont
  {Owens}}, \bibinfo {author} {\bibfnamefont {Aman}\ \bibnamefont
  {LaChapelle}}, \bibinfo {author} {\bibfnamefont {Brendan}\ \bibnamefont
  {Saxberg}}, \bibinfo {author} {\bibfnamefont {Brandon}\ \bibnamefont
  {Anderson}}, \bibinfo {author} {\bibfnamefont {Ruichao}\ \bibnamefont {Ma}},
  \bibinfo {author} {\bibfnamefont {Jonathan}\ \bibnamefont {Simon}}, \ and\
  \bibinfo {author} {\bibfnamefont {David~I.}\ \bibnamefont {Schuster}},\
  }\bibfield  {title} {\enquote {\bibinfo {title} {Quarter-flux {Hofstadter}
  lattice in qubit-compatible microwave cavity array},}\ }\href@noop {} {\
  }\Eprint {http://arxiv.org/abs/arXiv:1708.01651} {arXiv:1708.01651}
  \BibitemShut {NoStop}%
\bibitem [{\citenamefont {Ningyuan}\ \emph {et~al.}(2015)\citenamefont
  {Ningyuan}, \citenamefont {Owens}, \citenamefont {Sommer}, \citenamefont
  {Schuster},\ and\ \citenamefont {Simon}}]{ningyuan2015time}%
  \BibitemOpen
  \bibfield  {author} {\bibinfo {author} {\bibfnamefont {Jia}\ \bibnamefont
  {Ningyuan}}, \bibinfo {author} {\bibfnamefont {Clai}\ \bibnamefont {Owens}},
  \bibinfo {author} {\bibfnamefont {Ariel}\ \bibnamefont {Sommer}}, \bibinfo
  {author} {\bibfnamefont {David}\ \bibnamefont {Schuster}}, \ and\ \bibinfo
  {author} {\bibfnamefont {Jonathan}\ \bibnamefont {Simon}},\ }\bibfield
  {title} {\enquote {\bibinfo {title} {Time-and site-resolved dynamics in a
  topological circuit},}\ }\href {\doibase 10.1103/PhysRevX.5.021031}
  {\bibfield  {journal} {\bibinfo  {journal} {Phys. Rev. X}\ }\textbf {\bibinfo
  {volume} {5}},\ \bibinfo {pages} {021031} (\bibinfo {year}
  {2015})}\BibitemShut {NoStop}%
\bibitem [{\citenamefont {Wang}\ \emph {et~al.}(2009)\citenamefont {Wang},
  \citenamefont {Chong}, \citenamefont {Joannopoulos},\ and\ \citenamefont
  {Soljacic}}]{wang2009observation}%
  \BibitemOpen
  \bibfield  {author} {\bibinfo {author} {\bibfnamefont {Zheng}\ \bibnamefont
  {Wang}}, \bibinfo {author} {\bibfnamefont {Yidong}\ \bibnamefont {Chong}},
  \bibinfo {author} {\bibfnamefont {J.~D.}\ \bibnamefont {Joannopoulos}}, \
  and\ \bibinfo {author} {\bibfnamefont {Marin}\ \bibnamefont {Soljacic}},\
  }\bibfield  {title} {\enquote {\bibinfo {title} {Observation of
  unidirectional backscattering-immune topological electromagnetic states},}\
  }\href {\doibase 10.1038/nature08293} {\bibfield  {journal} {\bibinfo
  {journal} {Nature (London)}\ }\textbf {\bibinfo {volume} {461}},\ \bibinfo
  {pages} {772} (\bibinfo {year} {2009})}\BibitemShut {NoStop}%
\bibitem [{\citenamefont {Hafezi}\ \emph
  {et~al.}(2013{\natexlab{b}})\citenamefont {Hafezi}, \citenamefont {Mittal},
  \citenamefont {Fan}, \citenamefont {Migdall},\ and\ \citenamefont
  {Taylor}}]{hafezi2013imaging}%
  \BibitemOpen
  \bibfield  {author} {\bibinfo {author} {\bibfnamefont {Mohammad}\
  \bibnamefont {Hafezi}}, \bibinfo {author} {\bibfnamefont {S.}~\bibnamefont
  {Mittal}}, \bibinfo {author} {\bibfnamefont {J.}~\bibnamefont {Fan}},
  \bibinfo {author} {\bibfnamefont {A.}~\bibnamefont {Migdall}}, \ and\
  \bibinfo {author} {\bibfnamefont {J.~M.}\ \bibnamefont {Taylor}},\ }\bibfield
   {title} {\enquote {\bibinfo {title} {Imaging topological edge states in
  silicon photonics},}\ }\href {\doibase 10.1038/nphoton.2013.274} {\bibfield
  {journal} {\bibinfo  {journal} {Nat. Photon.}\ }\textbf {\bibinfo {volume}
  {7}},\ \bibinfo {pages} {1001} (\bibinfo {year}
  {2013}{\natexlab{b}})}\BibitemShut {NoStop}%
\bibitem [{\citenamefont {Rechtsman}\ \emph
  {et~al.}(2013{\natexlab{a}})\citenamefont {Rechtsman}, \citenamefont
  {Zeuner}, \citenamefont {Plotnik}, \citenamefont {Lumer}, \citenamefont
  {Podolsky}, \citenamefont {Dreisow}, \citenamefont {Nolte}, \citenamefont
  {Segev},\ and\ \citenamefont {Szameit}}]{rechtsman2013photonic}%
  \BibitemOpen
  \bibfield  {author} {\bibinfo {author} {\bibfnamefont {Mikael~C.}\
  \bibnamefont {Rechtsman}}, \bibinfo {author} {\bibfnamefont {Julia~M.}\
  \bibnamefont {Zeuner}}, \bibinfo {author} {\bibfnamefont {Yonatan}\
  \bibnamefont {Plotnik}}, \bibinfo {author} {\bibfnamefont {Yaakov}\
  \bibnamefont {Lumer}}, \bibinfo {author} {\bibfnamefont {Daniel}\
  \bibnamefont {Podolsky}}, \bibinfo {author} {\bibfnamefont {Felix}\
  \bibnamefont {Dreisow}}, \bibinfo {author} {\bibfnamefont {Stefan}\
  \bibnamefont {Nolte}}, \bibinfo {author} {\bibfnamefont {Mordechai}\
  \bibnamefont {Segev}}, \ and\ \bibinfo {author} {\bibfnamefont {Alexander}\
  \bibnamefont {Szameit}},\ }\bibfield  {title} {\enquote {\bibinfo {title}
  {Photonic {Floquet} topological insulators},}\ }\href {\doibase
  10.1038/nature12066} {\bibfield  {journal} {\bibinfo  {journal} {Nature
  (London)}\ }\textbf {\bibinfo {volume} {496}},\ \bibinfo {pages} {196}
  (\bibinfo {year} {2013}{\natexlab{a}})}\BibitemShut {NoStop}%
\bibitem [{\citenamefont {Li}\ \emph {et~al.}(2014)\citenamefont {Li},
  \citenamefont {Eggleton}, \citenamefont {Fang},\ and\ \citenamefont
  {Fan}}]{li2014photonic}%
  \BibitemOpen
  \bibfield  {author} {\bibinfo {author} {\bibfnamefont {Enbang}\ \bibnamefont
  {Li}}, \bibinfo {author} {\bibfnamefont {Benjamin~J.}\ \bibnamefont
  {Eggleton}}, \bibinfo {author} {\bibfnamefont {Kejie}\ \bibnamefont {Fang}},
  \ and\ \bibinfo {author} {\bibfnamefont {Shanhui}\ \bibnamefont {Fan}},\
  }\bibfield  {title} {\enquote {\bibinfo {title} {Photonic {Aharonov}--{Bohm}
  effect in photon--phonon interactions},}\ }\href {\doibase
  10.1038/ncomms4225} {\bibfield  {journal} {\bibinfo  {journal} {Nat.
  Commun.}\ }\textbf {\bibinfo {volume} {5}},\ \bibinfo {pages} {3225}
  (\bibinfo {year} {2014})}\BibitemShut {NoStop}%
\bibitem [{\citenamefont {Rechtsman}\ \emph
  {et~al.}(2013{\natexlab{b}})\citenamefont {Rechtsman}, \citenamefont
  {Zeuner}, \citenamefont {T{\"u}nnermann}, \citenamefont {Nolte},
  \citenamefont {Segev},\ and\ \citenamefont {Szameit}}]{rechtsman2013strain}%
  \BibitemOpen
  \bibfield  {author} {\bibinfo {author} {\bibfnamefont {Mikael~C.}\
  \bibnamefont {Rechtsman}}, \bibinfo {author} {\bibfnamefont {Julia~M.}\
  \bibnamefont {Zeuner}}, \bibinfo {author} {\bibfnamefont {Andreas}\
  \bibnamefont {T{\"u}nnermann}}, \bibinfo {author} {\bibfnamefont {Stefan}\
  \bibnamefont {Nolte}}, \bibinfo {author} {\bibfnamefont {Mordechai}\
  \bibnamefont {Segev}}, \ and\ \bibinfo {author} {\bibfnamefont {Alexander}\
  \bibnamefont {Szameit}},\ }\bibfield  {title} {\enquote {\bibinfo {title}
  {Strain-induced pseudomagnetic field and photonic {Landau} levels in
  dielectric structures},}\ }\href {\doibase 10.1038/nphoton.2012.302}
  {\bibfield  {journal} {\bibinfo  {journal} {Nat. Photon.}\ }\textbf {\bibinfo
  {volume} {7}},\ \bibinfo {pages} {153} (\bibinfo {year}
  {2013}{\natexlab{b}})}\BibitemShut {NoStop}%
\bibitem [{\citenamefont {Tzuang}\ \emph {et~al.}(2014)\citenamefont {Tzuang},
  \citenamefont {Fang}, \citenamefont {Nussenzveig}, \citenamefont {Fan},\ and\
  \citenamefont {Lipson}}]{tzuang2014non}%
  \BibitemOpen
  \bibfield  {author} {\bibinfo {author} {\bibfnamefont {Lawrence~D.}\
  \bibnamefont {Tzuang}}, \bibinfo {author} {\bibfnamefont {Kejie}\
  \bibnamefont {Fang}}, \bibinfo {author} {\bibfnamefont {Paulo}\ \bibnamefont
  {Nussenzveig}}, \bibinfo {author} {\bibfnamefont {Shanhui}\ \bibnamefont
  {Fan}}, \ and\ \bibinfo {author} {\bibfnamefont {Michal}\ \bibnamefont
  {Lipson}},\ }\bibfield  {title} {\enquote {\bibinfo {title} {Non-reciprocal
  phase shift induced by an effective magnetic flux for light},}\ }\href
  {\doibase 10.1038/nphoton.2014.177} {\bibfield  {journal} {\bibinfo
  {journal} {Nat. Photon.}\ }\textbf {\bibinfo {volume} {8}},\ \bibinfo {pages}
  {701} (\bibinfo {year} {2014})}\BibitemShut {NoStop}%
\bibitem [{\citenamefont {Mittal}\ \emph {et~al.}(2014)\citenamefont {Mittal},
  \citenamefont {Fan}, \citenamefont {Faez}, \citenamefont {Migdall},
  \citenamefont {Taylor},\ and\ \citenamefont
  {Hafezi}}]{mittal2014topologically}%
  \BibitemOpen
  \bibfield  {author} {\bibinfo {author} {\bibfnamefont {S.}~\bibnamefont
  {Mittal}}, \bibinfo {author} {\bibfnamefont {J.}~\bibnamefont {Fan}},
  \bibinfo {author} {\bibfnamefont {S.}~\bibnamefont {Faez}}, \bibinfo {author}
  {\bibfnamefont {A.}~\bibnamefont {Migdall}}, \bibinfo {author} {\bibfnamefont
  {J.~M.}\ \bibnamefont {Taylor}}, \ and\ \bibinfo {author} {\bibfnamefont
  {M.}~\bibnamefont {Hafezi}},\ }\bibfield  {title} {\enquote {\bibinfo {title}
  {Topologically robust transport of photons in a synthetic gauge field},}\
  }\href {\doibase 10.1103/PhysRevLett.113.087403} {\bibfield  {journal}
  {\bibinfo  {journal} {Phys. Rev. Lett.}\ }\textbf {\bibinfo {volume} {113}},\
  \bibinfo {pages} {087403} (\bibinfo {year} {2014})}\BibitemShut {NoStop}%
\bibitem [{\citenamefont {Sommer}\ \emph {et~al.}()\citenamefont {Sommer},
  \citenamefont {B{\"u}chler},\ and\ \citenamefont
  {Simon}}]{sommer2015quantum}%
  \BibitemOpen
  \bibfield  {author} {\bibinfo {author} {\bibfnamefont {Ariel}\ \bibnamefont
  {Sommer}}, \bibinfo {author} {\bibfnamefont {Hans~Peter}\ \bibnamefont
  {B{\"u}chler}}, \ and\ \bibinfo {author} {\bibfnamefont {Jonathan}\
  \bibnamefont {Simon}},\ }\bibfield  {title} {\enquote {\bibinfo {title}
  {Quantum crystals and {Laughlin} droplets of cavity {Rydberg} polaritons},}\
  }\href@noop {} {\ }\Eprint {http://arxiv.org/abs/arXiv:1506.00341}
  {arXiv:1506.00341} \BibitemShut {NoStop}%
\bibitem [{\citenamefont {Ningyuan}\ \emph {et~al.}(2016)\citenamefont
  {Ningyuan}, \citenamefont {Georgakopoulos}, \citenamefont {Ryou},
  \citenamefont {Schine}, \citenamefont {Sommer},\ and\ \citenamefont
  {Simon}}]{ningyuan2016observation}%
  \BibitemOpen
  \bibfield  {author} {\bibinfo {author} {\bibfnamefont {Jia}\ \bibnamefont
  {Ningyuan}}, \bibinfo {author} {\bibfnamefont {Alexandros}\ \bibnamefont
  {Georgakopoulos}}, \bibinfo {author} {\bibfnamefont {Albert}\ \bibnamefont
  {Ryou}}, \bibinfo {author} {\bibfnamefont {Nathan}\ \bibnamefont {Schine}},
  \bibinfo {author} {\bibfnamefont {Ariel}\ \bibnamefont {Sommer}}, \ and\
  \bibinfo {author} {\bibfnamefont {Jonathan}\ \bibnamefont {Simon}},\
  }\bibfield  {title} {\enquote {\bibinfo {title} {Observation and
  characterization of cavity {Rydberg} polaritons},}\ }\href {\doibase
  10.1103/PhysRevA.93.041802} {\bibfield  {journal} {\bibinfo  {journal} {Phys.
  Rev. A}\ }\textbf {\bibinfo {volume} {93}},\ \bibinfo {pages} {041802}
  (\bibinfo {year} {2016})}\BibitemShut {NoStop}%
\bibitem [{\citenamefont {Ma}\ \emph {et~al.}(2017)\citenamefont {Ma},
  \citenamefont {Owens}, \citenamefont {Houck}, \citenamefont {Schuster},\ and\
  \citenamefont {Simon}}]{ma2017autonomous}%
  \BibitemOpen
  \bibfield  {author} {\bibinfo {author} {\bibfnamefont {Ruichao}\ \bibnamefont
  {Ma}}, \bibinfo {author} {\bibfnamefont {Clai}\ \bibnamefont {Owens}},
  \bibinfo {author} {\bibfnamefont {Andrew}\ \bibnamefont {Houck}}, \bibinfo
  {author} {\bibfnamefont {David~I.}\ \bibnamefont {Schuster}}, \ and\ \bibinfo
  {author} {\bibfnamefont {Jonathan}\ \bibnamefont {Simon}},\ }\bibfield
  {title} {\enquote {\bibinfo {title} {Autonomous stabilizer for incompressible
  photon fluids and solids},}\ }\href {\doibase 10.1103/PhysRevA.95.043811}
  {\bibfield  {journal} {\bibinfo  {journal} {Phys. Rev. A}\ }\textbf {\bibinfo
  {volume} {95}},\ \bibinfo {pages} {043811} (\bibinfo {year}
  {2017})}\BibitemShut {NoStop}%
\bibitem [{Note1()}]{Note1}%
  \BibitemOpen
  \bibinfo {note} {See the Supplementary Information of Ref.~\cite
  {schine2015synthetic}}\BibitemShut {NoStop}%
\bibitem [{\citenamefont {Hayat}\ \emph {et~al.}(2012)\citenamefont {Hayat},
  \citenamefont {Lange}, \citenamefont {Rozema}, \citenamefont {Darabi},
  \citenamefont {van Driel}, \citenamefont {Steinberg}, \citenamefont {Nelsen},
  \citenamefont {Snoke}, \citenamefont {Pfeiffer},\ and\ \citenamefont
  {West}}]{hayat2012dynamic}%
  \BibitemOpen
  \bibfield  {author} {\bibinfo {author} {\bibfnamefont {Alex}\ \bibnamefont
  {Hayat}}, \bibinfo {author} {\bibfnamefont {Christoph}\ \bibnamefont
  {Lange}}, \bibinfo {author} {\bibfnamefont {Lee~A.}\ \bibnamefont {Rozema}},
  \bibinfo {author} {\bibfnamefont {Ardavan}\ \bibnamefont {Darabi}}, \bibinfo
  {author} {\bibfnamefont {Henry~M.}\ \bibnamefont {van Driel}}, \bibinfo
  {author} {\bibfnamefont {Aephraim~M.}\ \bibnamefont {Steinberg}}, \bibinfo
  {author} {\bibfnamefont {Bryan}\ \bibnamefont {Nelsen}}, \bibinfo {author}
  {\bibfnamefont {David~W.}\ \bibnamefont {Snoke}}, \bibinfo {author}
  {\bibfnamefont {Loren~N.}\ \bibnamefont {Pfeiffer}}, \ and\ \bibinfo {author}
  {\bibfnamefont {Kenneth~W.}\ \bibnamefont {West}},\ }\bibfield  {title}
  {\enquote {\bibinfo {title} {Dynamic {Stark} effect in strongly coupled
  microcavity exciton polaritons},}\ }\href {\doibase
  10.1103/PhysRevLett.109.033605} {\bibfield  {journal} {\bibinfo  {journal}
  {Phys. Rev. Lett.}\ }\textbf {\bibinfo {volume} {109}},\ \bibinfo {pages}
  {033605} (\bibinfo {year} {2012})}\BibitemShut {NoStop}%
\bibitem [{\citenamefont {Li}\ \emph {et~al.}(2013)\citenamefont {Li},
  \citenamefont {Dudin},\ and\ \citenamefont {Kuzmich}}]{li2013entanglement}%
  \BibitemOpen
  \bibfield  {author} {\bibinfo {author} {\bibfnamefont {L.}~\bibnamefont
  {Li}}, \bibinfo {author} {\bibfnamefont {Y.~O.}\ \bibnamefont {Dudin}}, \
  and\ \bibinfo {author} {\bibfnamefont {A.}~\bibnamefont {Kuzmich}},\
  }\bibfield  {title} {\enquote {\bibinfo {title} {Entanglement between light
  and an optical atomic excitation},}\ }\href {\doibase 10.1038/nature12227}
  {\bibfield  {journal} {\bibinfo  {journal} {Nature (London)}\ }\textbf
  {\bibinfo {volume} {498}},\ \bibinfo {pages} {466} (\bibinfo {year}
  {2013})}\BibitemShut {NoStop}%
\bibitem [{\citenamefont {Cancellieri}\ \emph {et~al.}(2014)\citenamefont
  {Cancellieri}, \citenamefont {Hayat}, \citenamefont {Steinberg},
  \citenamefont {Giacobino},\ and\ \citenamefont
  {Bramati}}]{cancellieri2014ultrafast}%
  \BibitemOpen
  \bibfield  {author} {\bibinfo {author} {\bibfnamefont {Emiliano}\
  \bibnamefont {Cancellieri}}, \bibinfo {author} {\bibfnamefont {Alex}\
  \bibnamefont {Hayat}}, \bibinfo {author} {\bibfnamefont {A.~M.}\ \bibnamefont
  {Steinberg}}, \bibinfo {author} {\bibfnamefont {Elisabeth}\ \bibnamefont
  {Giacobino}}, \ and\ \bibinfo {author} {\bibfnamefont {Alberto}\ \bibnamefont
  {Bramati}},\ }\bibfield  {title} {\enquote {\bibinfo {title} {Ultrafast
  {Stark}-induced polaritonic switches},}\ }\href {\doibase
  10.1103/PhysRevLett.112.053601} {\bibfield  {journal} {\bibinfo  {journal}
  {Phys. Rev. Lett.}\ }\textbf {\bibinfo {volume} {112}},\ \bibinfo {pages}
  {053601} (\bibinfo {year} {2014})}\BibitemShut {NoStop}%
\bibitem [{\citenamefont {Carusotto}\ \emph {et~al.}(2010)\citenamefont
  {Carusotto}, \citenamefont {Volz},\ and\ \citenamefont
  {Imamo{\u{g}}lu}}]{carusotto2010feshbach}%
  \BibitemOpen
  \bibfield  {author} {\bibinfo {author} {\bibfnamefont {Iacopo}\ \bibnamefont
  {Carusotto}}, \bibinfo {author} {\bibfnamefont {Thomas}\ \bibnamefont
  {Volz}}, \ and\ \bibinfo {author} {\bibfnamefont {A.}~\bibnamefont
  {Imamo{\u{g}}lu}},\ }\bibfield  {title} {\enquote {\bibinfo {title} {Feshbach
  blockade: {Single}-photon nonlinear optics using resonantly enhanced cavity
  polariton scattering from biexciton states},}\ }\href {\doibase
  10.1209/0295-5075/90/37001} {\bibfield  {journal} {\bibinfo  {journal}
  {Europhys. Lett.}\ }\textbf {\bibinfo {volume} {90}},\ \bibinfo {pages}
  {37001} (\bibinfo {year} {2010})}\BibitemShut {NoStop}%
\bibitem [{\citenamefont {Sommer}\ and\ \citenamefont
  {Simon}(2016)}]{sommer2016engineering}%
  \BibitemOpen
  \bibfield  {author} {\bibinfo {author} {\bibfnamefont {Ariel}\ \bibnamefont
  {Sommer}}\ and\ \bibinfo {author} {\bibfnamefont {Jonathan}\ \bibnamefont
  {Simon}},\ }\bibfield  {title} {\enquote {\bibinfo {title} {Engineering
  photonic {Floquet} {Hamiltonians} through {Fabry}--{P{\'e}rot} resonators},}\
  }\href {\doibase 10.1088/1367-2630/18/3/035008} {\bibfield  {journal}
  {\bibinfo  {journal} {New J. Phys.}\ }\textbf {\bibinfo {volume} {18}},\
  \bibinfo {pages} {035008} (\bibinfo {year} {2016})}\BibitemShut {NoStop}%
\bibitem [{\citenamefont {Darwin}(1927)}]{darwin1927free}%
  \BibitemOpen
  \bibfield  {author} {\bibinfo {author} {\bibfnamefont {Charles~G.}\
  \bibnamefont {Darwin}},\ }\bibfield  {title} {\enquote {\bibinfo {title}
  {Free motion in the wave mechanics},}\ }\href {\doibase
  10.1098/rspa.1927.0179} {\bibfield  {journal} {\bibinfo  {journal} {Proc. R.
  Soc. A}\ }\textbf {\bibinfo {volume} {117}},\ \bibinfo {pages} {258}
  (\bibinfo {year} {1927})}\BibitemShut {NoStop}%
\bibitem [{\citenamefont {Fock}(1928)}]{fock1928bemerkung}%
  \BibitemOpen
  \bibfield  {author} {\bibinfo {author} {\bibfnamefont {Vladimir}\
  \bibnamefont {Fock}},\ }\bibfield  {title} {\enquote {\bibinfo {title}
  {Bemerkung zur quantelung des harmonischen oszillators im magnetfeld},}\
  }\href {\doibase 10.1007/BF01390750} {\bibfield  {journal} {\bibinfo
  {journal} {Z. Phys.}\ }\textbf {\bibinfo {volume} {47}},\ \bibinfo {pages}
  {446} (\bibinfo {year} {1928})}\BibitemShut {NoStop}%
\bibitem [{\citenamefont {Fleischhauer}\ \emph {et~al.}(2005)\citenamefont
  {Fleischhauer}, \citenamefont {Imamoglu},\ and\ \citenamefont
  {Marangos}}]{fleischhauer2005electromagnetically}%
  \BibitemOpen
  \bibfield  {author} {\bibinfo {author} {\bibfnamefont {Michael}\ \bibnamefont
  {Fleischhauer}}, \bibinfo {author} {\bibfnamefont {Atac}\ \bibnamefont
  {Imamoglu}}, \ and\ \bibinfo {author} {\bibfnamefont {Jonathan~P.}\
  \bibnamefont {Marangos}},\ }\bibfield  {title} {\enquote {\bibinfo {title}
  {Electromagnetically induced transparency: {Optics} in coherent media},}\
  }\href {\doibase 10.1103/RevModPhys.77.633} {\bibfield  {journal} {\bibinfo
  {journal} {Rev. Mod. Phys.}\ }\textbf {\bibinfo {volume} {77}},\ \bibinfo
  {pages} {633} (\bibinfo {year} {2005})}\BibitemShut {NoStop}%
\bibitem [{\citenamefont {Fleischhauer}\ and\ \citenamefont
  {Lukin}(2000)}]{fleischhauer2000dark}%
  \BibitemOpen
  \bibfield  {author} {\bibinfo {author} {\bibfnamefont {Michael}\ \bibnamefont
  {Fleischhauer}}\ and\ \bibinfo {author} {\bibfnamefont {Mikhail~D.}\
  \bibnamefont {Lukin}},\ }\bibfield  {title} {\enquote {\bibinfo {title}
  {Dark-state polaritons in electromagnetically induced transparency},}\ }\href
  {\doibase 10.1103/PhysRevLett.84.5094} {\bibfield  {journal} {\bibinfo
  {journal} {Phys. Rev. Lett.}\ }\textbf {\bibinfo {volume} {84}},\ \bibinfo
  {pages} {5094} (\bibinfo {year} {2000})}\BibitemShut {NoStop}%
\bibitem [{\citenamefont {B{\'e}guin}\ \emph {et~al.}(2013)\citenamefont
  {B{\'e}guin}, \citenamefont {Vernier}, \citenamefont {Chicireanu},
  \citenamefont {Lahaye},\ and\ \citenamefont {Browaeys}}]{beguin2013direct}%
  \BibitemOpen
  \bibfield  {author} {\bibinfo {author} {\bibfnamefont {Lucas}\ \bibnamefont
  {B{\'e}guin}}, \bibinfo {author} {\bibfnamefont {Aline}\ \bibnamefont
  {Vernier}}, \bibinfo {author} {\bibfnamefont {Radu}\ \bibnamefont
  {Chicireanu}}, \bibinfo {author} {\bibfnamefont {Thierry}\ \bibnamefont
  {Lahaye}}, \ and\ \bibinfo {author} {\bibfnamefont {Antoine}\ \bibnamefont
  {Browaeys}},\ }\bibfield  {title} {\enquote {\bibinfo {title} {Direct
  measurement of the van der {Waals} interaction between two {Rydberg}
  atoms},}\ }\href {\doibase 10.1103/PhysRevLett.110.263201} {\bibfield
  {journal} {\bibinfo  {journal} {Phys. Rev. Lett.}\ }\textbf {\bibinfo
  {volume} {110}},\ \bibinfo {pages} {263201} (\bibinfo {year}
  {2013})}\BibitemShut {NoStop}%
\bibitem [{\citenamefont {Bienias}\ \emph {et~al.}(2014)\citenamefont
  {Bienias}, \citenamefont {Choi}, \citenamefont {Firstenberg}, \citenamefont
  {Maghrebi}, \citenamefont {Gullans}, \citenamefont {Lukin}, \citenamefont
  {Gorshkov},\ and\ \citenamefont {B{\"u}chler}}]{bienias2014scattering}%
  \BibitemOpen
  \bibfield  {author} {\bibinfo {author} {\bibfnamefont {P.}~\bibnamefont
  {Bienias}}, \bibinfo {author} {\bibfnamefont {S.}~\bibnamefont {Choi}},
  \bibinfo {author} {\bibfnamefont {O.}~\bibnamefont {Firstenberg}}, \bibinfo
  {author} {\bibfnamefont {M.~F.}\ \bibnamefont {Maghrebi}}, \bibinfo {author}
  {\bibfnamefont {M.}~\bibnamefont {Gullans}}, \bibinfo {author} {\bibfnamefont
  {Mikhail~D.}\ \bibnamefont {Lukin}}, \bibinfo {author} {\bibfnamefont
  {Alexey~Vyacheslavovich}\ \bibnamefont {Gorshkov}}, \ and\ \bibinfo {author}
  {\bibfnamefont {H.~P.}\ \bibnamefont {B{\"u}chler}},\ }\bibfield  {title}
  {\enquote {\bibinfo {title} {Scattering resonances and bound states for
  strongly interacting {Rydberg} polaritons},}\ }\href {\doibase
  10.1103/PhysRevA.90.053804} {\bibfield  {journal} {\bibinfo  {journal} {Phys.
  Rev. A}\ }\textbf {\bibinfo {volume} {90}},\ \bibinfo {pages} {053804}
  (\bibinfo {year} {2014})}\BibitemShut {NoStop}%
\bibitem [{\citenamefont {Peyronel}\ \emph {et~al.}(2012)\citenamefont
  {Peyronel}, \citenamefont {Firstenberg}, \citenamefont {Liang}, \citenamefont
  {Hofferberth}, \citenamefont {Gorshkov}, \citenamefont {Pohl}, \citenamefont
  {Lukin},\ and\ \citenamefont {Vuletic}}]{peyronel2012quantum}%
  \BibitemOpen
  \bibfield  {author} {\bibinfo {author} {\bibfnamefont {Thibault}\
  \bibnamefont {Peyronel}}, \bibinfo {author} {\bibfnamefont {Ofer}\
  \bibnamefont {Firstenberg}}, \bibinfo {author} {\bibfnamefont {Qi-Yu}\
  \bibnamefont {Liang}}, \bibinfo {author} {\bibfnamefont {Sebastian}\
  \bibnamefont {Hofferberth}}, \bibinfo {author} {\bibfnamefont {Alexey~V.}\
  \bibnamefont {Gorshkov}}, \bibinfo {author} {\bibfnamefont {Thomas}\
  \bibnamefont {Pohl}}, \bibinfo {author} {\bibfnamefont {Mikhail~D.}\
  \bibnamefont {Lukin}}, \ and\ \bibinfo {author} {\bibfnamefont {Vladan}\
  \bibnamefont {Vuletic}},\ }\bibfield  {title} {\enquote {\bibinfo {title}
  {Quantum nonlinear optics with single photons enabled by strongly interacting
  atoms},}\ }\href {\doibase 10.1038/nature11361} {\bibfield  {journal}
  {\bibinfo  {journal} {Nature (London)}\ }\textbf {\bibinfo {volume} {487}},\
  \bibinfo {pages} {57} (\bibinfo {year} {2012})}\BibitemShut {NoStop}%
\bibitem [{\citenamefont {Haldane}(1983)}]{haldane1983fractional}%
  \BibitemOpen
  \bibfield  {author} {\bibinfo {author} {\bibfnamefont {F.~Duncan~M.}\
  \bibnamefont {Haldane}},\ }\bibfield  {title} {\enquote {\bibinfo {title}
  {Fractional quantization of the {Hall} effect: A hierarchy of incompressible
  quantum fluid states},}\ }\href {\doibase 10.1103/PhysRevLett.51.605}
  {\bibfield  {journal} {\bibinfo  {journal} {Phys. Rev. Lett.}\ }\textbf
  {\bibinfo {volume} {51}},\ \bibinfo {pages} {605} (\bibinfo {year}
  {1983})}\BibitemShut {NoStop}%
\bibitem [{\citenamefont {Malinovsky}\ and\ \citenamefont
  {Krause}(2001)}]{malinovsky2001general}%
  \BibitemOpen
  \bibfield  {author} {\bibinfo {author} {\bibfnamefont {V.~S.}\ \bibnamefont
  {Malinovsky}}\ and\ \bibinfo {author} {\bibfnamefont {J.~L.}\ \bibnamefont
  {Krause}},\ }\bibfield  {title} {\enquote {\bibinfo {title} {General theory
  of population transfer by adiabatic rapid passage with intense, chirped laser
  pulses},}\ }\href {\doibase 10.1007/s100530170212} {\bibfield  {journal}
  {\bibinfo  {journal} {Eur. Phys. J. D}\ }\textbf {\bibinfo {volume} {14}},\
  \bibinfo {pages} {147} (\bibinfo {year} {2001})}\BibitemShut {NoStop}%
\bibitem [{\citenamefont {Yao}\ and\ \citenamefont
  {Padgett}(2011)}]{yao2011orbital}%
  \BibitemOpen
  \bibfield  {author} {\bibinfo {author} {\bibfnamefont {Alison~M.}\
  \bibnamefont {Yao}}\ and\ \bibinfo {author} {\bibfnamefont {Miles~J.}\
  \bibnamefont {Padgett}},\ }\bibfield  {title} {\enquote {\bibinfo {title}
  {Orbital angular momentum: origins, behavior and applications},}\ }\href
  {\doibase 10.1364/AOP.3.000161} {\bibfield  {journal} {\bibinfo  {journal}
  {Adv. Opt. Photon.}\ }\textbf {\bibinfo {volume} {3}},\ \bibinfo {pages}
  {161} (\bibinfo {year} {2011})}\BibitemShut {NoStop}%
\bibitem [{\citenamefont {Padgett}\ and\ \citenamefont
  {Bowman}(2011)}]{padgett2011tweezers}%
  \BibitemOpen
  \bibfield  {author} {\bibinfo {author} {\bibfnamefont {Miles}\ \bibnamefont
  {Padgett}}\ and\ \bibinfo {author} {\bibfnamefont {Richard}\ \bibnamefont
  {Bowman}},\ }\bibfield  {title} {\enquote {\bibinfo {title} {Tweezers with a
  twist},}\ }\href {\doibase 10.1038/nphoton.2011.81} {\bibfield  {journal}
  {\bibinfo  {journal} {Nat. Photon.}\ }\textbf {\bibinfo {volume} {5}},\
  \bibinfo {pages} {343} (\bibinfo {year} {2011})}\BibitemShut {NoStop}%
\bibitem [{\citenamefont {Landau}(1932)}]{landau1932a}%
  \BibitemOpen
  \bibfield  {author} {\bibinfo {author} {\bibfnamefont {Lev~D.}\ \bibnamefont
  {Landau}},\ }\bibfield  {title} {\enquote {\bibinfo {title} {A theory of
  energy transfer. {II}},}\ }\href {\doibase
  10.1016/B978-0-08-010586-4.50014-6} {\bibfield  {journal} {\bibinfo
  {journal} {Phys. Z. Sowjetunion}\ }\textbf {\bibinfo {volume} {2}},\ \bibinfo
  {pages} {46} (\bibinfo {year} {1932})}\BibitemShut {NoStop}%
\bibitem [{\citenamefont {Zener}(1932)}]{zener1932non}%
  \BibitemOpen
  \bibfield  {author} {\bibinfo {author} {\bibfnamefont {Clarence}\
  \bibnamefont {Zener}},\ }\bibfield  {title} {\enquote {\bibinfo {title}
  {Non-adiabatic crossing of energy levels},}\ }\href {\doibase
  10.1098/rspa.1932.0165} {\bibfield  {journal} {\bibinfo  {journal} {Proc. R.
  Soc. London A}\ }\textbf {\bibinfo {volume} {137}},\ \bibinfo {pages} {696}
  (\bibinfo {year} {1932})}\BibitemShut {NoStop}%
\bibitem [{\citenamefont {Vitanov}\ and\ \citenamefont
  {Garraway}(1996)}]{vitanov1996landau}%
  \BibitemOpen
  \bibfield  {author} {\bibinfo {author} {\bibfnamefont {N.~V.}\ \bibnamefont
  {Vitanov}}\ and\ \bibinfo {author} {\bibfnamefont {B.~M.}\ \bibnamefont
  {Garraway}},\ }\bibfield  {title} {\enquote {\bibinfo {title} {Landau-{Zener}
  model: {Effects} of finite coupling duration},}\ }\href {\doibase
  10.1103/PhysRevA.53.4288} {\bibfield  {journal} {\bibinfo  {journal} {Phys.
  Rev. A}\ }\textbf {\bibinfo {volume} {53}},\ \bibinfo {pages} {4288}
  (\bibinfo {year} {1996})}\BibitemShut {NoStop}%
\bibitem [{\citenamefont {Shore}(2008)}]{shore2008coherent}%
  \BibitemOpen
  \bibfield  {author} {\bibinfo {author} {\bibfnamefont {Bruce~W.}\
  \bibnamefont {Shore}},\ }\bibfield  {title} {\enquote {\bibinfo {title}
  {Coherent manipulations of atoms using laser light},}\ }\href@noop {}
  {\bibfield  {journal} {\bibinfo  {journal} {Acta Phys. Slovaca}\ }\textbf
  {\bibinfo {volume} {58}},\ \bibinfo {pages} {243} (\bibinfo {year}
  {2008})}\BibitemShut {NoStop}%
\bibitem [{\citenamefont {Vitanov}\ \emph {et~al.}(2001)\citenamefont
  {Vitanov}, \citenamefont {Halfmann}, \citenamefont {Shore},\ and\
  \citenamefont {Bergmann}}]{vitanov2001laser}%
  \BibitemOpen
  \bibfield  {author} {\bibinfo {author} {\bibfnamefont {Nikolay~V.}\
  \bibnamefont {Vitanov}}, \bibinfo {author} {\bibfnamefont {Thomas}\
  \bibnamefont {Halfmann}}, \bibinfo {author} {\bibfnamefont {Bruce~W.}\
  \bibnamefont {Shore}}, \ and\ \bibinfo {author} {\bibfnamefont {Klaas}\
  \bibnamefont {Bergmann}},\ }\bibfield  {title} {\enquote {\bibinfo {title}
  {Laser-induced population transfer by adiabatic passage techniques},}\ }\href
  {\doibase 10.1146/annurev.physchem.52.1.763} {\bibfield  {journal} {\bibinfo
  {journal} {Annu. Rev. Phys. Chem.}\ }\textbf {\bibinfo {volume} {52}},\
  \bibinfo {pages} {763} (\bibinfo {year} {2001})}\BibitemShut {NoStop}%
\bibitem [{\citenamefont {Rangelov}\ \emph {et~al.}(2010)\citenamefont
  {Rangelov}, \citenamefont {Vitanov},\ and\ \citenamefont
  {Shore}}]{rangelov2010rapid}%
  \BibitemOpen
  \bibfield  {author} {\bibinfo {author} {\bibfnamefont {A.~A.}\ \bibnamefont
  {Rangelov}}, \bibinfo {author} {\bibfnamefont {N.~V.}\ \bibnamefont
  {Vitanov}}, \ and\ \bibinfo {author} {\bibfnamefont {B.~W.}\ \bibnamefont
  {Shore}},\ }\bibfield  {title} {\enquote {\bibinfo {title} {Rapid adiabatic
  passage without level crossing},}\ }\href {\doibase
  10.1016/j.optcom.2009.11.080} {\bibfield  {journal} {\bibinfo  {journal}
  {Opt. Commun.}\ }\textbf {\bibinfo {volume} {283}},\ \bibinfo {pages} {1346}
  (\bibinfo {year} {2010})}\BibitemShut {NoStop}%
\bibitem [{\citenamefont {Vitanov}\ and\ \citenamefont
  {Shore}(2015)}]{vitanov2015designer}%
  \BibitemOpen
  \bibfield  {author} {\bibinfo {author} {\bibfnamefont {Nikolay~V.}\
  \bibnamefont {Vitanov}}\ and\ \bibinfo {author} {\bibfnamefont {Bruce~W.}\
  \bibnamefont {Shore}},\ }\bibfield  {title} {\enquote {\bibinfo {title}
  {Designer evolution of quantum systems by inverse engineering},}\ }\href
  {\doibase 10.1088/0953-4075/48/17/174008} {\bibfield  {journal} {\bibinfo
  {journal} {J. Phys. B}\ }\textbf {\bibinfo {volume} {48}},\ \bibinfo {pages}
  {174008} (\bibinfo {year} {2015})}\BibitemShut {NoStop}%
\bibitem [{Note2()}]{Note2}%
  \BibitemOpen
  \bibinfo {note} {See Supplemental Material at \protect \href
  {http://muellergroup.lassp.cornell.edu/twisted/}{\protect \nolinkurl
  {http://muellergroup.lassp.cornell.edu/twisted/}} for polariton-density
  animations showing examples of adiabatic and nonadiabatic creation of
  Laughlin states, generation of quasiholes, and braiding of
  quasiholes.}\BibitemShut {Stop}%
\bibitem [{\citenamefont {Amo}\ \emph {et~al.}(2010)\citenamefont {Amo},
  \citenamefont {Pigeon}, \citenamefont {Adrados}, \citenamefont {Houdr{\'e}},
  \citenamefont {Giacobino}, \citenamefont {Ciuti},\ and\ \citenamefont
  {Bramati}}]{amo2010light}%
  \BibitemOpen
  \bibfield  {author} {\bibinfo {author} {\bibfnamefont {A.}~\bibnamefont
  {Amo}}, \bibinfo {author} {\bibfnamefont {S.}~\bibnamefont {Pigeon}},
  \bibinfo {author} {\bibfnamefont {C.}~\bibnamefont {Adrados}}, \bibinfo
  {author} {\bibfnamefont {R.}~\bibnamefont {Houdr{\'e}}}, \bibinfo {author}
  {\bibfnamefont {E.}~\bibnamefont {Giacobino}}, \bibinfo {author}
  {\bibfnamefont {C.}~\bibnamefont {Ciuti}}, \ and\ \bibinfo {author}
  {\bibfnamefont {A.}~\bibnamefont {Bramati}},\ }\bibfield  {title} {\enquote
  {\bibinfo {title} {Light engineering of the polariton landscape in
  semiconductor microcavities},}\ }\href {\doibase 10.1103/PhysRevB.82.081301}
  {\bibfield  {journal} {\bibinfo  {journal} {Phys. Rev. B}\ }\textbf {\bibinfo
  {volume} {82}},\ \bibinfo {pages} {081301} (\bibinfo {year}
  {2010})}\BibitemShut {NoStop}%
\bibitem [{\citenamefont {Sanvitto}\ \emph {et~al.}(2011)\citenamefont
  {Sanvitto}, \citenamefont {Pigeon}, \citenamefont {Amo}, \citenamefont
  {Ballarini}, \citenamefont {De~Giorgi}, \citenamefont {Carusotto},
  \citenamefont {Hivet}, \citenamefont {Pisanello}, \citenamefont {Sala},
  \citenamefont {Guimaraes} \emph {et~al.}}]{sanvitto2011all}%
  \BibitemOpen
  \bibfield  {author} {\bibinfo {author} {\bibfnamefont {D.}~\bibnamefont
  {Sanvitto}}, \bibinfo {author} {\bibfnamefont {S.}~\bibnamefont {Pigeon}},
  \bibinfo {author} {\bibfnamefont {A.}~\bibnamefont {Amo}}, \bibinfo {author}
  {\bibfnamefont {D.}~\bibnamefont {Ballarini}}, \bibinfo {author}
  {\bibfnamefont {M.}~\bibnamefont {De~Giorgi}}, \bibinfo {author}
  {\bibfnamefont {I.}~\bibnamefont {Carusotto}}, \bibinfo {author}
  {\bibfnamefont {R.}~\bibnamefont {Hivet}}, \bibinfo {author} {\bibfnamefont
  {F.}~\bibnamefont {Pisanello}}, \bibinfo {author} {\bibfnamefont {V.~G.}\
  \bibnamefont {Sala}}, \bibinfo {author} {\bibfnamefont {P.~S.~S.}\
  \bibnamefont {Guimaraes}},  \emph {et~al.},\ }\bibfield  {title} {\enquote
  {\bibinfo {title} {All-optical control of the quantum flow of a polariton
  condensate},}\ }\href {\doibase 10.1038/nphoton.2011.211} {\bibfield
  {journal} {\bibinfo  {journal} {Nat. Photon.}\ }\textbf {\bibinfo {volume}
  {5}},\ \bibinfo {pages} {610} (\bibinfo {year} {2011})}\BibitemShut {NoStop}%
\bibitem [{\citenamefont {Knapp}\ \emph {et~al.}(2016)\citenamefont {Knapp},
  \citenamefont {Zaletel}, \citenamefont {Liu}, \citenamefont {Cheng},
  \citenamefont {Bonderson},\ and\ \citenamefont {Nayak}}]{knapp2016nature}%
  \BibitemOpen
  \bibfield  {author} {\bibinfo {author} {\bibfnamefont {Christina}\
  \bibnamefont {Knapp}}, \bibinfo {author} {\bibfnamefont {Michael}\
  \bibnamefont {Zaletel}}, \bibinfo {author} {\bibfnamefont {Dong~E.}\
  \bibnamefont {Liu}}, \bibinfo {author} {\bibfnamefont {Meng}\ \bibnamefont
  {Cheng}}, \bibinfo {author} {\bibfnamefont {Parsa}\ \bibnamefont
  {Bonderson}}, \ and\ \bibinfo {author} {\bibfnamefont {Chetan}\ \bibnamefont
  {Nayak}},\ }\bibfield  {title} {\enquote {\bibinfo {title} {The nature and
  correction of diabatic errors in anyon braiding},}\ }\href {\doibase
  10.1103/PhysRevX.6.041003} {\bibfield  {journal} {\bibinfo  {journal} {Phys.
  Rev. X}\ }\textbf {\bibinfo {volume} {6}},\ \bibinfo {pages} {041003}
  (\bibinfo {year} {2016})}\BibitemShut {NoStop}%
\bibitem [{Note3()}]{NoteX}%
  \BibitemOpen
  \bibinfo {note} {See the Supplementary Material of Ref.~\cite
  {umucalilar2013many}}\BibitemShut {NoStop}%
\bibitem [{\citenamefont {Umucalilar}\ \emph {et~al.}()\citenamefont
  {Umucalilar}, \citenamefont {Macaluso}, \citenamefont {Comparin},\ and\
  \citenamefont {Carusotto}}]{umucalilar2017observing}%
  \BibitemOpen
  \bibfield  {author} {\bibinfo {author} {\bibfnamefont {Rifat~Onur}\
  \bibnamefont {Umucalilar}}, \bibinfo {author} {\bibfnamefont {Elia}\
  \bibnamefont {Macaluso}}, \bibinfo {author} {\bibfnamefont {Tommaso}\
  \bibnamefont {Comparin}}, \ and\ \bibinfo {author} {\bibfnamefont {Iacopo}\
  \bibnamefont {Carusotto}},\ }\bibfield  {title} {\enquote {\bibinfo {title}
  {Observing anyonic statistics via time-of-flight measurements},}\ }\href@noop
  {} {\ }\Eprint {http://arxiv.org/abs/arXiv:1712.07940} {arXiv:1712.07940}
  \BibitemShut {NoStop}%
\bibitem [{\citenamefont {Nielsen}\ and\ \citenamefont
  {Chuang}(2000)}]{quantum2000nielsen}%
  \BibitemOpen
  \bibfield  {author} {\bibinfo {author} {\bibfnamefont {Michael~A.}\
  \bibnamefont {Nielsen}}\ and\ \bibinfo {author} {\bibfnamefont {Issac~L.}\
  \bibnamefont {Chuang}},\ }\href@noop {} {\emph {\bibinfo {title} {Quantum
  Computation and Quantum Information}}}\ (\bibinfo  {publisher} {Cambridge
  University Press, Cambridge, UK},\ \bibinfo {year} {2000})\BibitemShut
  {NoStop}%
\bibitem [{\citenamefont {Mair}\ \emph {et~al.}(2001)\citenamefont {Mair},
  \citenamefont {Vaziri}, \citenamefont {Weihs},\ and\ \citenamefont
  {Zeilinger}}]{mair2001entanglement}%
  \BibitemOpen
  \bibfield  {author} {\bibinfo {author} {\bibfnamefont {Alois}\ \bibnamefont
  {Mair}}, \bibinfo {author} {\bibfnamefont {Alipasha}\ \bibnamefont {Vaziri}},
  \bibinfo {author} {\bibfnamefont {Gregor}\ \bibnamefont {Weihs}}, \ and\
  \bibinfo {author} {\bibfnamefont {Anton}\ \bibnamefont {Zeilinger}},\
  }\bibfield  {title} {\enquote {\bibinfo {title} {Entanglement of the orbital
  angular momentum states of photons},}\ }\href {\doibase 10.1038/35085529}
  {\bibfield  {journal} {\bibinfo  {journal} {Nature (London)}\ }\textbf
  {\bibinfo {volume} {412}},\ \bibinfo {pages} {313} (\bibinfo {year}
  {2001})}\BibitemShut {NoStop}%
\end{thebibliography}

%

\end{document}